\documentclass[twocolumns,bibyear]{aa} 
\usepackage{graphicx,epstopdf}
\usepackage{url}
\usepackage{txfonts}
\usepackage{amsmath}
\def\Q{\ifmmode\mathcal{Q}\else$\mathcal{Q}$\fi}

\begin{document} 

   \title{The structure and characteristic scales of the \ion{H}{i} gas in galactic disks}

   \author{Sami Dib\inst{1}, Jonathan Braine\inst{2}, Maheswar Gopinathan\inst{3}, Maritza A. Lara-L\'{o}pez\inst{4}, Valery V. Kravtsov\inst{5}, Archana Soam\inst{6}, Ekta Sharma\inst{3,7}, Svitlana Zhukovska\inst{8}, Charles Aouad\inst{9}, Jos\'{e} Antonio Belinch\'{o}n\inst{10}, George Helou\inst{11}, Di Li\inst{12,13,14}}
 
   \institute{Max Planck Institute for Astronomy, K\"{o}nigstuhl 17, 69117, Heidelberg, Germany\\
      \email{sami.dib@gmail.com}
      \and
              Laboratoire d'Astrophysique de Bordeaux, Universit\'{e} de Bordeaux, CNRS, B18N, all\'{e}e Geoffroy Saint-Hilaire, 33615, Pessac, France
     \and
            Indian Institute of Astrophysics (IIA), Sarjapur Road, Koramangala, Bangalore 560034, India
       \and        
             Armagh Observatory and Planetarium, College Hill, Armagh BT61 9DG, United Kingdom 
       \and       
             Sternberg Astronomical Institute, Lomonosov Moscow State University, University Avenue 13, 119899 Moscow, Russia 
       \and 
              SOFIA Science Center, Universities Space Research Association, NASA Ames Research Center, M.S. N232-12, Moffett Field, CA 94035, USA
       \and 
              Department of Physics and Astrophysics, University of Delhi, Delhi 110007, India     
       \and       
              School of Physics and Astronomy, University of Exeter, Stocker Road, Exeter EX4 4QL, United Kingdom   
       \and 
               Astrophysics Research Institute, Liverpool John Moores University, IC2, Liverpool Science Park, 146 Brownlow Hill, Liverpool L3 5RF, United Kingdom                               
       \and 
               Departamento de Matem\'{a}ticas, Universidad de Atacama Av. Copayapu 485, Copiap\'{o}, Atacama, Chile
       \and 
               IPAC, California Institute of Technology, Pasadena, CA, 91125, USA       
       \and
               CAS Key Laboratory of FAST, National Astronomical Observatories, Chinese Academy of Sciences, Beijing 100101, People's Republic of China 
      \and         
               University of Chinese Academy of Sciences, Beijing 100049, People's Republic of China 
       \and 
                NAOC-UKZN Computational Astrophysics Centre, University of KwaZulu-Natal, Durban 4000, South Africa                          
               }
          
\authorrunning{Dib et al.}
\titlerunning{Structure of the \ion{H}{i} gas in galactic disks}
         
 
\abstract{The spatial distribution of the \ion{H}{i} gas in galactic disks holds important clues about the physical processes that shape the structure and dynamics of the interstellar medium (ISM). The structure of the ISM could be affected by a variety of perturbations internal and external to the galaxy, and the unique signature of each of these perturbations could be visible in the structure of interstellar gas. In this work, we quantify the structure of the \ion{H}{i} gas in a sample of 33 nearby galaxies taken from the HI Nearby Galaxy Survey (THINGS) using the delta-variance ($\Delta$-variance) spectrum. The THINGS galaxies display a large diversity in their spectra, but there are a number of recurrent features. In many galaxies, we observe a bump in the spectrum on scales of a few to several hundred parsec. We find the characteristic scales associated with the bump to be correlated with the galactic star formation rate (SFR) for values of the SFR $\gtrsim 0.5$ M$_{\odot}$ yr$^{-1}$ and also with the median size of the \ion{H}{i} shells detected in these galaxies. We interpret this characteristic scale as being associated with the effects of feedback from supernova explosions. On larger scales, we observe in most galaxies two self-similar, scale-free regimes. The first regime, on intermediate scales ($\lesssim 0.5 R_{25}$), is shallow, and the power law that describes this regime has an exponent in the range [0.1-1] with a mean value of $0.55$ that is compatible with the density field that is generated by supersonic turbulence in the cold phase of the \ion{H}{i} gas. The second power law is steeper, with a range of exponents between 0.5 and 2.3 and a mean value of $\approx 1.5$. These values are associated with subsonic to transonic turbulence, which is characteristic of the warm phase of the \ion{H}{i} gas. The spatial scale at which the transition between the two self-similar regimes occurs is found to be $\approx 0.5 R_{25}$ , which is very similar to the size of the molecular disk in the THINGS galaxies. Overall, our results suggest that on scales $\lesssim 0.5 R_{25}$, the structure of the ISM is affected by the effects of supernova explosions. On larger scales ($\gtrsim 0.5 R_{25}$), stellar feedback has no significant impact, and the structure of the ISM is determined by large-scale processes that govern the dynamics of the gas in the warm neutral medium, such as the flaring of the \ion{H}{i} disk at large galactocentric radii and the effects of ram pressure stripping.}


 \maketitle

%

\section{Introduction}\label{introduction}

The interstellar medium (ISM) in the Milky Way and in external galaxies exhibits a scale-free nature that can extend over many physical scales. This is observed for the neutral \ion{H}{i} gas (e.g., Green 1993; Stanimirovi\'{c} 1999; Elmegreen 2001; Dickey et al. 2001; Dib 2005; Dib et al. 2005; Begum et al. 2006;, Zhang et al. 2012; Dutta et al. 2013, Martin et al. 2015; Nandankumar \& Dutta 2020) and in the molecular phase (e.g., Stutzki et al. 1998; Heyer \& Brunt 2004; Dib et al. 2008; Elia et al. 2018; Dib \& Henning 2019; Yahia et al. 2021). This self-similarity is also observed in the spatial distribution of young clusters in galactic disks (e.g., Elmegreen et al. 2006; Gouliermis et al. 2017; Grasha et al. 2019).

It is well established that turbulence, which is ubiquitously observed in all phases of the gas, is one of the primary regulators of the ISM structure and dynamics of local disk galaxies. It is therefore responsible for setting the self-similar behavior of many of the physical quantities that are used to describe it (e.g., Elmegreen \& Scalo 2004). In the warm ($T\approx 10^{4}$ K) neutral medium (WNM), turbulence is transonic or possibly subsonic, while in the cold ($T\approx 100$K) neutral medium (CNM), it is supersonic. Turbulent motions in the WNM and CNM phases can be sustained by a variety of instabilities and energy and momentum injection mechanisms, both internal and external to the galaxy. The spatial scales associated with the fastest-growing modes of these instabilities and those associated with direct energy and momentum injection mechanisms can break the self-similarity of the gas. Some of these scales could be detected as characteristic scales in the ISM (Dib et al. 2009; Eden et al. 2020; Dib et al. 2020). Internal processes include stellar feedback from massive stars, that is, ionizing radiation, radiation pressure, stellar winds, and supernova explosions, which impart significant amounts of energy and momentum to the ISM on intermediate scales, that is, $\approx 50-1000$ pc (e.g., Heiles 1979, Ehlerova \& Palous 1996; de Avillez \& Breitschwerdt 2005; Dib et al. 2006; Hodge \& Deshpande 2006; Shetty \& Ostriker 2008; Dib et al. 2011,2013; Gent et al. 2013; Agertz et al. 2013; Hony et al. 2015; Suad et al. 2019; Chamandy \& Shukurov 2020; Pokhrel et al. 2020, Bacchini et al. 2020). Large-scale gravitational instabilities due to the combined action of gas and stars (Jog \& Solomon 1984; Elmegreen 2011; Shadmehri \& Khajenabi 2012; Dib et al. 2017; Marchuk 2018; Marchuk \& Sotnikova 2018) can also drive turbulence in galactic disks. Other internal mechanisms of the galaxy that can perturb the self-similar nature of the gas and shape its spatial structure include stellar spiral density waves (e.g., Lin \& Shu 1966; Guibert 1974;  Adler \& Westpfahl 1996; Tosaki et al. 2007; Khoperskov \& Bertin 2015; Wang et al. 2015), the Parker instability (e.g., Parker 1967; Franco et al. 2002; Hanasz \& Lesch 2003; Rodrigues et al. 2016; Mouschovias et al. 2009; Heintz et al. 2020), and the impact of high-velocity clouds on the galactic disk (e.g., Santill\'{a}n et al. 1999; Boomsma et al. 2008; Heitsch \& Putman 2009; Park et al. 2016). External mechanisms can also impart energy and momentum to the gas on galactic scales. They include ram pressure stripping (e.g., Clemens et al. 2000; Marcolini et al. 2003; Vollmer et al. 2004; Freeland et al. 2010; Steyrleithner et al. 2020) and tidal stripping in interacting systems (e.g., Combes et al. 1988; Marziani et al. 2003; Mayer et al. 2006; Holwerda et al. 2013; Lipnicki et al. 2018; Fattahi et al. 2018). In galaxy mergers, galactic disks can experience strong compressions due to tides, and these compressions can significantly affect the structure and dynamical properties of the gas in the interacting galactic disks (e.g., Renaud et al. 2009).  

In this paper, we quantify the structure of the \ion{H}{i} gas for a number of nearby galaxies using the $\Delta$-variance spectrum (Stutzki et al. 1998; Ossenkopf et al. 2008). The $\Delta$-variance spectrum is another expression of the power spectrum, and it has been employed successfully to characterize the self-similar structure of the gas as well as to uncover the existence of characteristic scales (e.g., Elmegreen et al. 2001; Dib et al. 2020). In addition to quantifying the structure of the \ion{H}{i} gas, we aim to relate features that are observed in the $\Delta$-variance spectra to physical processes that may affect the spatial distribution of the gas. In \S.~\ref{obsdata} we summarize the sample of galaxies we used in this work, which are taken from the THINGS survey of nearby galaxies (Walter et al. 2008). The $\Delta$-variance method is discussed in \S.~\ref{deltavar}, and its application to the THINGS maps is presented in \S.~\ref{results}. In \S.~\ref{interpret} we interpret our results using simple models and results from a cosmological zoom-in simulation of a star-forming disk galaxy. We also explore the correlations that exist between characteristic scales detected in the $\Delta$-variance spectrum of the galaxies and their star formation rate (SFR). In \S.~\ref{discussion} we discuss our results and compare them to previous work, in particular, to results obtained using the identification of \ion{H}{i} shells in the THINGS sample of galaxies. In \S.~\ref{conclusions} we discuss our results and conclude. 

\begin{figure*}
\centering
\includegraphics[width=0.27\textwidth]{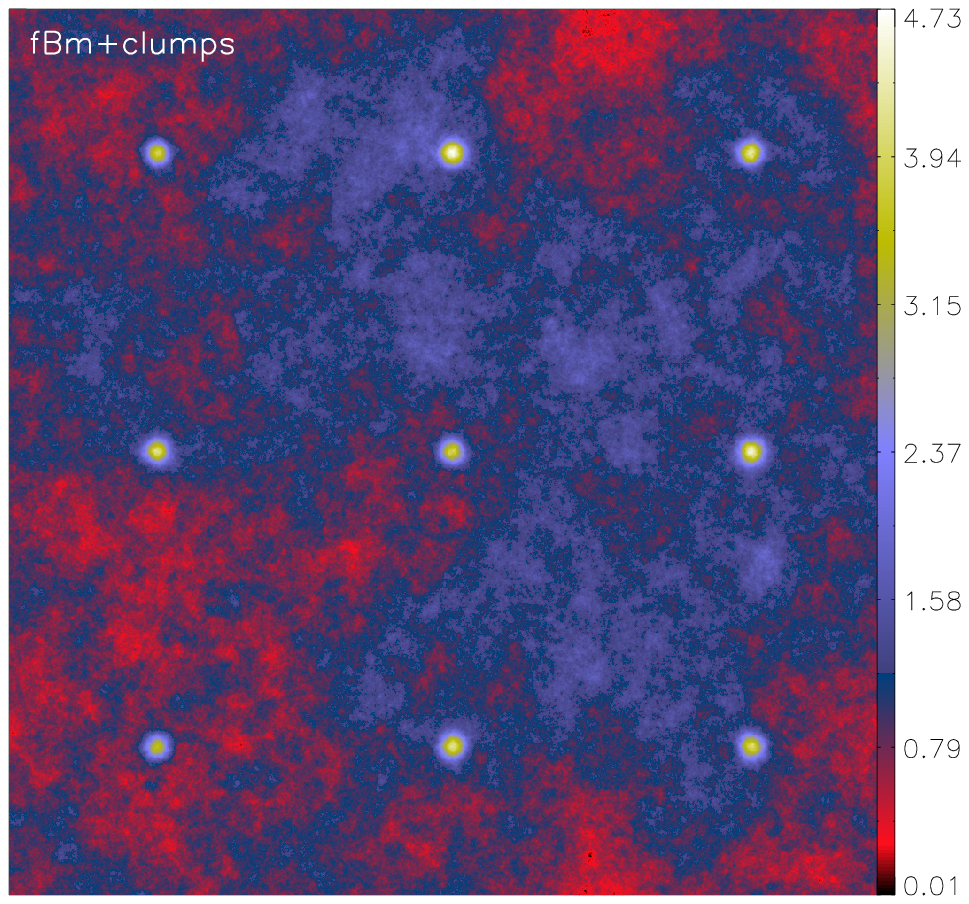}
\hspace{0.3cm}
\includegraphics[width=0.27\textwidth]{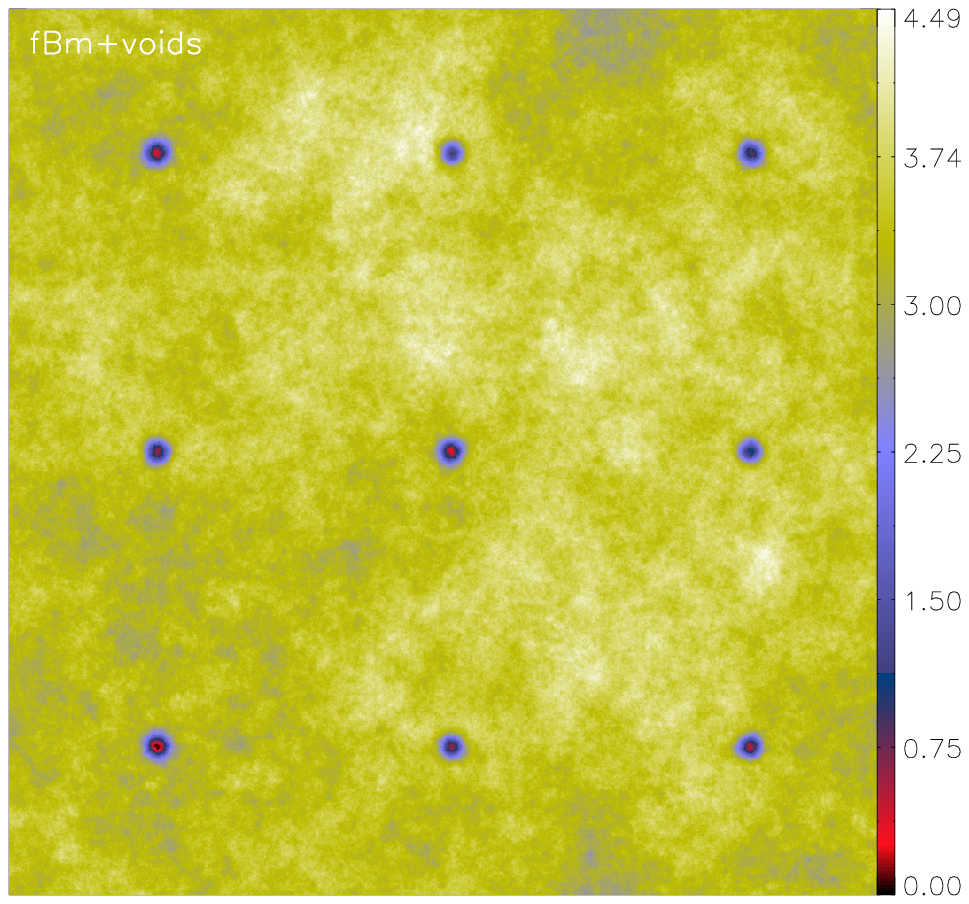}
\hspace{0.3cm}
\includegraphics[width=0.27\textwidth]{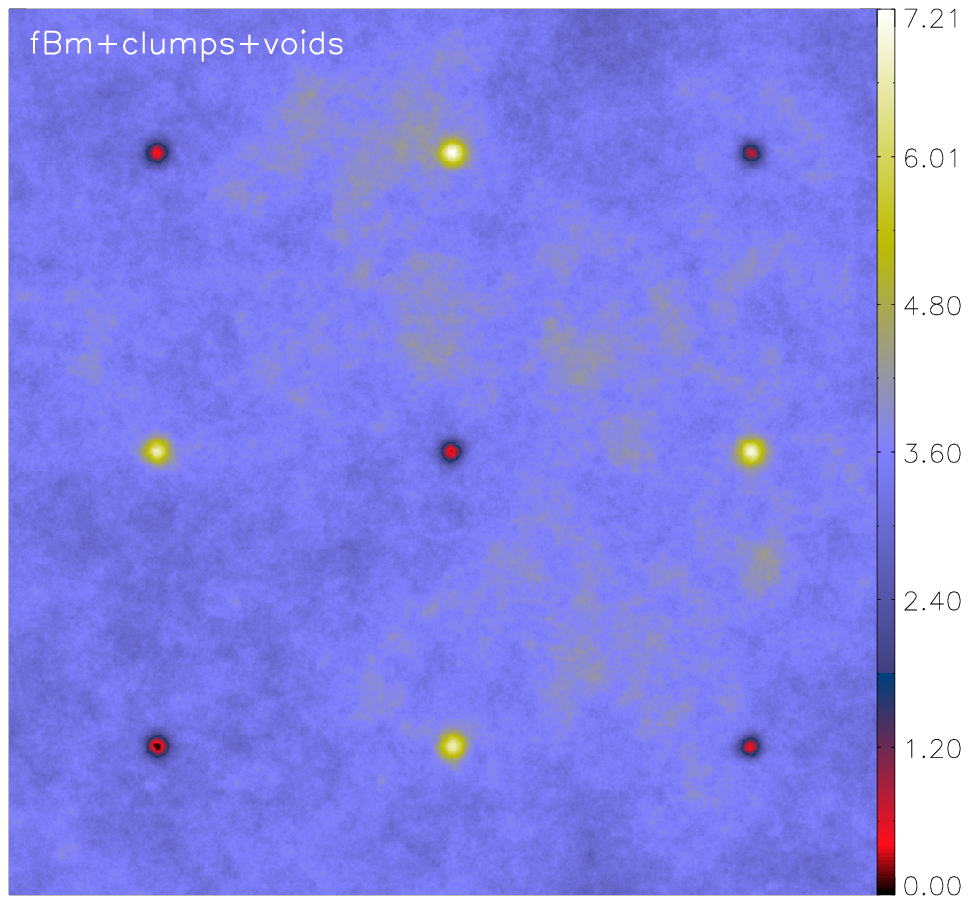}\\
\vspace{0.3cm}
\includegraphics[width=0.27\textwidth]{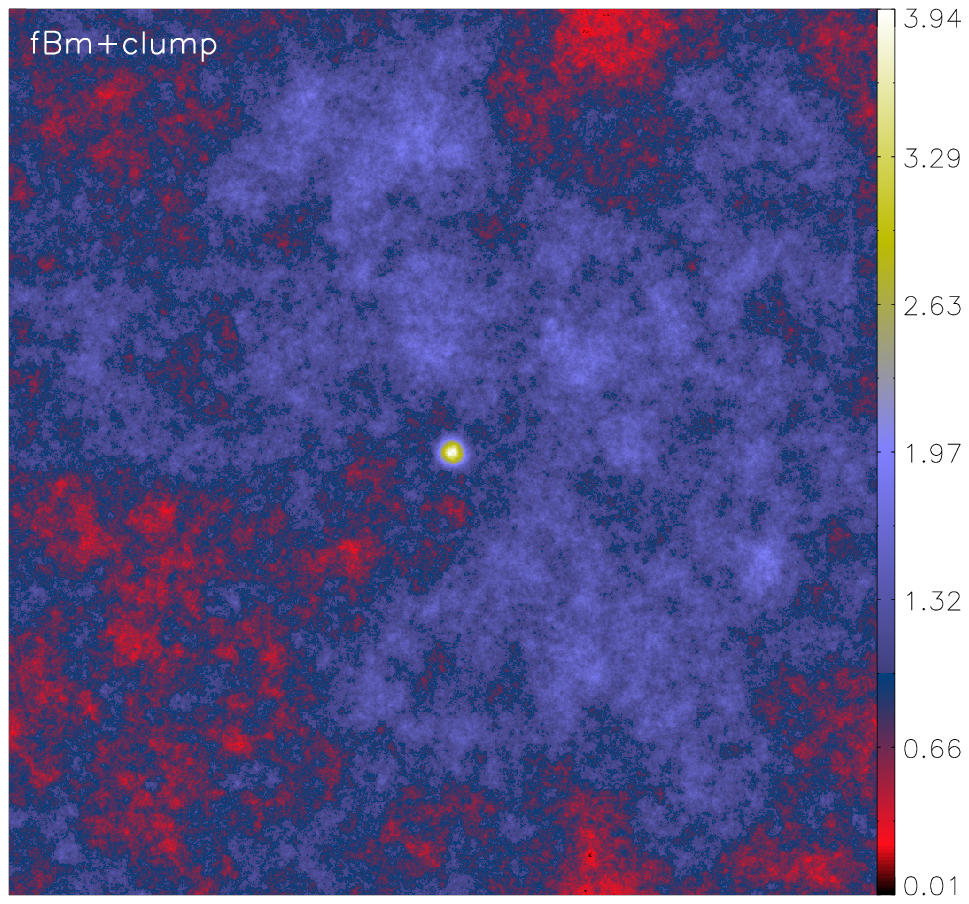}
\hspace{0.3cm}
\includegraphics[width=0.27\textwidth]{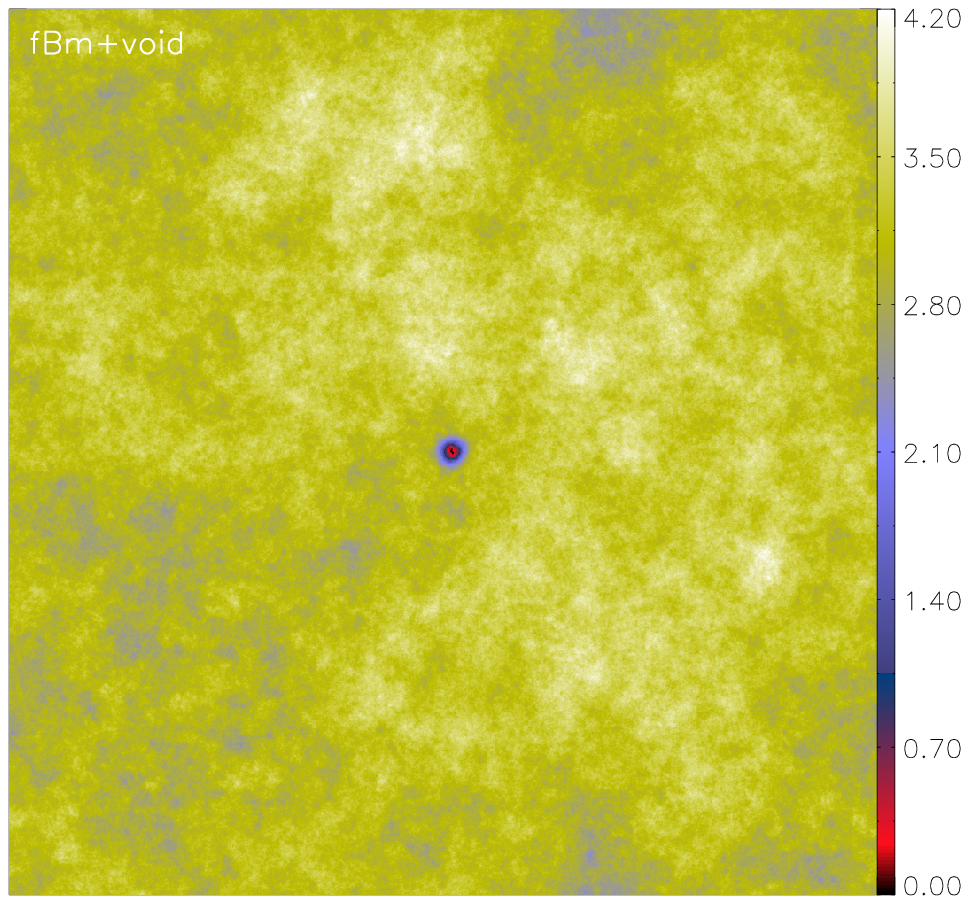}
\hspace{0.3cm}
\includegraphics[width=0.27\textwidth]{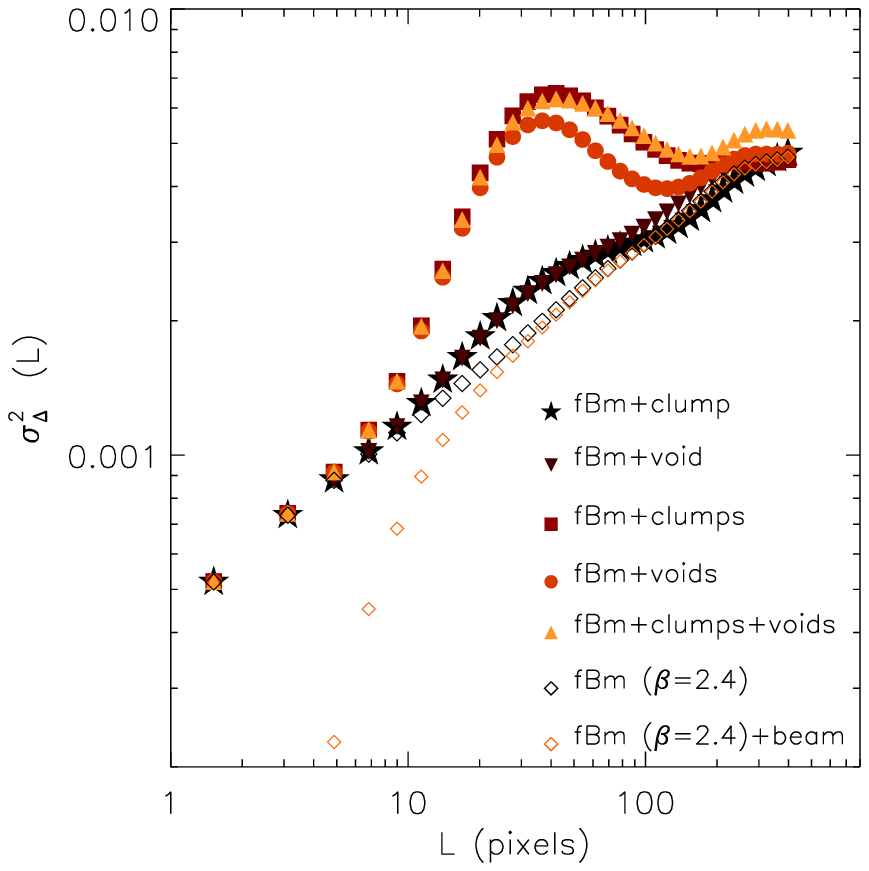}
\caption{2D Gaussian structures injected on top of an fBm image with $\beta= 2.4$. he fBms are shifted to positive values by adding an arbitrary constant and normalized by their mean values. The maps have a resolution of $1000\times1000$ pixels. The 2D Gaussian functions all have an aspect ratio ($f=\sigma_{1}/\sigma_{2}=1$) and a contrast between the peak of the Gaussian and the mean value in the map of $\delta_{c}=3$. The standard deviations of the 2D Gaussians are $\sigma_{1}=\sigma_{2}=10$ pixels. All maps are normalized to their mean value. The maps correspond to the case of a single 2D Gaussian (bottom left), a number of 2D Gaussians (top left), an inverted 2D Gaussian (bottom mid), a number of inverted 2D Gaussians (top middle), and a mix of 2D Gaussians and inverted 2D Gaussians (top right). The corresponding $\Delta$-variance functions calculated for each case are displayed in the bottom right subpanel, and they are compared to the $\Delta$-variance function of the underlying fBm image as well to the case of the same fBm smoothed with a Gaussian beam whose FWHM is 6 pixels.}
\label{fig1}
\end{figure*}

\section{Data: The HI Nearby Galaxy Survey }\label{obsdata}

We used the moment-0 (integrated intensity) \ion{H}{i} maps from The \ion{H}{i} Nearby Galaxy Survey (THINGS; Walter et al. 2008)\footnote{https://www2.mpia-hd.mpg.de/THINGS/Overview.html}. THINGS is a homogeneous survey in the 21 cm \ion{H}{i} emission line for 34 nearby galaxies. The observations, performed with the NRAO Very Large Array (VLA), have an angular resolution of $\approx 6\arcsec$. At the distances of these galaxies ($D_{gal} \approx 2-15$ Mpc), this corresponds to spatial resolutions of a few to several hundred parsecs. The galaxies were mapped with various configurations, and the integrated \ion{H}{i} intensity maps have a total $1024\times1024$ or $2048\times2048$ pixels. Each pixel represents an angular size of $1\arcsec$ to $1.5\arcsec$, depending on the galaxy. The sample of galaxies spans a range of morphological types, metallicities, total \ion{H}{i} mass, and star formation rates extending from low-mass, metal-poor, only weakly star-forming dwarf galaxies to metal-rich massive spiral galaxies with high star formation rates. Galaxies in the THINGS survey have a wide range of inclinations, and it is imperative to correct for the effect of inclination in order to minimize the projection effects. We deprojected all galaxies using the inclinations measured by de Blok et al. (2008) using the \ion{H}{i} data alone ($i_\ion{H}{i}$). For the few galaxies (NGC 1569, NGC 3077, and NGC 4449) for which no such measurement is reported in de Blok et al. (2008), we used values of the inclination of the optical disk that are reported in the LEDA database (Paturel et al. 2003). For the position angles ($PA$) needed to deproject the maps, we used the values listed in Walter et al. (2008). The adopted inclinations are listed in Tab.~\ref{tab1}. For NGC 3031, we removed the very central nuclear \ion{H}{i} ring (the $50\times50$ inner pixels) from the original data. This ring dominates the signal. We also discarded the first $600$ pixels in each direction as they are affected by edge effects. 

\begin{figure*}
\centering
\includegraphics[width=0.85\textwidth]{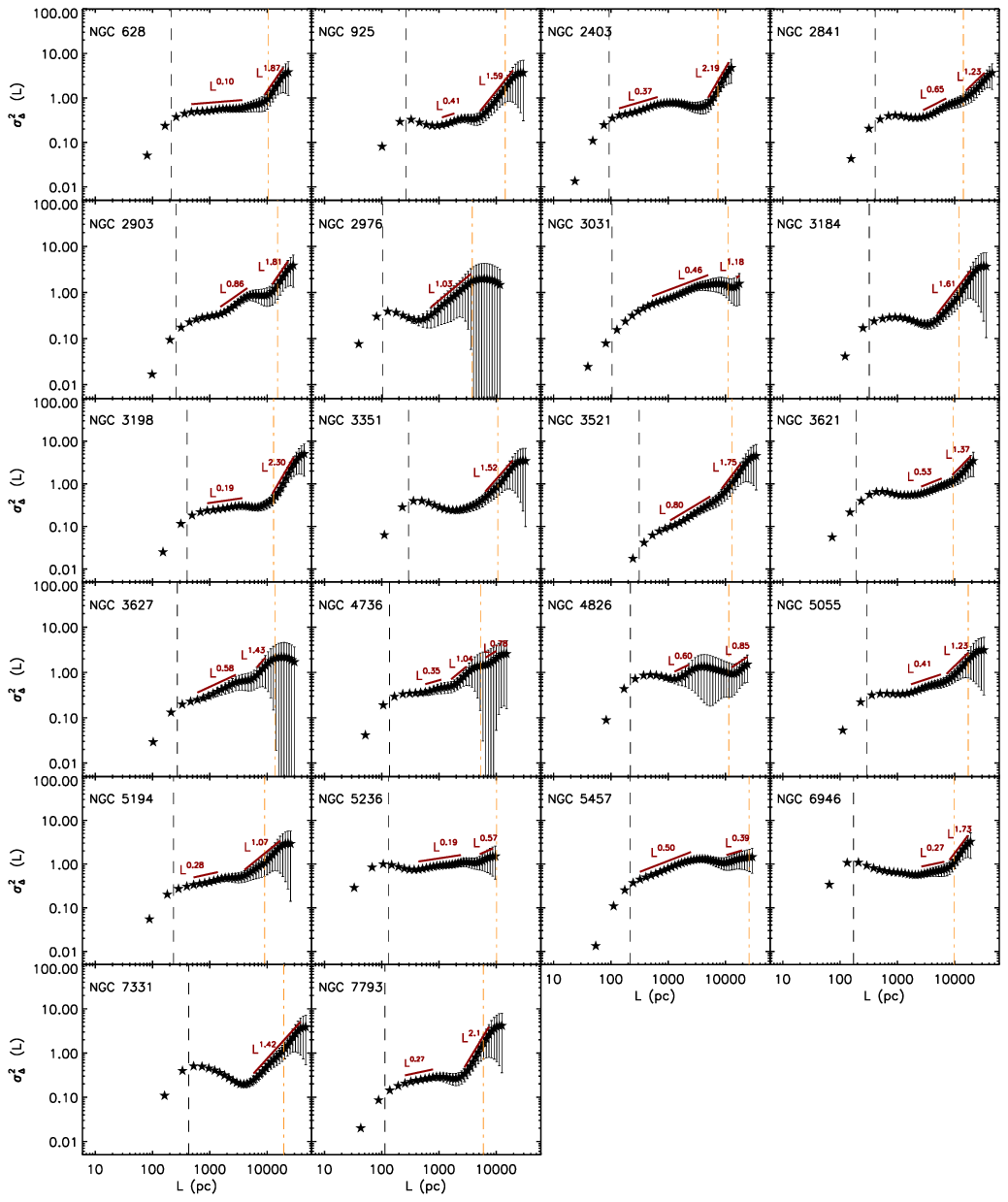}
\vspace{0.8cm}
\caption{Delta-variance spectra for the galaxies classified as spirals in the THINGS sample. The vertical dashed black line in each subpanel indicates the spatial resolution for each galaxy, and the vertical dash-dotted orange line corresponds to the optical radius of the galaxy. The values of $R_{25}$ are taken from Walter et al. (2008). The spectra are normalized by their respective mean values.}
\label{fig2}
\end{figure*}

\begin{figure*}
\centering
\includegraphics[width=0.85\textwidth]{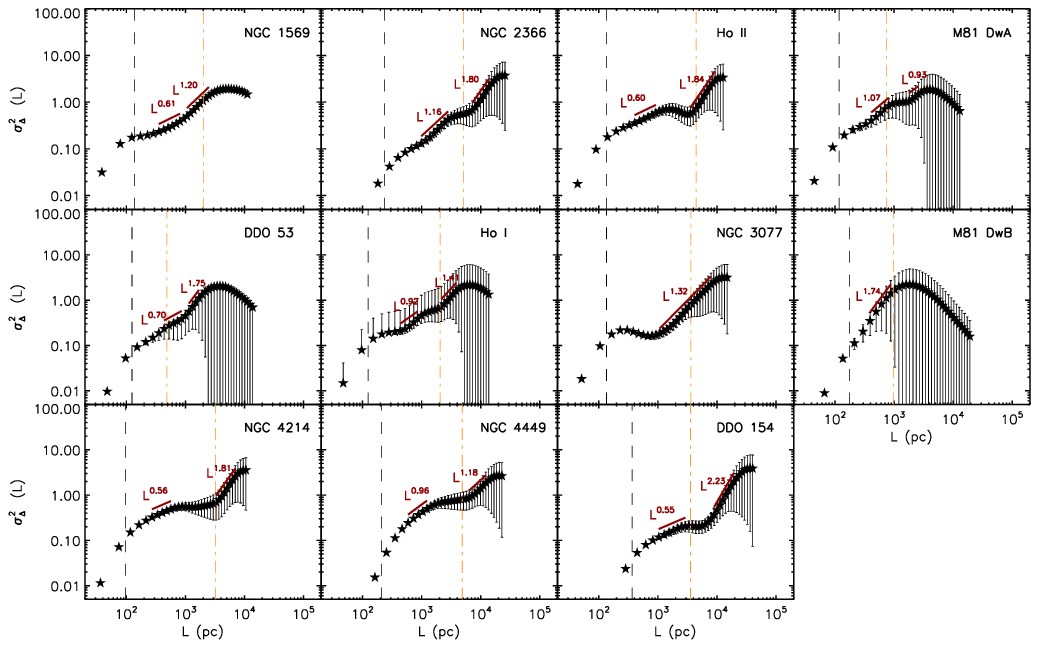}
\vspace{0.8cm}
\caption{Delta-variance spectra for the galaxies classified as dwarfs in the THINGS sample. The vertical dashed black line in each subpanel indicates the spatial resolution for each galaxy, and the vertical dash-dotted orange line corresponds to the optical radius of the galaxy. The values of $R_{25}$ are adopted from Walter et al. (2008). The spectra are normalized by their respective mean values.}
\label{fig3}
\end{figure*}

\section{Method: $\Delta$ Variance spectrum}\label{deltavar}

In order to quantify the structure of the \ion{H}{i} gas, we used the $\Delta$-variance spectrum method, originally introduced in Stutzki et al. (1998) and Zielinsky et al. (1999). In this work, we used an improved version of the method presented in Ossenkopf et al. (2008)\footnote{The IDL package for calculating the $\Delta$-variance can be found at \url{https://hera.ph1.uni-koeln.de/~ossk/Myself/deltavariance.html}}. For a two-dimensional field  $A(x,y)$, the $\Delta$-variance on a scale $L$ is defined as being the variance of the convolution of $A$ with a filter function $\sun_{L}$ , such that

\begin{equation}
\sigma_{\Delta}^{2}(L)=\frac{1}{2\pi} \langle (A * \sun_{L})^{2}  \rangle_{x,y}.
\label{eq1}
\end{equation}

For the filter function, Ossenkopf et al. (2008) recommend the use of a Mexican hat function, which is defined as

\begin{equation}
\sun_{L} \left(r\right)= \frac{4}{\pi L^{2}} e^{\frac{r^{2}} {(L/2)^{2}}} - \frac{4}{\pi L^{2} (v^{2} -1)} \left[ e^{\frac{r^{2}}{(vL/2)^{2}}} -e^{\frac{r^{2}}{(L/2)^{2}}}\right],
\label{eq2}
\end{equation}

where the two terms on the right side of Eq.~\ref{eq2} represent the core and the annulus of the Mexican-hat function, respectively, and $v$ is the ratio of their diameters (we used a value of $v=1.5$). For a faster and more efficient computation of Eq.~\ref{eq1}, Ossenkopf et al. (2008) performed the calculation as a multiplication in Fourier space, and thus, the $\Delta$-variance is given by

\begin{equation}
\sigma_{\Delta}^{2}(L)=\frac{1}{2 \pi} \int \int P \left| \bar{\sun}_{L} \right |^{2} dk_{x} dk_{y},
\label{eq3}
\end{equation}  

where $P$ is the power spectrum of $A$, and $\bar{\sun}_{L}$ is the Fourier transform of the filter function. If $\beta$ is the exponent of the power spectrum, then a relation exists between the slope of the $\Delta$-variance and $\beta$ (St\"{u}tzki et al. 1998). This is given by

\begin{equation}
\sigma_{\Delta}^{2}(L) \propto L^{\alpha} \propto L^{\beta-2}
\label{eq4}
.\end{equation}

The slope of the $\Delta$-variance can be inferred from the range of spatial scales over which it displays a self-similar behavior. It can be tied to the value of $\beta$. Characteristic scales are scales at which there are breaks of the self-similarity and that show up in the $\Delta$-variance plots as break points or inflection points. The error bars of the $\Delta$-variance are computed from the counting error determined by the finite number of statistically independent measurements in a filtered map and the variance of the variances, that is, the fourth moment of the filtered map. The $\Delta$-variance has been employed to analyze the structure of observed molecular clouds (e.g., Bensch et al. 2001; Sun et al. 2006; Rowles \& Froebrich 2011; Elia et al. 2014; Dib et al. 2020) as well as simulated molecular clouds (e.g., Ossenkopf et al. 2001; Bertram et al. 2015). Elmegreen et al. (2001) in the only work that used the $\Delta$-variance spectrum to characterize the structure of the \ion{H}{i} gas. However, their study was limited to the Large Magellanic Cloud (LMC). 

\begin{figure}
\centering
\includegraphics[width=\columnwidth,]{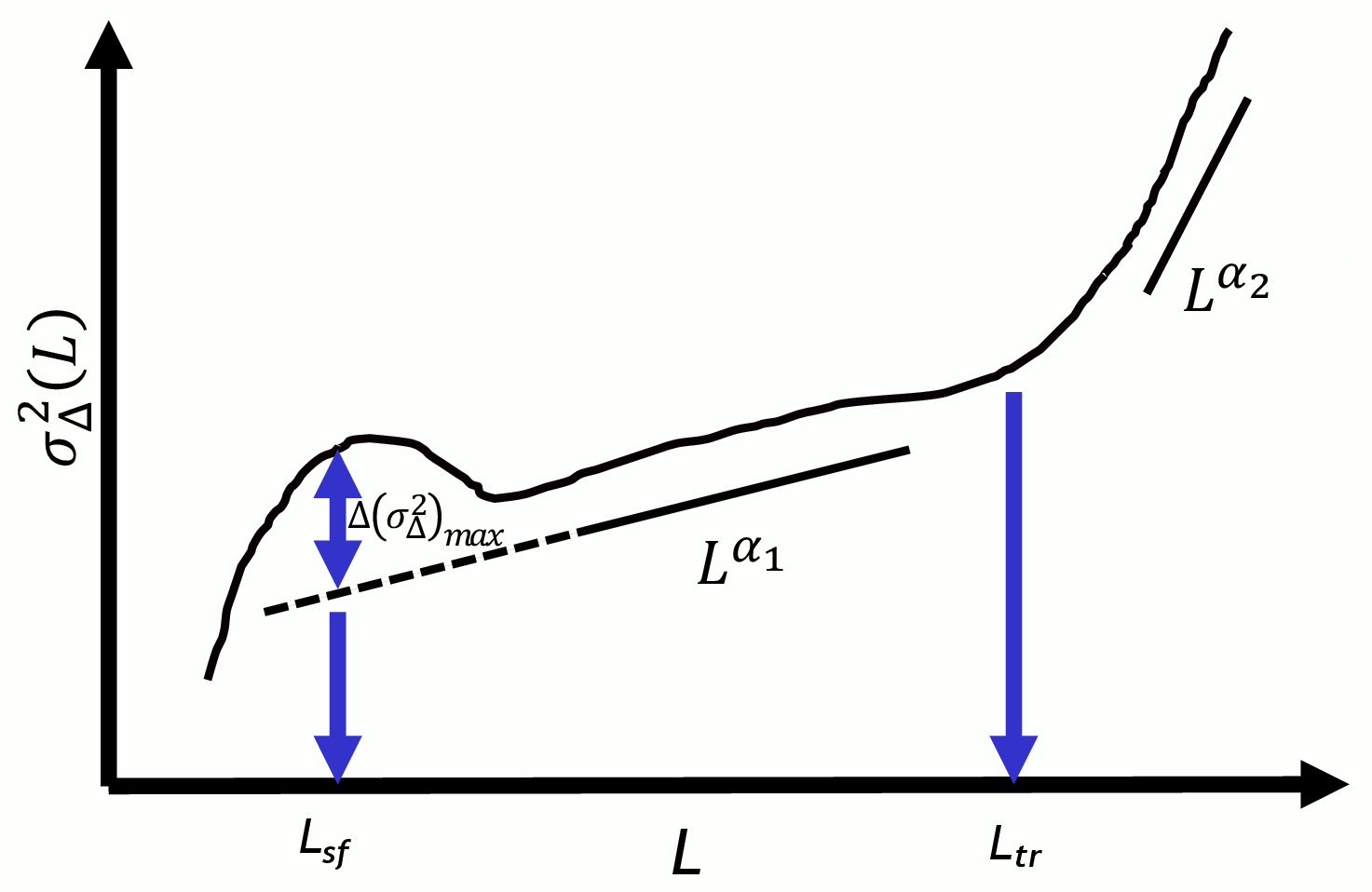}
\caption{Schematic figure representing the shape of the $\Delta$-variance spectrum for the THINGS galaxies. In some galaxies, some of the features of the spectrum such as the bump at small scales or the presence of two distinct power laws are not observed. The dashed line represents the extrapolation of the first power law down to smaller scales. The quantity $\Delta\left(\sigma_{\Delta}^{2}\right)$ represents the maximum deviation between the bump and the extrapolated power law. The value of $L_{sf}$ represents the physical scale at which this maximum deviation occurs. As illustrated, this scales does not correspond to the peak of the bump and is generally smaller. The quantity $L_{tr}$ is the physical scales at which a transition is observed between the first and second power law.}
\label{fig4}
\end{figure}

\begin{figure*}
\centering
\includegraphics[width=0.49\textwidth]{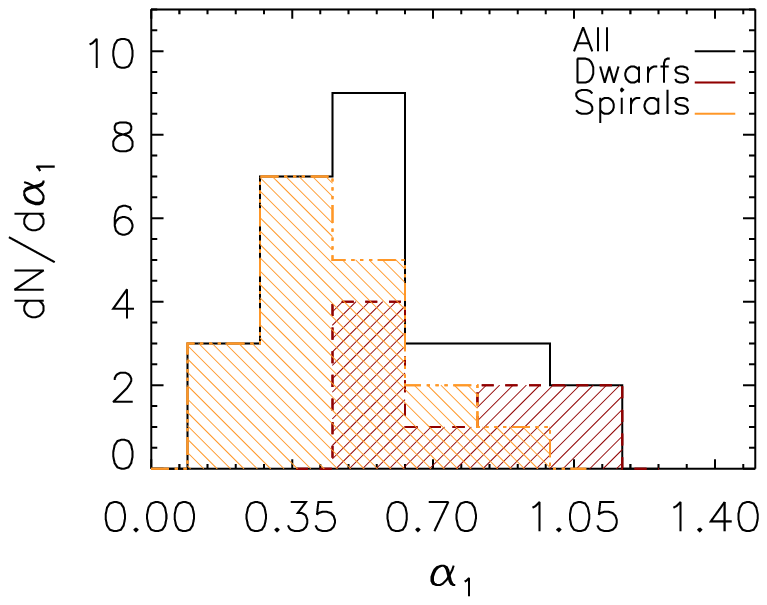}
\includegraphics[width=0.49\textwidth]{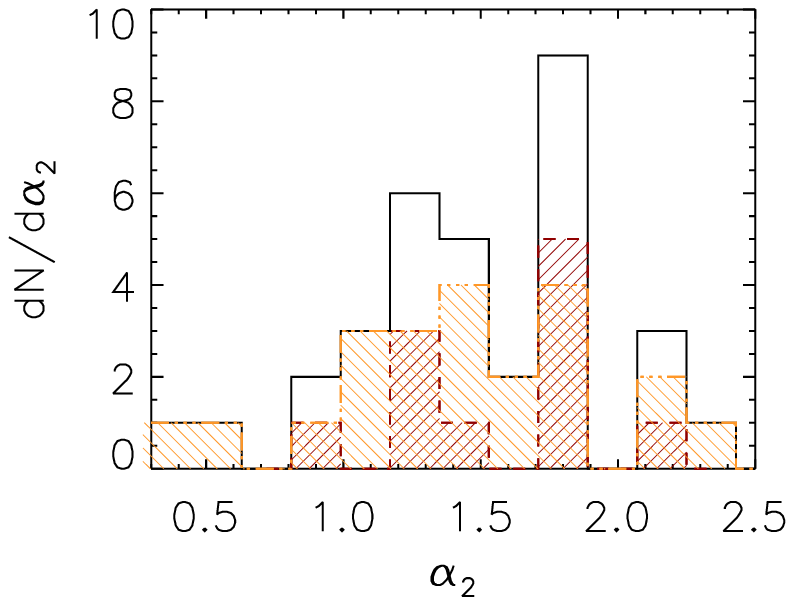}\\
\includegraphics[width=0.49\textwidth]{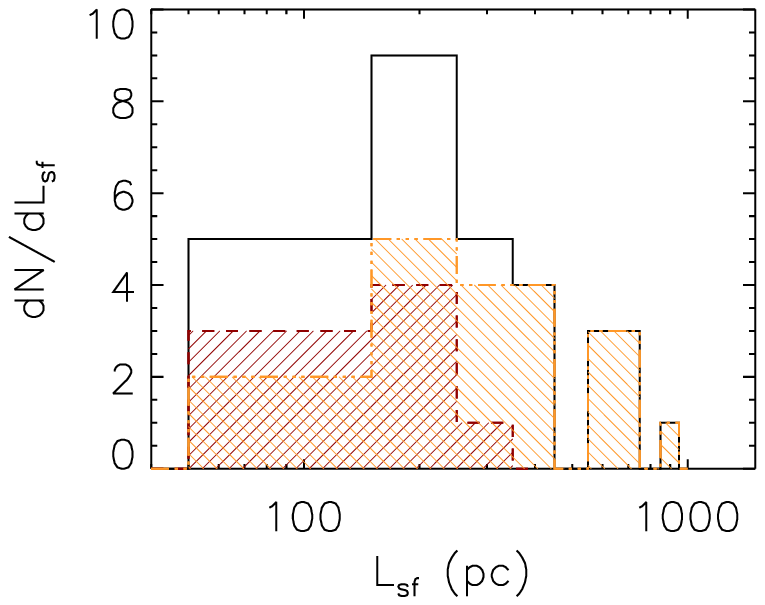}
\includegraphics[width=0.49\textwidth]{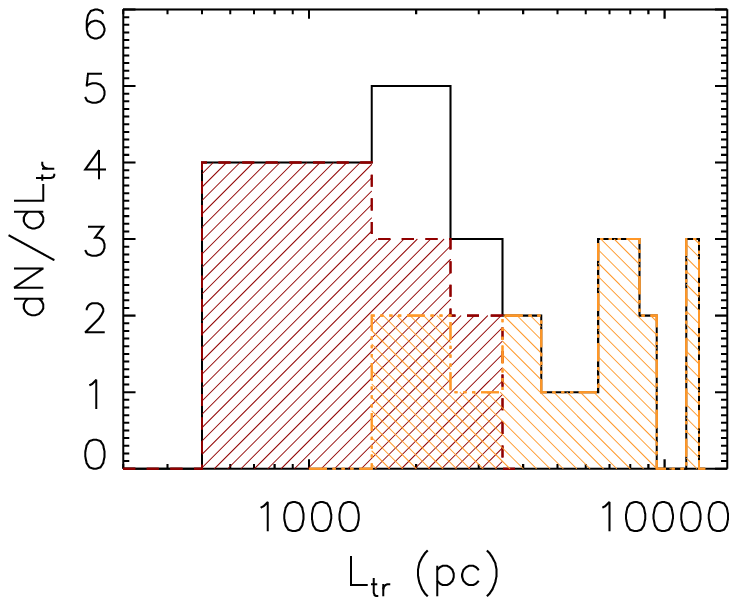}
\caption{Distribution of the characteristic scales and exponents of the self-similar regimes extracted from the the $\Delta$-variance spectra of the THINGS galaxies. The subpanels display the distribution of the characteristic scale $L_{sf}$ (bottom left), the distribution of the transition point between the two self-similar regimes, $L_{tr}$ (bottom right), the distribution of the exponent of the first power law found on intermediate spatial scales (top left), and the distribution of the exponent of the second power law found at large spatial scales (top right).}
\label{fig5}
\end{figure*}

Coherent (i.e., nonhierarchical) structures in a larger self-similar medium generate a bump in the $\Delta$-variance spectrum. This is true regardless of whether the structure is an overdensity (i.e., a clump in a column density map) or a region of low column density (i.e., a hole or void). The reason is that the $\Delta$-variance measures the variance of an image over a given scale. This issue was discussed in detail in Dib et al. (2020). Here, we show a limited number of examples for completeness. Figure \ref{fig1} displays five realizations of a fractal Brownian motion (fBm) image with an exponent of $\beta=2.4$, on which we superimpose a single clump (bottom left), a number of identical clumps (top left), a single void (bottom middle), a number of identical voids (top middle), and a mixture of clumps and voids (top right). The size of each image is $1000\times1000$ pixels, and the clumps and voids are represented by 2D generalized Gaussian functions and inverted 2D Gaussians, respectively. In the examples displayed in Fig.~\ref{fig1}, the standard deviation of the Gaussian functions in each direction is 10 pixels. The $\Delta$-variance spectra for all cases are displayed in the bottom right subpanel of Fig.~\ref{fig1}. Additionally, we show the $\Delta$-variance spectrum of the pure fBm image. In the latter case, the $\Delta$-variance spectrum is a power law with an exponent $\alpha=\beta-2=0.4$. As demonstrated in Dib et al. (2020), coherent structures in a self-similar medium increase the $\Delta$-variance on all spatial scales. However, the point of maximum deviation from the spectrum of the underlying fBm (i.e, the peak of the bump) occurs on scales that are $\approx 4\sqrt{\sigma_{1} \sigma_{2}}$ as this is where most of the signal of the 2D Gaussian lies and where $\sigma_{1}$ and $\sigma_{2}$ are the standard deviations in both directions. It is important to mention here that the scale at which the maximum deviation ($\Delta(\sigma_{\Delta}^{2})_{max}$) occurs between the $\Delta$-variance spectrum in the presence of added structure and the spectrum of the underlying fBm does not necessarily correspond to the position of the peak, and it is generally smaller (see the schematic representation in Fig.~\ref{fig4}). 

The results discussed so far relate to the case of a pure fBm image and to cases in which discrete coherent structures are overlaid on the fBm image. We have also calculated a case of a pure fBm (with $\beta=2.4$) in which the image was smeared with a Gaussian beam whose full width at half maximum (FWHM) is $D_{beam}=6$ pixels. The $\Delta$-variance spectrum corresponding to this case is also displayed in Fig.~\ref{fig1} (lower right subpanel). The effect of the reduced resolution is to cause a depression in the $\Delta$-variance spectrum on scales $\lesssim 2 D_{beam}$. The effects of beam smearing can extend to scales larger than $D_{beam}$. However, on scales $\gtrsim 1.5 D_{beam}$, the effects of the beam do not exceed a few tens down to a few percent on larger scales. This effect has been presented and discussed in earlier studies using the $\Delta$-variance (e.g., Bensch et al. 2001; Dib et al. 2020). The important aspect of this is that the effects of beam smearing do not generate any spurious bump or other features in the $\Delta$-variance spectrum. 

\begin{figure}
\centering
\includegraphics[width=\columnwidth]{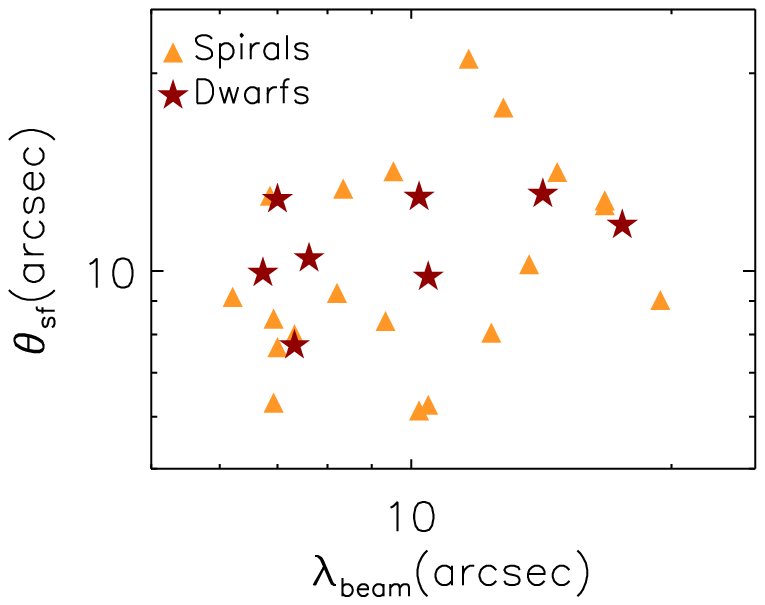}\\
\includegraphics[width=\columnwidth]{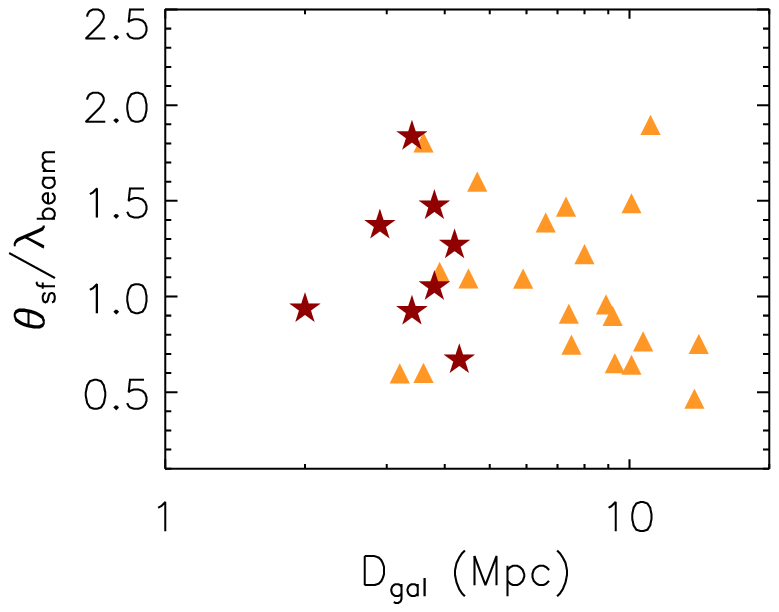}
\caption{Angular size of $L_{sf}$ plotted against the projection-corrected angular beam size $\lambda_{beam}$ (top subpanel) and the ratio of these two quantities plotted against the distance of the galaxy (bottom subpanel).}
\label{fig6}
\end{figure}  

\section{Results}\label{results}

\begin{table*}
\begin{center}
\caption{Inclinations and characteristics of the THINGS galaxy $\Delta$-variance spectra.}
\begin{tabular}{lcccccccccc}
\hline
\hline
galaxy      &  $i$~($^{\circ}$) & $\alpha_{1}$   & $\sigma_{\alpha_{1}}$ & $\alpha_{2}$   & $\sigma_{\alpha_{2}}$   &  range $\alpha_{1}$ (kpc) & range $\alpha_{2}$ (kpc) &    $L_{sf}$ (pc)   &  $\sigma_{L_{sf}} (pc)$ & $L_{tr}$ (kpc)     \\ 
\hline
         NGC 628  &  15   &    0.10     &  0.006   &     1.87    &   0.054     &    [0.4,4]         &     [9,20]      &  323 & 51  & 7 \\
         NGC 925   &  50  &    0.41     &  0.033   &     1.59    &   0.028     &    [1,2]            &     [5,20]      &  374  & 100 &  4 \\
         NGC 2403  & 55 &    0.37     &  0.015   &     2.19    &   0.055     &    [0.12,0.7]    &     [5,12]      &  97     & 9 &  4 \\
         NGC 2841   & 69  &    0.65     &  0.031   &     1.23    &   0.053     &    [2.5,8]         &     [15,35]    &  860  & 164 & 12 \\
         NGC 2903   & 66 &    0.86     &  0.025   &      1.81   &   0.033     &    [1.5,4.5]      &     [12,25]    &  610  & 79 &  9 \\
         NGC 2976    &  54 &    -     &   -   &    1.03       &     0.006            &     -      &      [0.7,4.5]         &  107  & 12 & - \\
         NGC 3031    &  59 &     0.46    &  0.010   &     1.18    &    0.057    &     [0.5,5]        &      [15,25]    &  367 & 24 &  12  \\
         NGC 3184    & 29 &      -         &      -       &     1.61    &   0.044     &      -               &      [4.5,20]   &  700  & 80 & - \\
         NGC 3198    & 72 &    0.19     &  0.006   &     2.30    &   0.066      &    [0.8,4]        &     [12,30]     &  604 & 62 & 7.5 \\
         NGC 3351    & 39 &     -          &     -       &      1.52   &    0.047     &         -            &      [6,20]       &  243  & 22  & - \\
         NGC 3521    & 69 &    0.80     &   0.015  &     1.75    &    0.053     &    [1,6]           &     [8,20]       &  664  & 48 & 7 \\
         NGC 3621    & 62 &    0.53     &   0.015  &     1.37    &    0.061     &    [2.5,6.5]     &     [8.5,20]    &  567  & 245 & 7.5 \\
         NGC 3627    &  61 &    0.58     &   0.011  &      1.43    &  0.011       &    [0.5,3]       &      [6,10]       &  363 & 45 &  7 \\
         NGC 4736     & 44 &    0.35     &   0.011  &      1.04/0.78 &     0.075/0.079            &      [0.5,1.2]  \& [1.5,3]   &  [6,10]   &  304 & 36  & 2 \\
         NGC 4826     & 64 &    0.60     &   0.081  &      0.85    &   0.085     &    [1.2,2.5]    &       [12,25]     &  372 & 41 & 12 \\
         NGC 5055     & 51 &    0.41     &   0.018  &      1.23    &   0.047     &    [1.5,6]       &       [6.5,18]    &  694 & 120 & 6 \\
         NGC 5194    & 30 &    0.28     &   0.016  &      1.07    &   0.037     &    [0.4,1.5]    &       [4,18]       &  328  & 45 & 3.3 \\
         NGC 5236     & 31 &    0.19     &   0.007  &       0.57   &   0.025     &    [0.4,2.5]     &      [5,9]          &  167 & 31 & 4.5  \\
         NGC 5457     & 30 &    0.50     &   0.006  &       0.39   &   0.014     &    [0.25,2.5]   &      [10,20]      &  226  & 18 & 9 \\
         NGC 6946     & 35 &    0.27     &   0.005  &       1.73   &   0.039     &    [2.5,7]         &     [7.5,18]     &  229 & 42 & 8 \\
         NGC 7331     & 77 &      -         &   -         &        1.42  &   0.024      &      -               &     [5.5,40]     &  270  & 72 & - \\
         NGC 7793     & 43 &     0.27    &   0.008  &       2.14   &   0.033     &    [0.25,0.85]  &    [2.5,8]       & 175  & 17 & 1.8 \\         
 \hline 
         NGC 1569    & 55 &      0.61   &     0.029 &       1.20   &   0.014     &  [0.2,0.5]        &     [0.6,1.5]    & 95      & 13 & 0.5 \\
         NGC 2366     & 65 &     1.16   &     0.029 &       1.80   &   0.054     &  [0.4,1.2]        &     [3,6]          & 216   & 18 & 1.5\\
         Ho II              & 31 &     0.60   &     0.013  &      1.84   &   0.055      &  [0.3,0.9]        &     [3,8]          & 212   & 20 & 3 \\
         M81A            & 27 &      1.07   &     0.015 &       0.93   &  0.053      &  [0.35,0.8]      &     [1.5,2.5]     &  -      & - & 1.2 \\
         DDO 53        & 33  &      0.70  &     0.030  &      1.75   &   0.019      &  [0.3,0.8]       &      [0.9,1.5]    & -        & - & 0.9 \\
         Ho I               & 27 &      0.92  &      0.084 &       1.41  &   0.028      &  [0.35,0.8]       &     [1.8,3.5]   & 183   & 19 & 1 \\
         NGC 3077     & 35  &       -     &        -       &        1.32  &   0.019      &      -                &     [0.8,7]      & 142   & 18 & - \\
         M81 B           & 28 &         -     &       -       &        1.74   &    0.067     &  -                    &    [0.3,0.8]   &  -       &  - & - \\
         NGC 4214    & 38 &       0.56  &      0.010 &       1.81  &   0.054      &  [0.2,0.5]        &     [2.5,6]      & 147  &  13 & 1.5 \\
         NGC 4449    & 54 &      0.96  &      0.024 &       1.18  &   0.055      &  [0.3,0.8]        &     [3.5,8]      & 264  & 14 & 3.2 \\
         DDO 154      & 70 &      0.55  &      0.017 &       2.23  &   0.022      &  [0.3-1]           &     [3,7]         & 245  & 20 & 1.8 \\
\hline 
\end{tabular}
\tablefoot{Columns represent the (1) name of the galaxy, (2) adopted inclination, (3) value of the exponent of the first power low $\alpha_{1}$, (4) $1\sigma$ uncertainty on $\alpha_{1}$, (5) value of the exponent of the second power law $\alpha_{2}$, (6) 1$\sigma$ uncertainty on $\alpha_{2}$, (7) spatial range over which the first power law was fitted, (8) spatial range over which the second power law was fitted, (9) position of the characteristic scales, $L_{sf}$, (10) $1\sigma$ uncertainty on $L_{sf}$, and the (11) position of the transition between the first and second power laws $L_{tr}$. The upper and lower groups of galaxies are the spirals and dwarfs, respectively.}
\label{tab1}
\end{center}
\end{table*}

We calculated the $\Delta$-variance spectrum for all galaxies in the THINGS sample that are available in the online database (33 galaxies) after deprojecting them. The results are displayed in Fig.~\ref{fig2} and Fig.~\ref{fig3} for spiral and dwarf galaxies, respectively. The $\Delta$-variance spectra for the spiral and dwarf galaxies exhibit a variety of features. However, some features are the same in many galaxies. In some galaxies, a bump in the spectrum is observed on scales of a few to several hundred parsec ($\approx 100-850$ pc). With the exception of NGC 2077, this feature is more commonly observed in spiral galaxies than in dwarfs. A second feature that can be observed in the $\Delta$-variance spectra of the majority of THINGS galaxies in Fig.~\ref{fig2} and Fig.~\ref{fig3} is the existence of two self-similar regimes where the $\Delta$-variance can be described by two power laws, $\sigma_{\Delta}^{2} \propto L^{\alpha_{1}}$ on intermediate spatial scales (i.e., one to several kpc) and $\sigma_{\Delta}^{2} \propto L^{\alpha_{2}}$ on larger scales (i.e., a few to several kpc). Figure~\ref{fig4} shows a schematic sketch of the common features that are observed in the $\Delta$-variance spectra of the THINGS galaxies. We determined the boundaries of each self-similar regime by visual inspection and avoided any overlap with other features of the spectra (i.e., bumps, dips, and inflection points). We also used a different approach in which we computed the first-order derivative of the spectrum in order to evaluate in which range it is constant. We find that there is no particular advantage in following this approach because the value of the slope is never exactly a constant, and the range at which the spectrum starts to deviate from a power law is also not unambiguously determined. We find that the visual inspection with a careful selection of the ranges on which the spectrum is assumed to be a power law is as accurate as an automated selection.

We fit the self-similar regimes and determined for each galaxy the values of $\alpha_{1}$ and $\alpha_{2}$. In a few galaxies (i.e., NGC 2976, NGC 3184, NGC 3351, and NGC 7331), only one self-similar regime can be identified, and given its steepness and the involved spatial ranges, we categorized it as being described by the second power-law function, whose exponent is $\alpha_{2}$. Following the logic described in \S.~\ref{deltavar}, we determined the position at which the maximum deviation occurs, $L_{sf}$, using the following procedure: We extrapolated the power law that comes after the bump down to spatial scales where the bump is observed, and we measured the difference between the observed spectrum and the extrapolation of the first power law on scales where the bump is located (see Fig.~\ref{fig4}). The upper limit for the scales that were considered in this subtraction is the lower limit of where the first power law is assumed to be valid. The difference in the spectra was then fit with a Gaussian function, and the position of the peak of the Gaussian function was assumed to be the position where the maximum deviation occurs. In all cases, we found that the value of $L_{sf}$ is lower than the position of the peak of the bump. A note of caution is probably due. By performing the extrapolation of the first power law down to physical scales where the bump is observed and by subtracting it from the observed spectra on those scales, we assumed that the first power is the underlying slope of the spectra if there are no discrete coherent structures in the \ion{H}{i} on these scales. This is not entirely guaranteed, and there is a possibility that the process that generated the bump in the spectrum (both position and width) can also affect the value of the first power-law slope in the spectra. On the other hand, it is important to stress that the values of $L_{sf}$ measured in this way are very close to the position of the bump in each spectrum. This implies that any misassumption on what the true slope of the spectrum might be at small spatial scales has a very minor effect on our results and conclusions. We also point out that the wing on the left-hand side of the bump is likely to be affected by the resolution of the observations, and as a consequence, both the amplitude and width of the bump are reduced. However, this should affect the position of the bump only marginally (see the application to the simulated galaxy in \S.~\ref{simul}). Finally, the position of the transition point between the first and second power law (when present), $L_{tr}$, is estimated by eye, and given the high uncertainties that affect the spectra at these large scales, there is little ground to expect that any automated procedure will yield more accurate estimates. In some galaxies, this transition appears as an inflection point (e.g., NGC 3521 and NGC 3621), whereas in others, a dip can be observed (e.g., NGC 7793 and Ho II). We defined the value of $L_{tr}$ as being either the position of the inflection point or the deepest position in the dip, when present. A conservative estimate of the uncertainty on $L_{tr}$ is about 10$\%$. The values of $L_{sf}$, $\alpha_{1}$, $\alpha_{2}$, and $L_{tr}$ for all of the THINGS galaxies are reported in Tab.~\ref{tab1}, along with the uncertainties measured for $L_{sf}$, $\alpha_{1}$ , and $\alpha_{2}$ and the spatial ranges over which every power-law fit was performed.         

Most \ion{H}{i} holes are unlikely to be circular as they are affected by local inhomogeneities in the local velocity and density field and by the effects of galactic shear (Dib et al. 2006, Bagetakos et al. 2011, Ohlin et al. 2019, Aouad et al. 2020), and thus, $L_{sf}$ is a measure of the effective radius. Figure ~\ref{fig5} (bottom left panel) displays the distribution of $L_{sf}$ ($dN/dL_{sf}$) for the spiral and dwarf galaxies and for the combined sample. In dwarf galaxies, $L_{sf} \lesssim 250$ pc, while in spiral galaxies, $L_{sf}$ can be as large as $\approx 850$ pc. The distributions of $\alpha_{1}$ and $\alpha_{2}$ ($\left(dN/d\alpha_{1}\right)$ and $\left(dN/d\alpha_{2}\right)$, respectively) are displayed in Fig.~\ref{fig5} (top subpanels). The mean values of $\alpha_{1}$ are $0.79\pm0.23$, $0.43\pm0.20$, and $0.55\pm0.27$ for the dwarfs, spirals, and for the entire sample, respectively. The $\Delta$-variance spectra are steeper on large spatial scales, and the mean values of $\alpha_{2}$ are $1.56\pm0.38$, $1.42\pm0.04$, and $1.47\pm0.46$ for the dwarfs, spirals, and the entire sample, respectively. The distributions of the transition scale between the first and second power laws in the spectra, ($dN/dL_{tr}$), are displayed in the bottom right subpanel of Fig.~\ref{fig5}. The distribution of $L_{tr}$ in dwarf galaxies peaks at $\approx 1$ kpc and extends to $\approx 3$ kpc. For spirals, the distribution of $L_{tr}$ is broader, with values that fall in the range of 2 to 12 kpc (Tab.~\ref{fig1}).

As discussed above and in \S.~\ref{deltavar}, the shape of the $\Delta$-variance spectrum might be affected by resolution effects on sizes of about the beam size and smaller. In most galaxies, the peak of the bump and its right wing are well resolved, whereas the left wing is more affected by the resolution of the observations. We have shown that resolution issues cannot cause the occurrence of a bump similar to the one that is observed in many galaxies in Fig.~\ref{fig2} and Fig.~\ref{fig3}. Here, we test further the effects of resolution by comparing the angular size of $L_{sf}$ ($\theta_{SF}$) with the inclination-corrected angular beam size, $\lambda_{beam}$. Similarly to what was presented in Li et al. (2021), we calculated these two quantities as being

\begin{equation}
\theta_{sf}=\frac{L_{sf}}{D_{gal}},
\label{eq5}
\end{equation}

and 

\begin{equation}
\lambda_{beam} = \frac{6\arcsec}{{\rm cos}(i)},
\end{equation}

where $6\arcsec$ is the angular resolution of the beam, $D_{gal}$ is the distance of the galaxy, and $i$ is the inclination angle. The values of $D_{gal}$ were taken from Walter et al. (2008) and the adopted values of $i$ are those described in \S.~\ref{obsdata}. The top subpanel in Fig.~\ref{fig6} indicates that there is no obvious correlation between $\theta_{sf}$ and $\lambda_{beam}$ , and the bottom subpanel shows that the ratio $\left(\theta_{sf}/\lambda_{beam}\right)$ is not dependent on the distance to the galaxy. The Pearson correlation coefficient for the points in the ($\theta_{sf}-\lambda_{beam}$) scatter plot is $P\approx-0.0015$. This clearly indicates that there is no linear correlation between these two quantities and that the determination of $L_{sf}$ is largely unaffected by the beam size. Furthermore, the ratio $\left(\theta_{sf}/\lambda_{beam}\right)$ is not constant and varies by a factor of $\approx 4$ at given distance. This rules out that $L_{sf}$ could be an artifact of the data reduction that would be affecting the THINGS data on small spatial scales.

\section{Interpretation}\label{interpret}
\begin{figure*}
\centering
\includegraphics[width=0.26\textwidth]{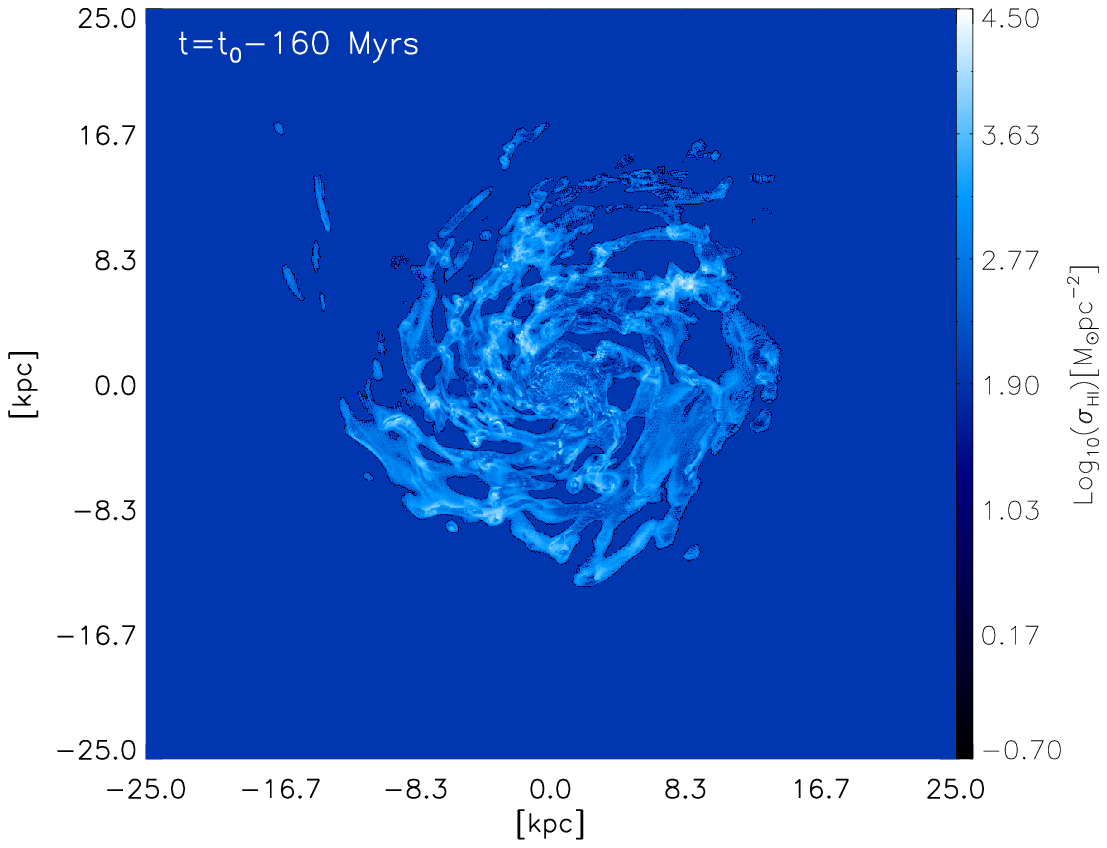}
\hspace{1.5cm}
\includegraphics[width=0.26\textwidth]{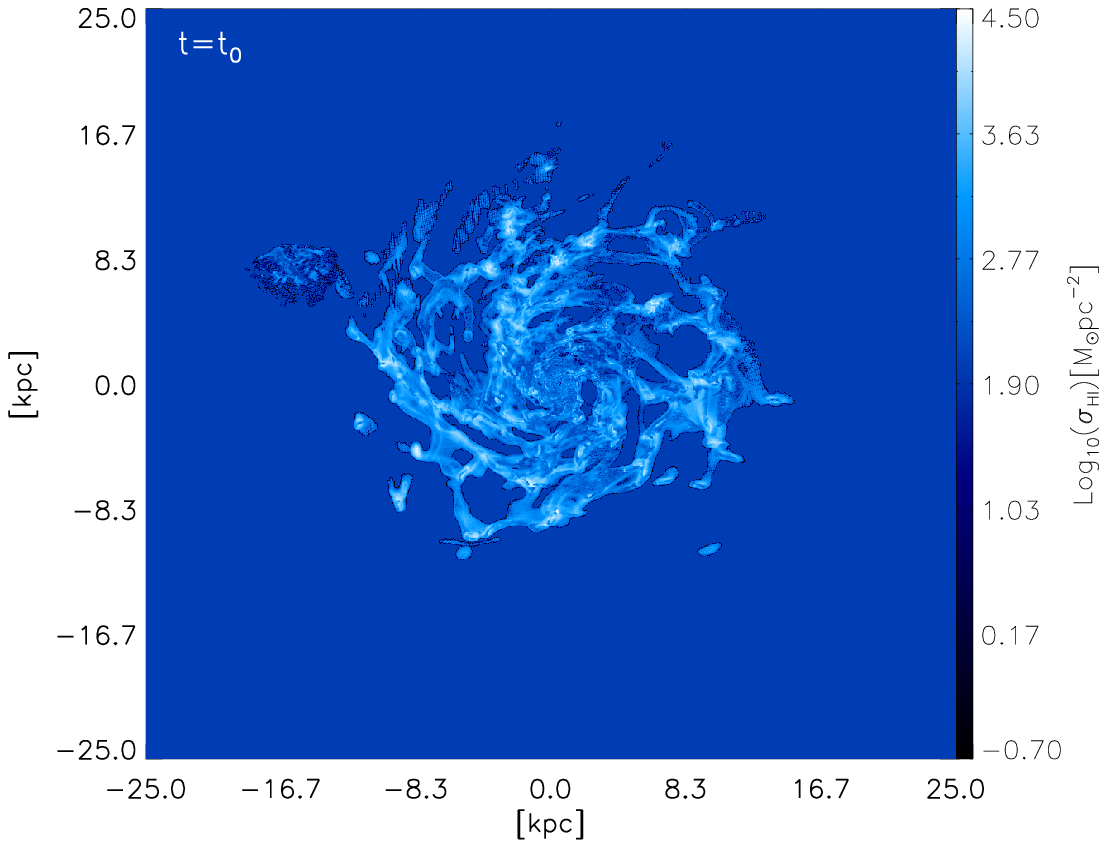}
\hspace{1.5cm}
\includegraphics[width=0.26\textwidth]{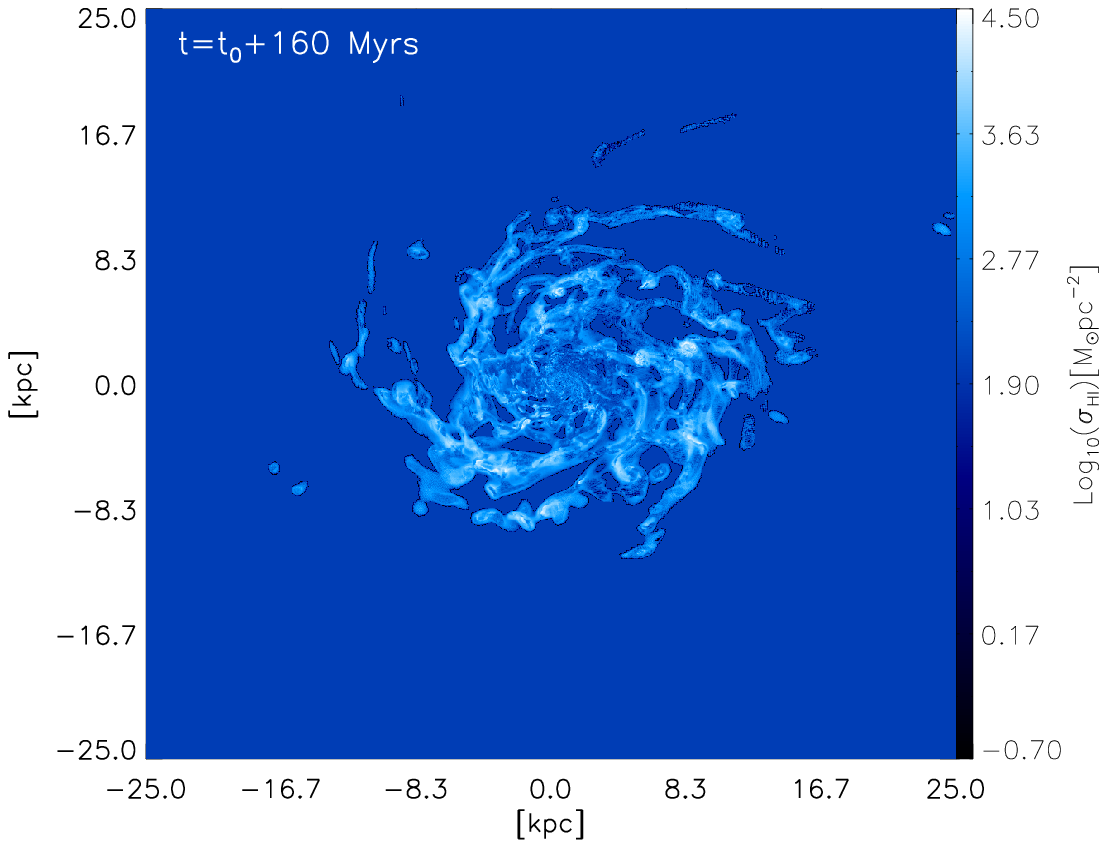}
\vspace{0.5cm}
\caption{Face-on view of the \ion{H}{I} surface density of the simulated galaxy at $t_{0}$ (middle panel, corresponding to redshift $z\approx 0.3$), $t_{0}$-160 Myr (left panel), and $t_{0}+160$ Myr (right panel).}
\label{fig7}
\end{figure*}

\begin{figure}
\centering
\includegraphics[width=\columnwidth]{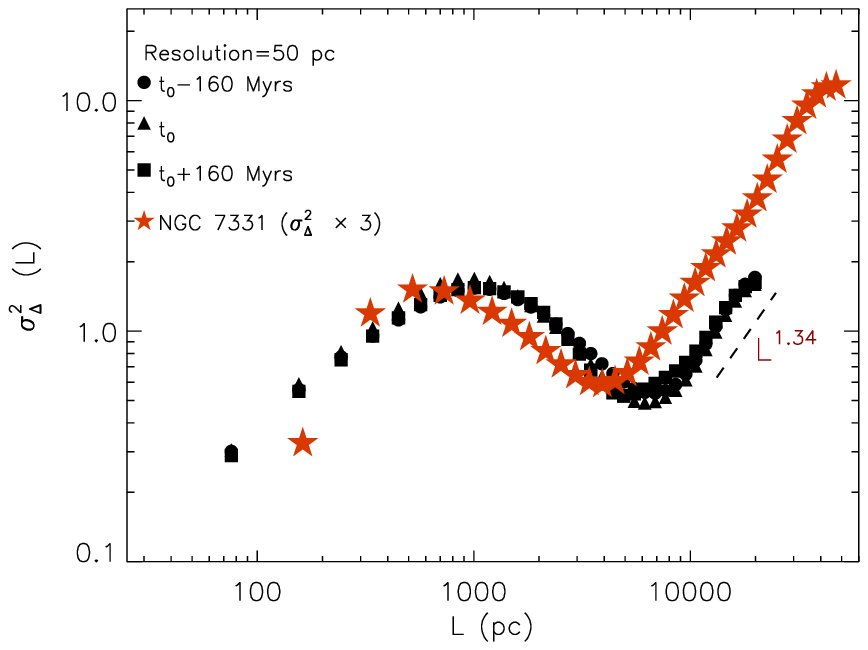}
\caption{$\Delta$-variance spectra of the simulated galaxy at different epochs. The spectra are calculated using the full resolution of the grid, which is 50 pc. The spectra are all normalized to their respective mean values. The models are compared to the case of the galaxy NGC 7331, where the $\Delta$-variance of NGC 7331 has been multiplied by a factor of 3. The dashed line shows a fit to the $\Delta$-variance spectrum of the models in the scale range $[10-20]$ kpc and at $t=t_{0}$.}
\label{fig8}
\end{figure}  

The bumps in the $\Delta$-variance spectra that are observed in most spiral galaxies and in some of the dwarf galaxies at small spatial scales (i.e., a few to several hundred parsec; see Fig.~\ref{fig2} and Fig.~\ref{fig3}) might be due to large \ion{H}{i} complexes and to \ion{H}{i} holes that are created either by feedback from massive stars or by other mechanisms, such as large-scale thermal and gravitational instabilities (Kim et al. 1999; Dib \& Burkert 2004,2005; Silich et al. 2006; Weisz et al. 2009; Bagetakos et al. 2011, Cannon et al. 2012). While \ion{H}{i} complexes may make a certain contribution to the bump, it is unlikely that they are the main source of the signal that is observed at these scales. This is simply because \ion{H} {i} complexes are themselves self-similar in nature and are a consequence of the large-scale turbulence cascade. On the other hand, supernova remnants and bubbles are filled mostly with hot rarefied gas and are devoid of any significant \ion{H}{i} gas emission (e.g., Walter et al. 2008, Bagetakos et al. 2011). Thus, \ion{H}{i} holes are more similar to the coherent structures described in \S.~\ref{deltavar}, and their sizes (or distribution of sizes) can have a direct imprint on the shape of the $\Delta$-variance spectrum. The bump can be described by three quantities: its amplitude, the position of the point of maximum deviation from the underlying power law ($L_{sf}$), and its width. The scale at which the bump joins the first power law is related to the \ion{H}{i} hole separation, as shown in Dib et al. (2020) and in Fig.~\ref{fig1}. Both the amplitude of the bump and the value of $L_{sf}$ could be related to the SFR, which given a certain IMF, sets the frequency of type II supernova explosions in the disks. However, given that the first power-law slope might itself be perturbed by the existence of the bump, the value of $L_{sf}$ is likely to have a tighter correlation with the SFR than the bump amplitude. 

In order to validate our results and gain more insight into the features we observe in the $\Delta$-variance spectra of the THINGS galaxies, we measured the $\Delta$-variance spectrum of a simulated galaxy. We tested the validity of the star formation activity related scenario by exploring the connection between the features observed in the $\Delta$-variance spectra of the THINGS galaxies and indicators of their star formation activity. We also explore the possible origin of the broken power law that is observed in the $\Delta$-variance spectra and interpret the transition point between the two self-similar regimes.
  
\subsection{Insight from numerical simulations of whole galaxies}\label{simul}

\begin{figure*}
\centering
\includegraphics[width=0.26\textwidth]{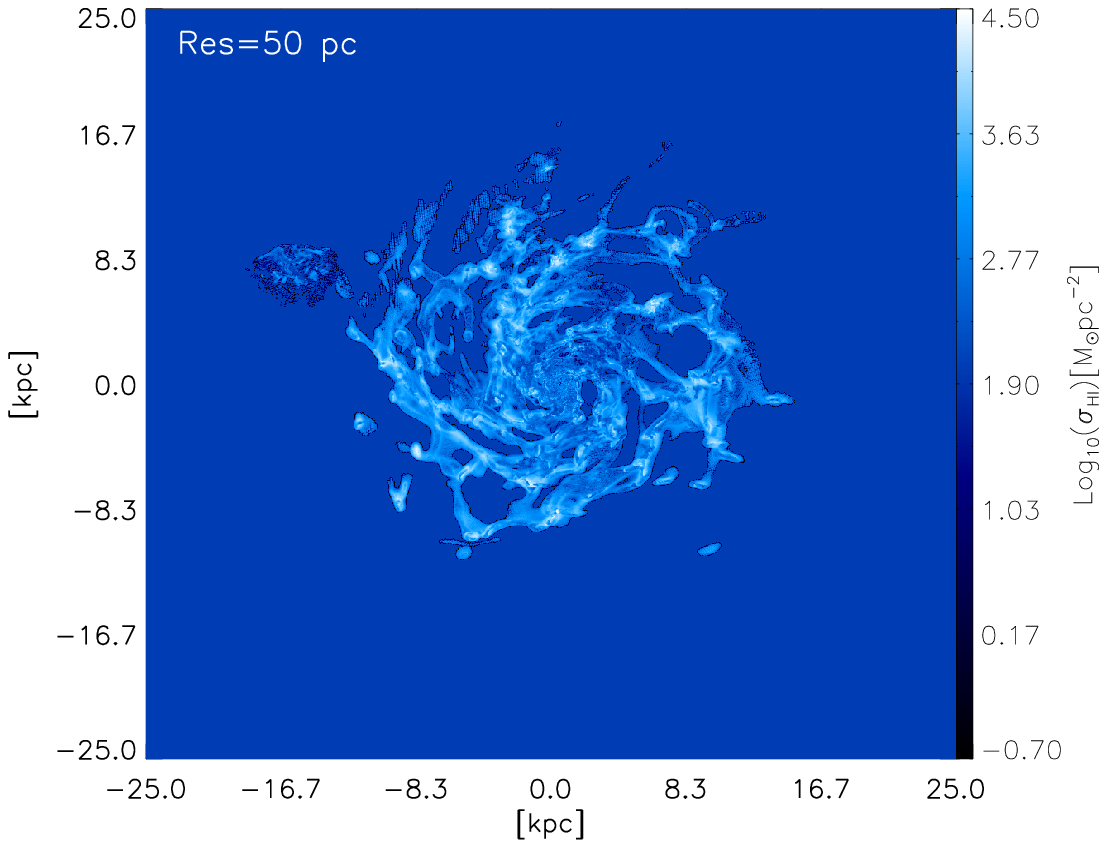}
\hspace{1.5cm}
\includegraphics[width=0.26\textwidth]{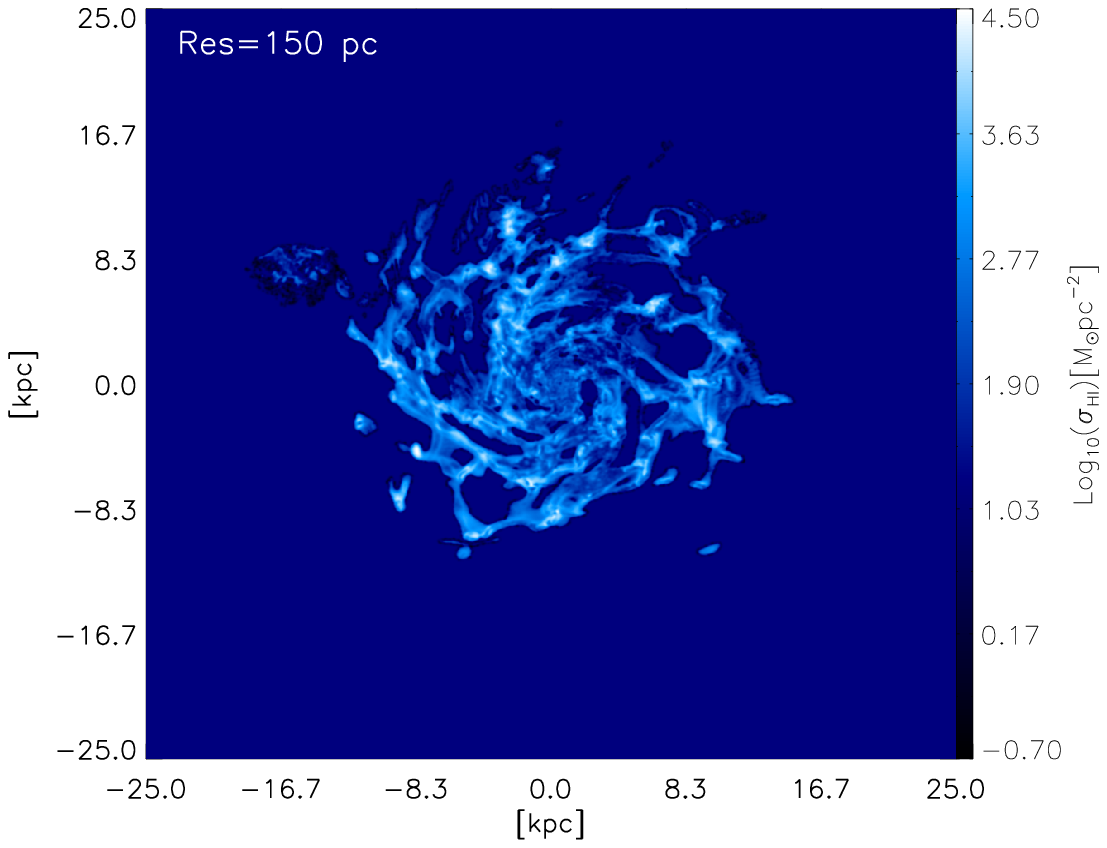}
\hspace{1.5cm}
\includegraphics[width=0.26\textwidth]{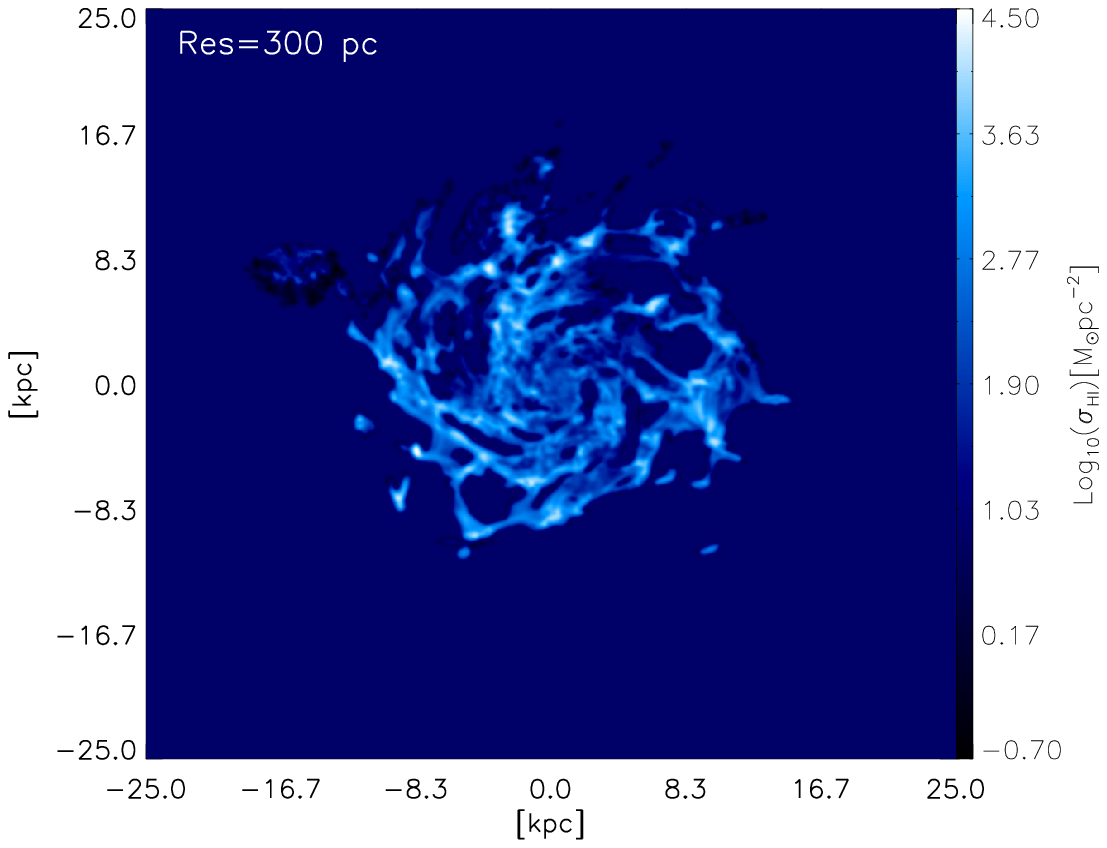}
\vspace{0.5cm}
\caption{Face-on view of the \ion{H}{I} surface density of the simulated galaxy at $t_{0}$ at the original resolution of 50 pc (left panel) and degraded resolutions of 150 pc (middle panel) and 300 pc (right panel).}
\label{fig9}
\end{figure*}

\begin{figure}
\centering
\includegraphics[width=\columnwidth]{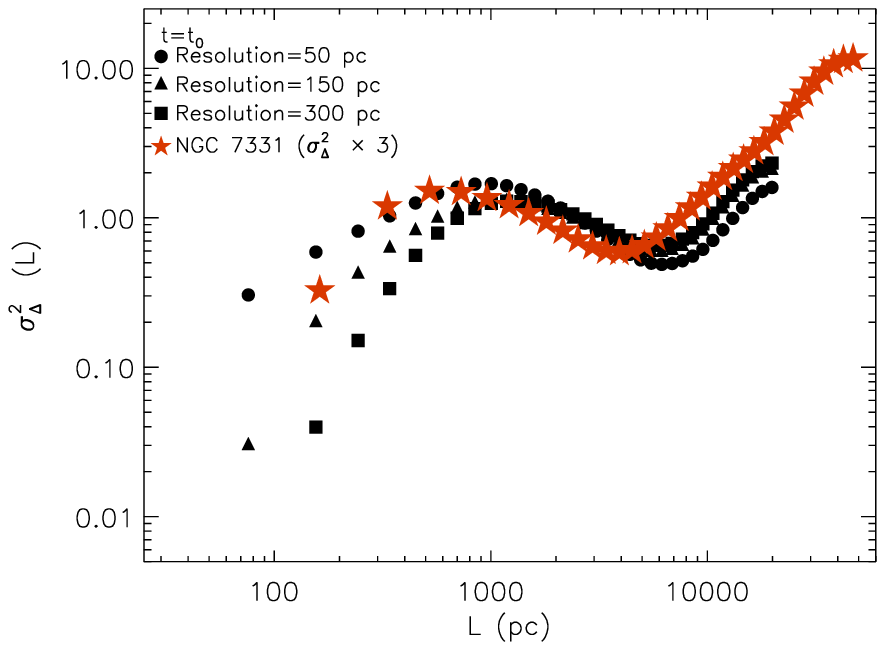}
\caption{$\Delta$-variance spectra of the simulated galaxy for the same snapshot, but with for a different spatial resolution. The spectra are all normalized to their respective mean values. The models are compared to the case of the galaxy NGC 7331, where the $\Delta$-variance of NGC 7331 has been multiplied by a factor of 3.}
\label{fig10}
\end{figure}  

We used the VINTERGATAN cosmological zoom-in simulation of a Milky Way-like galaxy (Agertz et al. 2021, Renaud et al. 2021a,b). The simulation reaches a resolution of $20$ pc in the densest medium and includes prescriptions for ultraviolet background radiation, atomic and molecular cooling lines in the form of tabulated data of the Sutherland \& Dopita (1993) and Rosen and Bregman (1995) cooling curves, and a prescription for star formation. Stellar feedback from massive stars is accounted for in the form of stellar winds, radiation pressure, and type II and type Ia supernovae (see Agertz et al. 2021 for details). The global properties of the simulated galaxy agree with measurements of the Galactic mass, the surface density profiles of its baryonic components, the rotation curve, and the chemical bimodality of the stellar populations. 

The analysis shown here was conducted at a look-back time of 3.5 Gyr, corresponding to a redshift of $z\approx0.3$ (we refer to the time of the canonical snapshot with time $t=t_{0}$), when the stellar mass of the galaxy was $\approx 6\times10^{10}$ M$_{\odot}$ and the mass of the atomic gas component was $\approx 5\times10^{9}$ M$_{\odot}$. At this epoch, the effects of the last major merger had faded, and the galaxy was in its phase of secular evolution, with an SFR of $\approx 9$ M$_{\odot}$ yr$^{-1}$. In order to measure the mass of \ion{H}{i} in each simulation cell, we first solved the local Saha equation using the cell temperature and density. This allowed us to obtain the mass of the ionized gas. By subtracting it from the total gas mass, we measured the total mass of neutral gas. Using the prescription of Krumholz et al. (2009)\footnote{The implementation of the Krumholz et al. (2009) method to compute $f_{\rm H_{2}}$ in the version of the RAMSES code used in this work is described in detail in Agertz \& Kravtsov (2015) (Eqs. 2, 3, and 6). We used $\left( \sigma_{d,-21} /R_{-16.5}\right)=1$, where $\sigma_{d,-21}$ is the dust cross-section per hydrogen nucleus to radiation at 1000 $\AA$ normalized to $10^{-21}$ cm$^{-2}$, and $R_{-16.5}$ is the rate for H$_{2}$ formation on dust grains, normalized to the Milky Way value of $10^{-16.5}$ cm$^{3}$ s$^{-1}$ (Wolfire et al. 2008). Both quantities are directly proportional to the dust abundance and thus to the gas-phase metallicity, which is tracked in each cell of the simulation. As in Agertz \& Kravtsov (2015), we adopted a value of $3$ for the parameter $\phi_{\rm CNM}$ and calculate the optical depth $\tau_{c}$ that appears in Eq. 3 of Agertz \& Kravtsov (2015) as $\tau_{c}=\rho_{cell} s_{cell} \sigma_{d}$, where $s_{cell}$ and $\rho_{cell}$ are the size of the cell and its density, respectively.}, we then computed the mass of molecular hydrogen at the local gas metallicity and removed it from the neutral gas mass, which gives the mass of atomic hydrogen. The surface density maps of \ion{H}{i} were then computed in face-on projections of the simulation cells, were remapped at a uniform resolution of 50 pc, and covered a surface area of 50 kpc x 50 kpc. For each snapshot, we also generated a number of \ion{H}{i} column density maps with various inclinations. The effect of inclination on the $\Delta$-variance spectrum is discussed in Appendix~\ref{appa}.

\begin{figure*}
\centering
\includegraphics[width=0.49\textwidth]{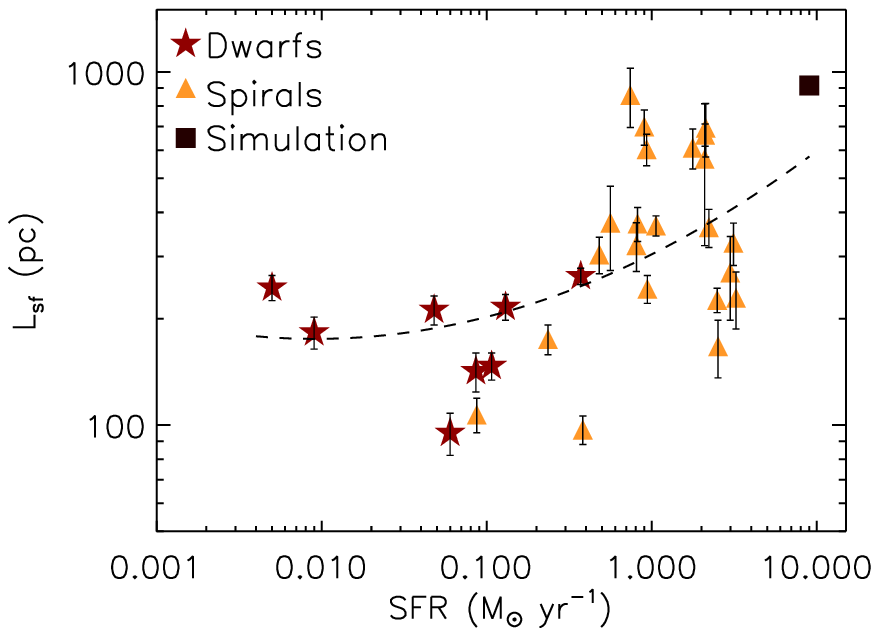}
\includegraphics[width=0.49\textwidth]{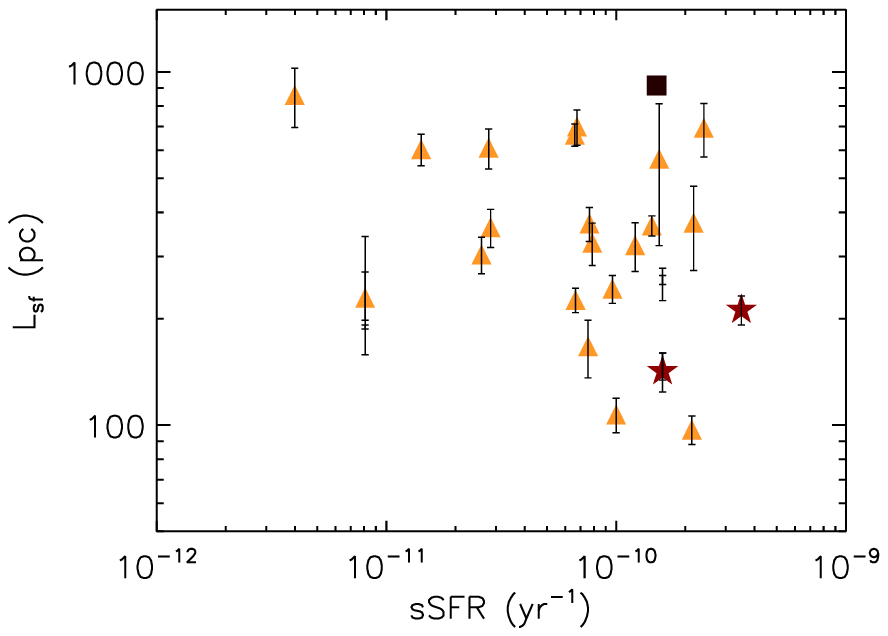}\\
\includegraphics[width=0.49\textwidth]{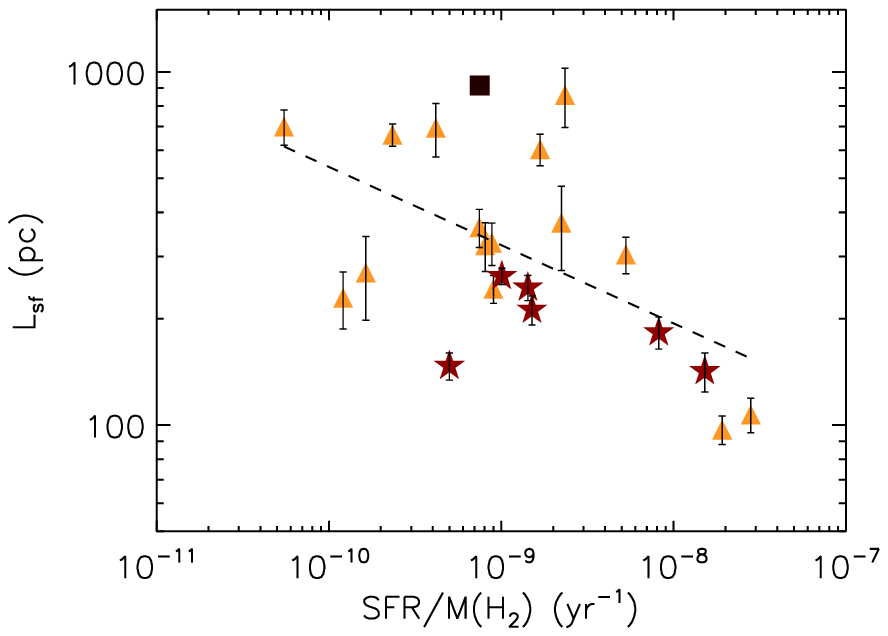}
\includegraphics[width=0.49\textwidth]{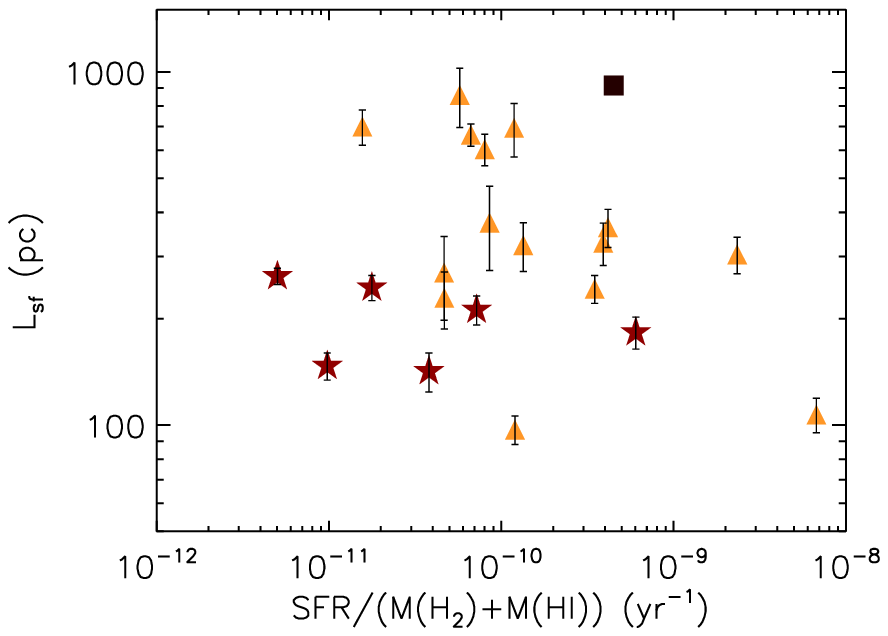}
\caption{Scatter plots between the characteristic scale of star formation $L_{sf}$ with the galaxy integrated (SFR, top left), the specific star formation rate (sSFR, top right), the star formation efficiency relative to the molecular gas content SFE$_{m}={\rm SFR}/M({\rm H_{2}})$ (bottom left), and the SFE relative to the total gas content SFE$_{g}={\rm SFR}/(M({\rm H_{2})}+M(\ion{H}{i}))$ (bottom right).}
\label{fig11}
\end{figure*}

Figure \ref{fig7} displays the column density maps of the face-on simulated galaxy at $t=t_{0}$ (middle panel), at $t=t_{0}-160$ Myr (left panel) and at $t=t_{0}+160$ Myr (right panel). The spatial resolution in all three maps is $50$ pc. The corresponding $\Delta$-variance spectra for these three cases are displayed in Fig.~\ref{fig8}. Figure \ref{fig8} provides clear evidence that the galaxy is in its phase of secular evolution because the $\Delta$-variance spectra of the three snapshots are nearly identical. The spectra display a prominent bump at $\approx 0.8-1$ kpc and are similar to those that are observed in some of the THINGS galaxies, such as NGC 2976, NGC 3351, and NGC 7331, whose spectra also display a prominent bump and an absence of a first power law on intermediate spatial scales. In Fig.~\ref{fig8} we compare the $\Delta$-variance spectra of the models to that of the galaxy NGC 7731 ($L_{sf}=305$ pc, $\alpha_{2}=1.42\pm 0.024$). The spectra are very similar, but we had to multiply the spectrum of NGC 7331 by a factor of $3$. This factor is simply due to the difference in the mean surface density of the \ion{H}{i} maps between the simulation galaxy and NGC 7331.  A fit to the self-similar regime of the models in the scale range $[10-20]$ kpc yields a value of the slope $\alpha_{2}=1.34\pm0.08,$ which is very similar to the value derived for NGC 7331. Applying the same procedure as in the observations, we derive a value of $L_{sf}\approx 890$ pc from the $\Delta$-variance spectra of the simulation. Because most THINGS galaxies have a spatial resolution higher than 50 pc, we generated \ion{H}{i} surface density maps in which the 50 pc resolution map is convolved with a beam whose FWHM is 150 pc and 300 pc. The three maps with different resolution for the fiducial timestep (i.e., $t=t_{0}$) are displayed in Fig.~\ref{fig9}. The corresponding $\Delta$-variance spectra for these cases are displayed in Fig.~\ref{fig10}. The loss of resolution does not affect the value of the power law that characterizes the shape of the spectrum at large spatial scales. However, the reduced resolution affects the shape of the bump and its position. Increased smoothing reduces the variance in the map and thus the amplitude of the bump decreases, and as discussed earlier, the left-hand wing of the bump also becomes increasingly affected and the bump becomes narrower. The position of the bump, and consequently, the value of $L_{sf}$ are also shifted to higher values ($L_{sf} \approx 965$ pc at the resolution of 300 pc). However, this effect, which is present in the $\Delta$-variance spectra of the observed galaxies shown in Fig.~\ref{fig2} and Fig.~\ref{fig3}, is not dramatic. The increase in the position of the bump and in the value of $L_{sf}$ is only $\approx 10\%$.   
  
\subsection{Relation between the $\Delta$-variance spectra and galactic star formation}\label{correlation}

If the bump that is observed in the $\Delta$-variance spectra is connected with stellar feedback, in particular, with supernova explosions that can carve holes in the \ion{H}{i} gas, then we might expect a correlation between the value of $L_{sf}$ and the global galactic SFR. This correlation is expected on the basis of the known empirical relation between the global SFR of a galaxy and the maximum mass of clusters that form within it (Weidner et al. 2004; G\'{o}nzalez-L\'{o}pezlira et al. 2012,2013; Schulz et al. 2015 ). More massive clusters will statistically harbor more massive stars, and the bubble that can form when massive stars explode as supernova would therefore be larger, leading to a correlation between the characteristic size of \ion{H}{i} holes in galaxies and the global SFR. Figure~\ref{fig11} (top left panel) displays the value of $L_{sf}$ plotted as a function of the global galactic SFR. The values of the SFRs for the ensemble of the THINGS galaxies are taken from Leroy et al. (2008) and Walter et al. (2008). Figure~\ref{fig11} shows that for low values of the SFR (i.e., SFR $\lesssim 0.5$ M$\odot$ yr$^{-1}$), $L_{sf}$ is quasi-constant or weakly dependent on the SFR. This is expected if $\ion{H}{i}$ holes in these galaxies are caused by one (or a few) supernovae remnants. For higher SFRs (SFR $\gtrsim 0.5$ M$_{\odot}$ yr$^{-1}$), we observe a correlation between the SFR and $L_{sf}$. We also include the point of the simulated galaxy smoothed at a resolution that is close to the observational resolution of the observations (i.e., 300 pc). An empirically motivated second-order polynomial fit to the $L_{sf}$-SFR data yields the following result:

\begin{figure}
\centering
\includegraphics[width=\columnwidth]{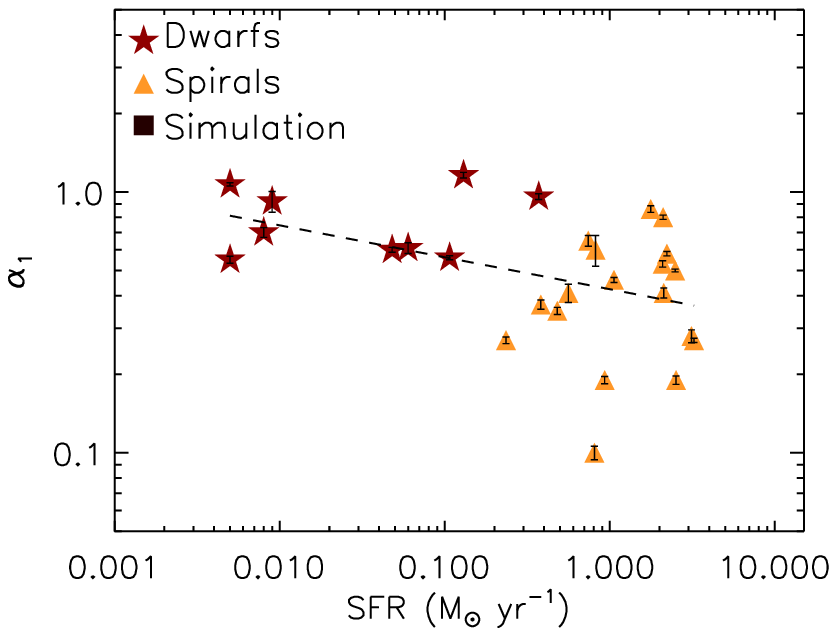}\\
\includegraphics[width=\columnwidth]{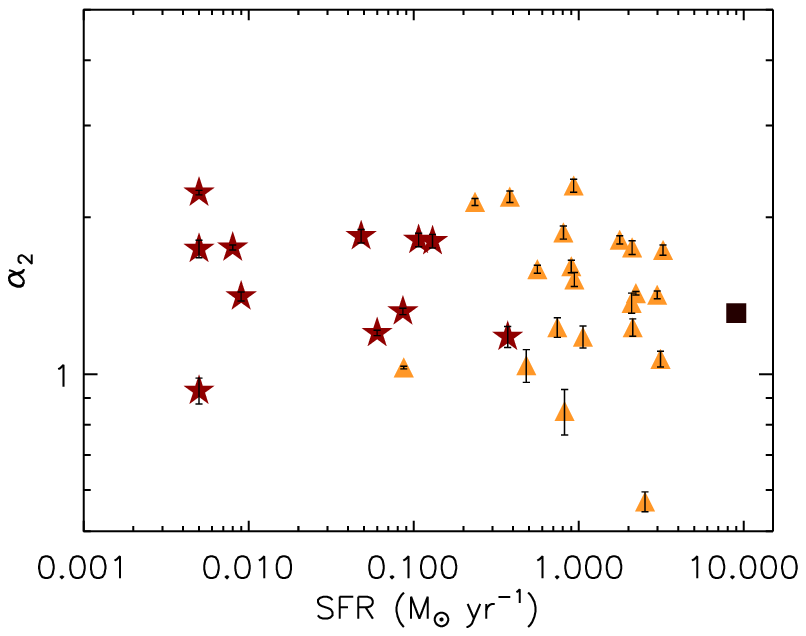}
\caption{Relation between the galactic SFR and the slope of the first self-similar regime in the $\Delta$-variance spectrum, $\alpha_{1}$ (top subpanel), and the slope of the second self-similar regime (bottom subpanel).}
\label{fig12}
\end{figure}  

\begin{equation}
 {\rm log}(L_{sf})=2.48+0.23~{\rm log}({\rm SFR})+0.05~({\rm log}({\rm SFR}))^{2},
 \label{eq7}
\end{equation}     

\noindent and the fit is overplotted on the data in Fig.~\ref{fig11} (dashed line, top left subpanel). Figure~\ref{fig11} also displays the values of the $L_{sf}$ plotted against the specific star formation rate sSFR=SFR$/M_{*}$ (top right subpanel) and the star formation efficiency (SFE) per unit time with respect to the H$_{2}$ mass reservoir, SFE$_{m}={\rm SFR}/M(\rm H_{2}$) (bottom left subpanel) and with respect to the total gas reservoir SFE$_{g}={\rm SFR}/(M({\rm H_{2}})+M(\ion{H}{i}))$, where $M_{*}$ is the total stellar mass in the galaxy, and $M(\rm {H_{2}}$) and $M({\ion{H}{i}})$ are the total mass in H$_{2}$ and \ion{H}{i}, respectively. The masses of \ion{H}{i} and H$_{2}$ are taken from Leroy et al. (2008) (Table 4). For the \ion{H}{i}, this estimate is based on the integration of the \ion{H}{i} surface density from Walter et al. (2008) within 1.5$R_{25}$, where $R_{25}$ is the galactic isophotal radius. The same procedure was adopted by Leroy et al. (2008) to derive the H$_{2}$ masses. The H$_{2}$ surface densities are those obtained by the HERACLES survey (Leroy et al. 2009), the BIMA SONG survey (Helfer et al. 2003) by Walter et al. (2001) for NGC 3077, and by Bolatto et al. (2008) for NGC 4449. Leroy et al. (2008) adopted a constant CO-to-H$_{2}$ conversion factor of $X_{\rm CO}=2\times10^{20}$ cm$^{-2}$ (K km$^{-1}$ s$^{-1})^{-1}$ , which is about the mean value found in the Milky Way. The $1\sigma$ uncertainty on this quantity is $\approx$ 30$\%$ (Bolatto et al. 2013).

Because the galactic SFR and stellar mass M$_{*}$ are correlated (e.g., Lara-L\'{o}pez et al. 2013), $L_{sf}$ has the same dependence on $M_{*}$ as on the SFR (see details in Appendix~\ref{appb}). Hence, $L_{sf}$ is independent of the sSFR, but the scatter between the two quantities is large due to the combined uncertainties on the SFR and $M_{*}$. The correlation between $L_{sf}$ and SFE$_{g}$ is unclear as well because this last quantity is only loosely related to the star-forming gas. However, a weak anticorrelation is observed between $L_{sf}$ and SFE$_{m}$. A power-law fit to the $L_{sf}$-SFE$_{m}$ data points yields the following result:  

\begin{equation}
L_{sf}= 2.95~{\rm SFE}_{m}^{-0.22}.
\label{eq8}
\end{equation}   

The anticorrelation between $L_{sf}$ and $SFE_{m}$ is largely due to the low-metallicity galaxies, which have a higher SFE$_{m}$. Whether low-metallicity galaxies, such as the dwarf galaxies in the THINGS sample, have a higher SFE$_{m}$ is still a matter of debate. The difficulty in measuring SFE$_{m}$ in subsolar metallicity galaxies is the determination of the appropriate ratio between the column density of H$_{2}$ and the CO intensity $\left(N({\rm H_{2}}) /I_{\rm CO}   \right)$ ($X_{\rm CO}$ factor). The $X_{\rm CO}$ factor is expected to increase as the metallicity decreases due notably to increased photodissociation of CO (Bolatto et al. 2013 and references therein). Thus, a CO flux at low metallicity corresponds to a higher H$_{2}$ mass as compared to the same flux from a higher metallicity environment.  A careful comparison of the CO intensity with the column density of H$_{2}$ as estimated via dust emission in the nearby galaxies M33 (Gardan et al. 2007; Braine et al. 2010; Gratier et al. 2017) and NGC 6822 (Gratier et al. 2010) led to a higher $X_{\rm CO}$ than in the Milky Way, but nonetheless to a higher SFE than in large spirals (e.g., Murgia et al 2002). An obvious reason for this is that the conversion of \ion{H}{i} into H$_{2}$ occurs at higher density when fewer dust grains are present (Hollenbach et al 1971; Braine et al 2001), as is the case in low-metallicity environments. This reduces the free-fall time in the molecular component\footnote{The conversion rate is proportional to $Z~n_{\ion{H}{i}}^2$ where $Z$ is the metallicity and $n_{\ion{H}{i}}$ is the number density of neutral hydrogen. This means that as $Z$ decreases, the rate of conversion from \ion{H}{i} into H$_{2}$ occurs at a higher density. This is valid under the assumption that the dust number density $n_{dust} \propto n_{H}$ , which is correct down to $Z \approx 0.1 Z_{\odot}$). The free-fall time is $t_{ff} \propto n_{H}^{-0.5}$ , which implies that $t_{ff} \propto Z^{0.25}$.}. A second, more subtle, mechanism is that the weaker stellar winds in low-metallicity environments expel gas less efficiently from protocluster-forming clouds, such that a higher stellar mass can be formed for a given molecular gas mass, which leads to a higher SFE$_{m}$ (Dib et al. 2011; Dib 2011, Dib et al 2013). These effects are cumulative.

\subsection{Two self-similar regimes and the transition point}\label{transition}

With the exception of a few galaxies (NGC 2976, NGC 3184, NGC 3351, and NGC 7331), the $\Delta$-variance spectrum of most galaxies in the THINGS sample displays two distinct power-law regimes. On  scales $\lesssim 0.5 R_{25}$, the $\Delta$-variance can be described by a power law with an exponent that varies between $0.1$ and $1.16$ and whose mean value is $\alpha_{1} \approx 0.5$ . On larger scales ($\gtrsim 0.5 R_{25}$), it is described by a second power law, whose exponent can be as large as 2.3 and has a mean value of $\approx 1.5$ (Tab.~\ref{tab1}). A similar result was obtained for the LMC by Elmegreen et al. (2001). In terms of the exponent of the power spectrum, this would correspond to exponents of $\beta_{1} \approx 2.5$ and $\beta_{2} \approx 3.5$. A detailed explanation of the specific values of $\alpha_{1}$ and $\alpha_{2}$, and consequently of $L_{tr}$, for each individual galaxy in the THINGS sample is beyond the scope of this work. This would require comparisons with numerical simulations of a cosmological volume that resembles the ensemble of nearby galaxies. Intuitively, the shape of the $\Delta$-variance spectra that are observed for the THINGS samples of galaxies might be thought to be the result of an exponential disk. While we show in Appendix~\ref{appc} that an exponential disk could indeed generate a spectrum with a broken power law, we discard this hypothesis on the basis that no exponential disks are observed in \ion{H}{i}. This fact was noted earlier by other authors (e.g., Casasola et al. 2017). Instead, \ion{H}{i} disks are observed to be nearly flat (i.e., nearly constant column density), with radial variations by a factor of $\approx 2$ and in some cases a depression toward the inner regions of the galaxy, where most of the hydrogen gas becomes molecular (Walter et al. 2008; Leroy et al. 2008). 

The mean values of $\alpha_{1}$ and $\alpha_{2}$ are so different that the structure of the ISM in the range of spatial scales they represent must originate from different physical processes or correspond to different phases of the gas with different compression levels. The values of $\alpha_{1}$ we find in this work, with a mean value of $\approx 0.5$, are very similar to those found for molecular clouds using either molecular transitions or cold dust emission (e.g., Stutzki et al. 1998; Bensch et al. 2001; Miville-Desch\^{e}nes et al. 2010; Dib et al. 2020). They are also similar to those found in \ion{H}{i} seen in absorption which is tracing the cold ($\approx 100$ K) component of the \ion{H}{i} gas (Deshpande et al. 2000). These values are also consistent with those found in numerical simulations of supersonic magnetohydrodynamic turbulence where the gas is compressed into smaller pockets (e.g., Kowal et al. 2007; Dib et al. 2008)\footnote{Values of $\alpha \lesssim 1$ (i.e., $\beta \lesssim 3$) are consistent with optically thin media, whereas in optically thick media, the value of $\beta$ saturates at a value of $3$ (Lazarian \& Pogosyan 2000).}. In contrast, the range of values of $\alpha_{2}$ is consistent with the values observed for the \ion{H}{i} in emission toward diffuse regions that are dominated by the warm component of the \ion{H}{i} gas both in the Galaxy (e.g., Miville-Desch\^{e}nes et al. 2003; Chepurnov et al. 2010) and in the LMC (Elmegreen et al. 2001). The mean value of $\alpha_{2} \approx 1.5$, which corresponds to an exponent of the power spectrum of $\beta \approx 3.5,$ is consistent with the picture in which turbulence in the warm neutral medium (WNM) is subsonic to transonic (e.g., Burkhart et al. 2013). 

As discussed above, a bump on scales of a few to several hundreds of parsecs perturbs the underlying self-similar regime on these scales. Although we have avoided any overlap with the bump when we performed the fit for the first power law, we explored whether any correlation exists between the SFR and the exponent of the first power law, $\alpha_{1}$. Figure~\ref{fig12} (top panel) shows a weak anticorrelation between the SFR and $\alpha_{1}$. Weakly star-forming dwarf galaxies have a systematically steeper spectrum on spatial scales that are covered by the first self-similar regime. A power-law fit to the SFR-$\alpha_{1}$ data points yields 

\begin{equation}
\alpha_{1}= 0.42~{\rm SFR}^{-0.12}.
\label{eq9}
\end{equation}   

The SFR-$\alpha_{1}$ anticorrelation suggests that the star formation activity in galaxies shapes the structure of the gas distribution on scales larger than those associated with the sizes of individual supernova remnants or larger superbubbles, up to the transition scale $L_{tr}$. Lower values of $\alpha_{1}$ imply that more substructure is present in galaxies with a higher star formation rate. In contrast, the exponent of the power law that describe the second self-similar regime ($\alpha_{2}$) is independent of the SFR (Fig.~\ref{fig12}, middle panel), indicating that the dynamics of the gas on large scales, and consequently its structure, are shaped by processes that act on scales larger than those associated with feedback from supernova explosions. 

\begin{figure}
\centering
\includegraphics[width=\columnwidth]{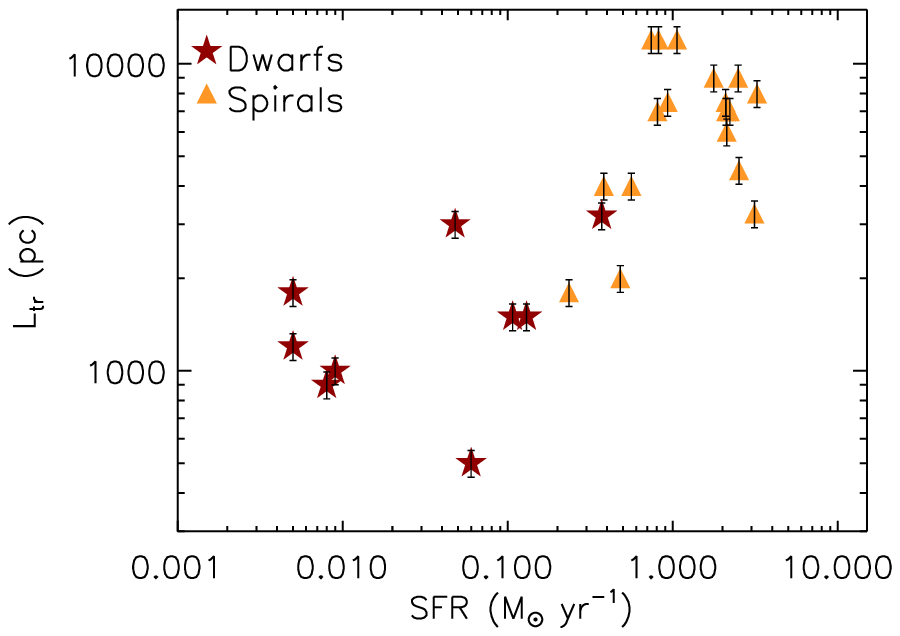} \\
\includegraphics[width=\columnwidth]{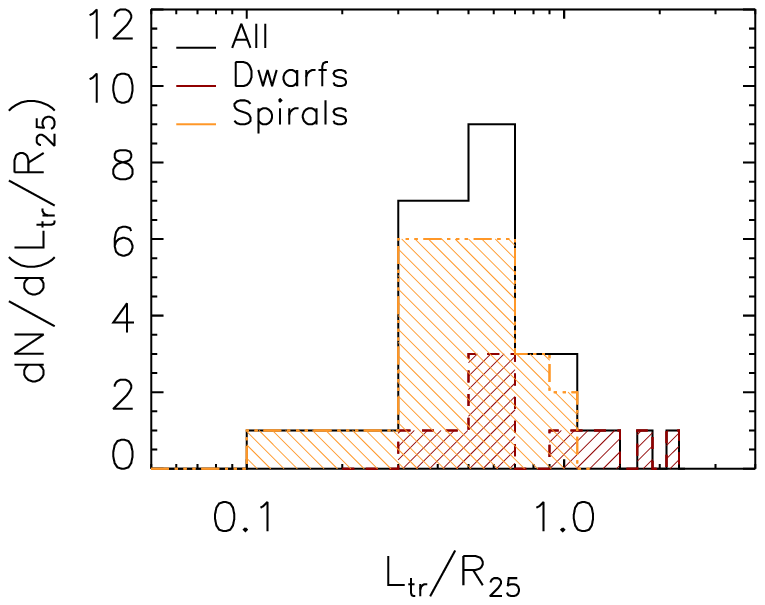}
\caption{Top panel: Relation between the galactic SFR and the position of the transition point in the $\Delta$-variance spectrum. Bottom panel: Distribution of the ratio of the transition point between the two self-similar regimes ($L_{tr}$) and the galactic optical radius ($R_{25}$).}
\label{fig13}
\end{figure}  

\begin{figure*}
\centering
\includegraphics[width=0.3\textwidth]{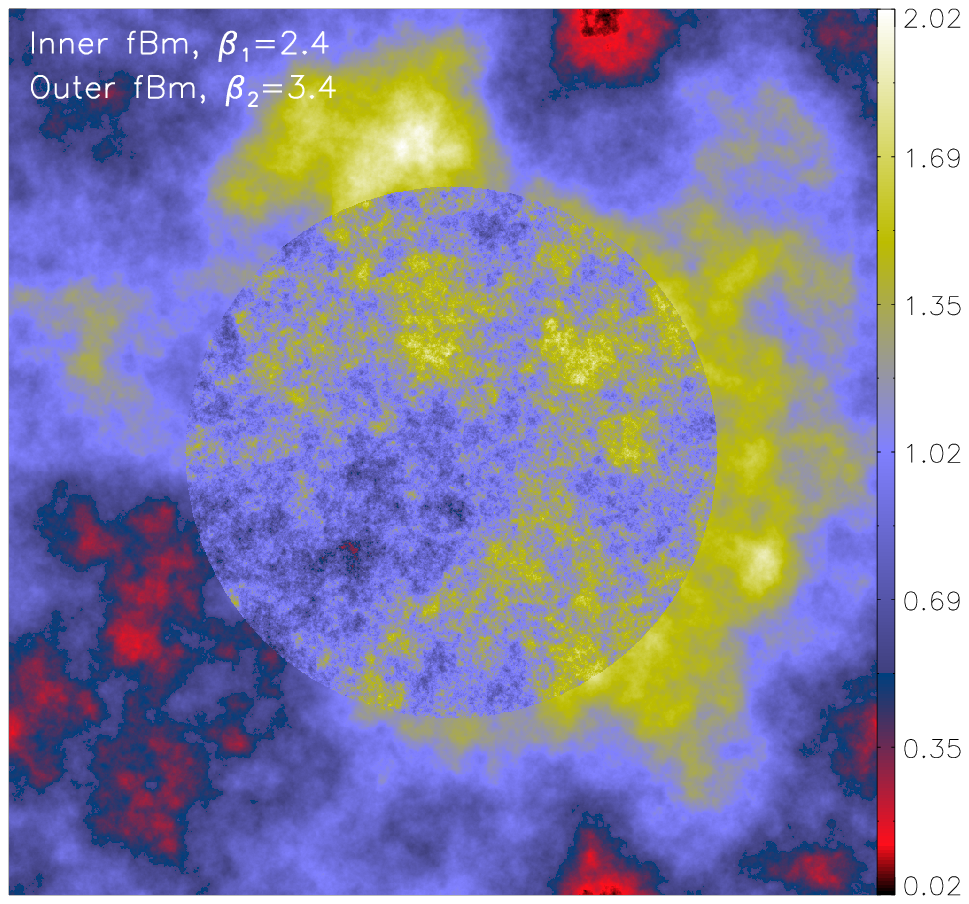}
\hspace{0.3cm}
\includegraphics[width=0.3\textwidth]{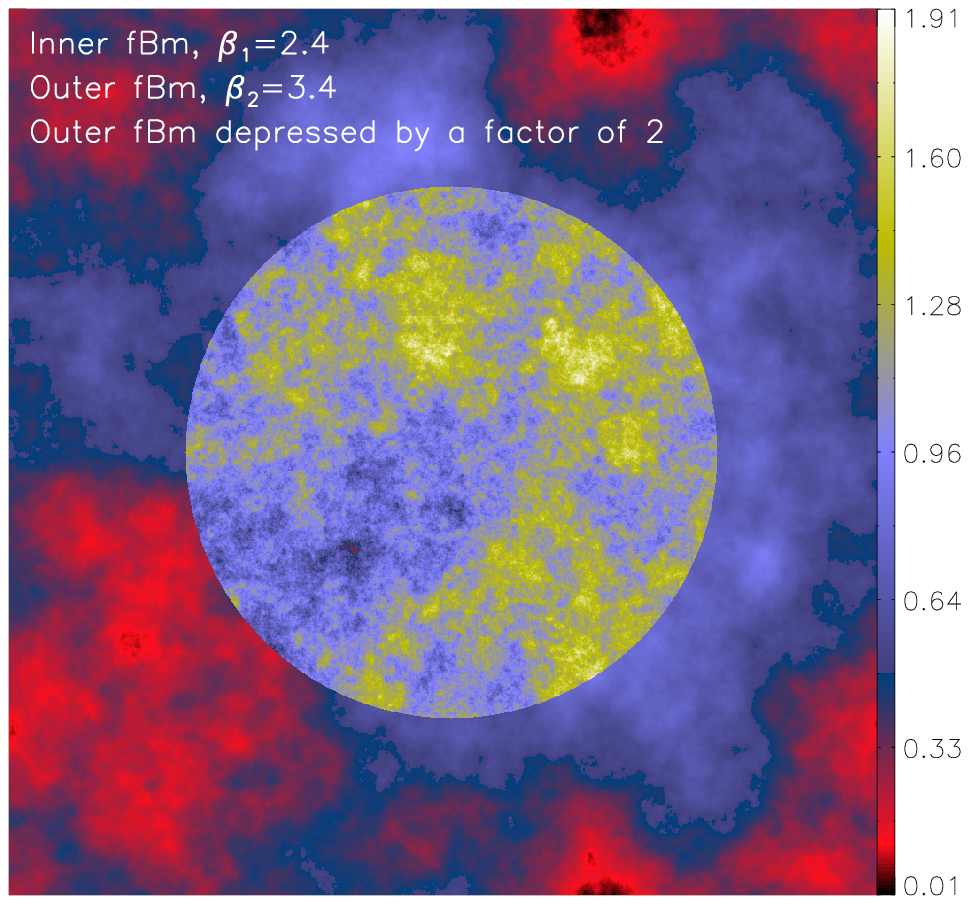}
\hspace{0.3cm}
\includegraphics[width=0.3\textwidth]{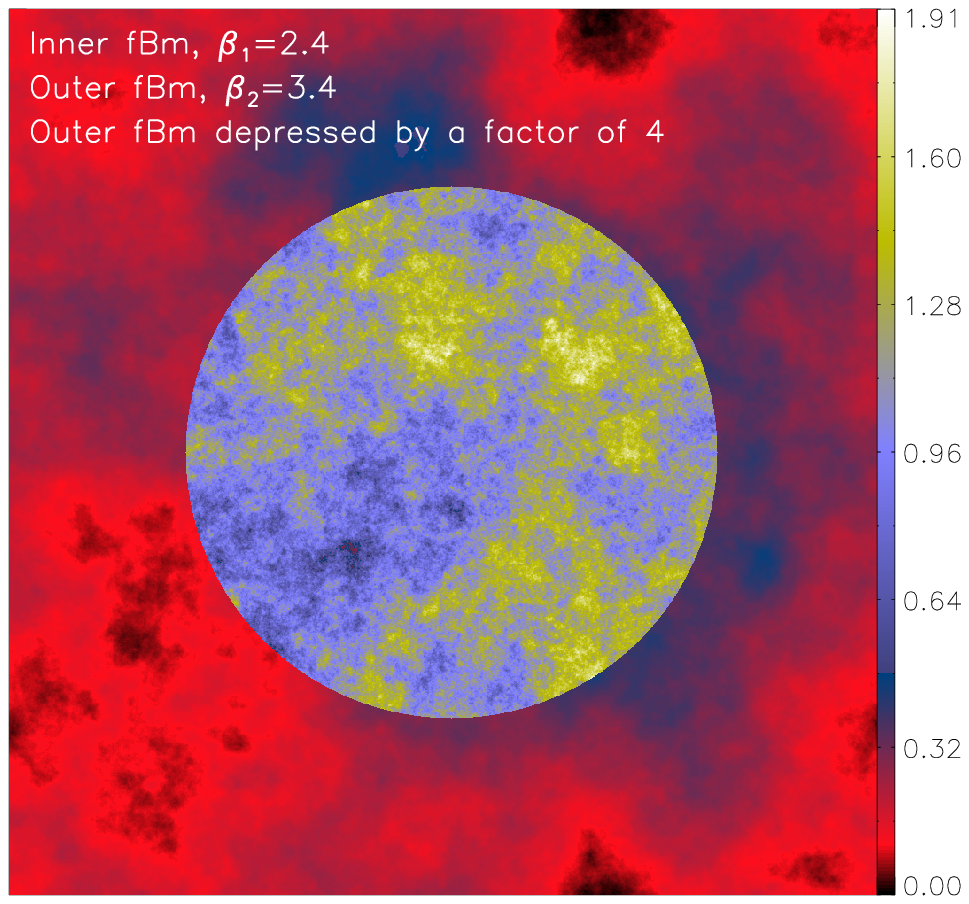}\\
\includegraphics[width=0.32\textwidth]{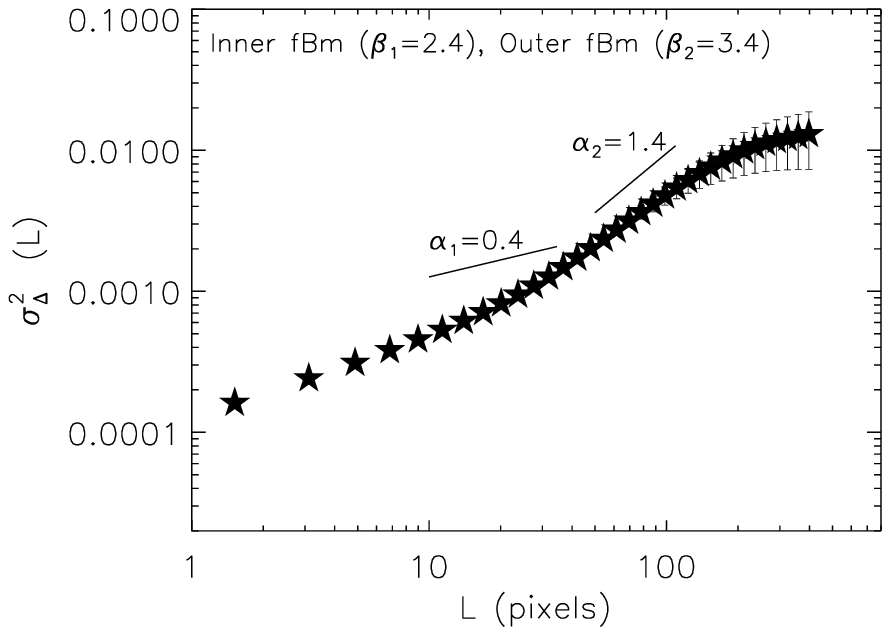}
\includegraphics[width=0.32\textwidth]{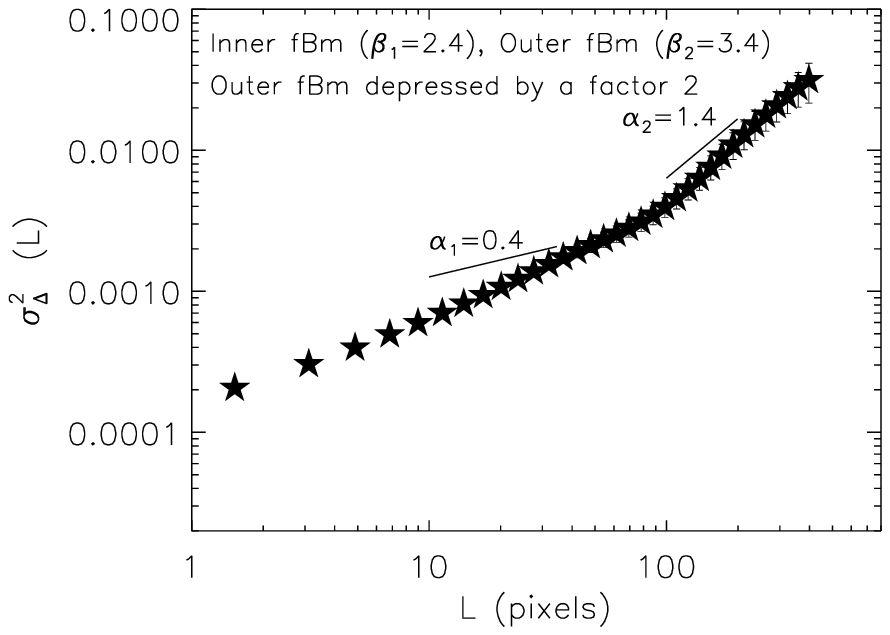}
\includegraphics[width=0.32\textwidth]{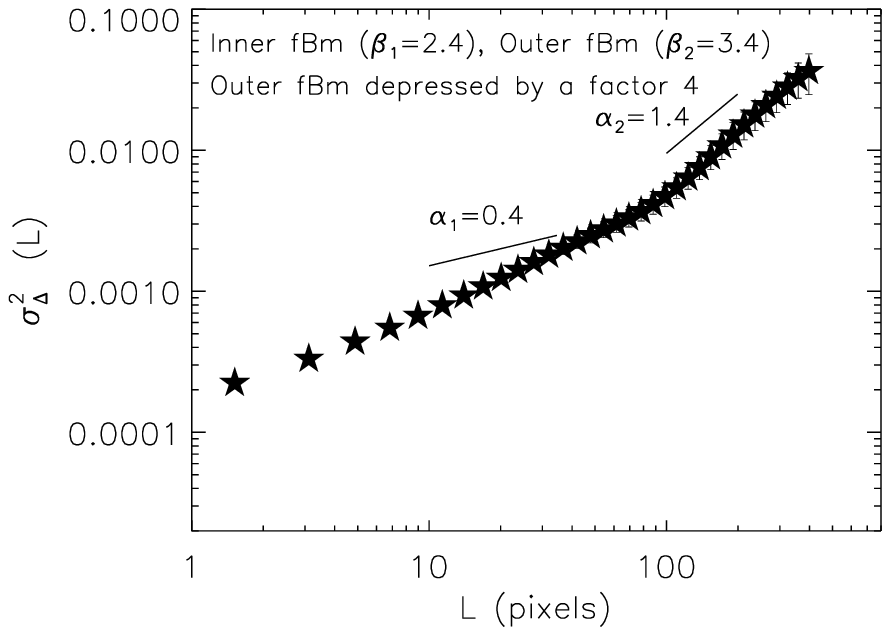}
\caption{Maps of two fBm images with different values of $\beta$. The fBms  are shifted to positive values by adding an arbitrary constant, and they are normalized by their mean values. The maps have a resolution of $1000\times1000$ pixels. The inner fBm, residing within a circle with a radius of 300 pixels, has $\beta=2.4,$ and the outer fBm has $\beta=3.4$. The top left subpanel corresponds to the fiducial case, and the middle and right top subpanels correspond to cases in which the inner and outer fBms have been divided by a factor of 2 and 4, respectively. The lower subpanels display the corresponding $\Delta$-variance spectra. The values of $\alpha=0.4$ and $1.4$ are not fits to the spectra, but are shown as a reference to guide the eye.}      
\label{fig14}
\end{figure*}

The transition between the two regimes is observed in Fig.~\ref{fig2} and Fig.~\ref{fig3} as a dip or an inflection point. As stated in \S.~\ref{results}, we adopted as the value of $L_{tr}$ the position of the inflection point or the deepest position of the dip, when present. The values of the derived $L_{tr}$ are reported in Tab.~\ref{tab1}, and the distributions of $L_{tr}$ for dwarf and spiral galaxies are displayed in Fig.~\ref{fig5} (bottom right panel). The top panel of Fig.~\ref{fig13} displays the dependence of $L_{tr}$ on the galactic SFR. The correlation between the $L_{tr}$ and the galactic SFR (and with $M_{*}$, see Appendix~\ref{appb}) is clear. Figure \ref{fig13} (bottom panel) displays the distribution of the ratio of $L_{tr}$ to the galactic optical radius, $R_{25}$. While there is some scatter, most values of this ratio lie around $(L_{tr}/R_{25})\approx 0.4-0.5$. Interestingly, the value of $\approx (0.4-0.5) R_{25}$ is very similar to the size of the molecular disk in the THINGS galaxies (Leroy et al. 2008). Our results do not support the idea that $L_{tr}$ is connected to the scale height of the \ion{H}{i} disk, $h_{\ion{H}{i}}$. Recent derivations of $h_{\ion{H}{i}}$ for a number of the THINGS galaxies clearly indicate that the \ion{H}{i} disks are flared and the values of $h_{\ion{H}{i}}$ vary radially from about $\lesssim 100$ pc in the inner region of the disk to $\approx$ 1 kpc in its outer region (Bacchini et al. 2019; Patra 2020a,b). These values are lower than any of the values of $L_{tr}$ derived in this work. 

Because the exponential disk scenario can be discarded, a different physical mechanism must cause the broken power law and the transition point that are observed in the \ion{H}{i} $\Delta$-variance spectra of most THINGS galaxies. The evidence gathered from the distributions of $\alpha_{1}$, $\alpha_{2}$, and $L_{tr}$ and the anticorrelation between $\alpha_{1}$ and the galactic SFR appears to point out to the following scenario: The $\Delta$-variance spectrum in the first self-similar regime is dominated by emission from the CNM component of the \ion{H}{i} gas. The range of values found for $\alpha_{1}$ around a mean value of $\approx 0.5$ is compatible with the occurrence of compressible supersonic turbulence that governs the dynamics of cold gas, that is, the cold component of the \ion{H}{i} gas (e.g., Bensch et al. 1998; Bertram et al. 2015; Dib et al. 2020). The anticorrelation between $\alpha_{1}$ and the SFR is compatible with the idea that the dynamics of the gas is increasingly dominated by compressive motions for an increasing SFR, leading to a shallower spectrum (i.e., a lower value of $\alpha_{1}$). The reason is that compressive turbulence for the same Mach number can compress gas to higher overdensities than solenoidal modes (e.g., Federrath et al. 2008). On large spatial scales (i.e., $\gtrsim 0.5 R_{25}$), the signal is dominated by the contribution from the external regions of the galaxy where the WNM phase of the \ion{H}{i} is larger. At large galactocentric radius, where the \ion{H}{i} is more flared, the gas is more easily affected by ram pressure stripping and the heating by the extragalactic background UV field. The combination of these processes keeps the gas warm and diffuse, and thus its dynamics is governed by subsonic to mildly supersonic turbulence with little connection to the galactic SFR.

The signature of the \ion{H}{i} gas in the CNM phase extends only to scales $\lesssim L_{tr} \approx \left(0.4-0.5\right) R_{25}$ because most of the cold \ion{H}{i} resides in the inner region of the galaxy, on scales smaller than and up to the size of the molecular disk, and there is little or no cold \ion{H}{i} in the outer regions of the galaxy (e.g., Braun 1997; Zhang et al. 2012). In contrast, the warm component of the \ion{H}{i} gas is likely to be present everywhere in the disk, but in smaller proportion (in terms of total local mass) in the inner regions and dominant in the outer regions. The warm \ion{H}{i} dominates the emission on larger scales and has less substructure on smaller scales. It can still contribute to the signal on small scales, but most of the variance on those scales is dominated by the cold component. In order to illustrate this idea, we show in Fig.~\ref{fig14} (left subpanel) a toy model in which the inner parts of the disk are described by a fBm whose $\beta=2.4$, characteristic of cold gas, and the outer regions are described by as second fBm with a steeper spectrum ($\beta=3.4$), which is characteristic of warm gas. Both fBms are normalized by their mean values. The size of the map is $1000\times1000$ pixels and the inner fBm is contained within a region whose radius is 300 pixels. The corresponding $\Delta$-variance spectrum of this model displays a broken power law with a transition point located at a scale of $\approx 50-100$ pixels. The break point is smaller than the imposed transition radius of 300 pixels because the low-intensity regions in both fBm have the same values, and thus the outer fBm has a non-negligible contribution to the signal on small scales because it covers a large surface area. Fig.~\ref{fig14} (middle and right subpanels) shows that depressing the value of the outer fBm by a certain factor (here 2 and 4, respectively) results in a spectrum in which the transition between the two regimes in the $\Delta$-variance spectra is sharper and the transition point nearer to the radius of the inner fBm. Despite being overly simplistic in comparison to a real galactic disk in which the stable cold and warm gas phases can locally coexist with gas in the unstable regime (e.g., Dib et al. 2006), this toy model shows that a dominant cold \ion{H}{i} component in the inner region of a galactic disk and a warm \ion{H}{i} component that dominates the emission in the outer regions of the disk can explain the broken power law that is observed in the $\Delta$-variance spectra of the \ion{H}{i} 21 cm emission line in the THINGS galaxies.

\section {Discussion and connection to previous work}\label{discussion}

\begin{figure}
\centering
\includegraphics[width=\columnwidth]{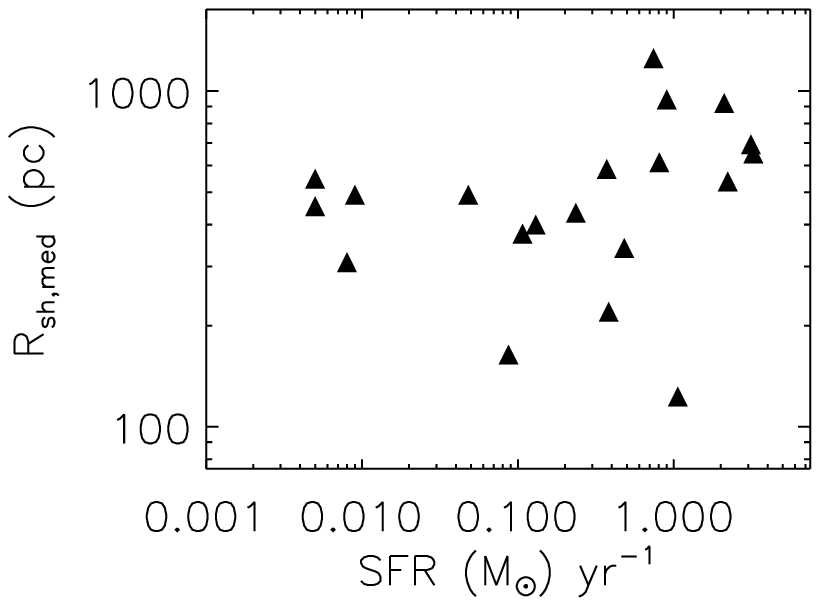} \\
\includegraphics[width=\columnwidth]{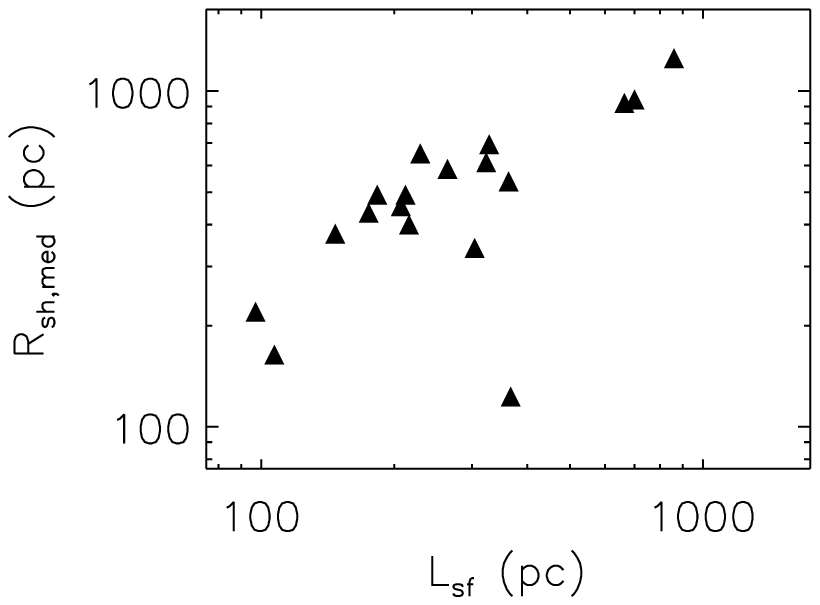}
\caption{Top panel: Relationship between the galactic SFR and the median size of the \ion{H}{i} shells in the catalogue of Bagetakos et al. (2011). Bottom panel: Comparison of the median sizes of the \ion{H}{i} shells in the catalogue of Bagatekos et al. (2011) with the values of $L_{sf}$ derived in this work.}
\label{fig15}
\end{figure}  

Several other studies have explored the structure of the diffuse ISM on the scale of entire galaxies either using the \ion{H}{i} 21 cm line emission or other tracers of the diffuse gas, such as dust mid- to far-infrared emission (e.g., Koch et al. 2020). These studies can be sorted into two main categories. On the one hand, there are studies that used isotropic methods in order to characterize the structure of the \ion{H}{i} gas distribution in galactic disks, such as the calculation of the power spectra of the \ion{H}{i} intensity map or of the line-of-sight velocity fluctuations (e.g., Begum et al. 2006; Dutta \& Bharadwaj 2013; Dutta et al. 2013; Szotkowski et al. 2019; Nandakumar \& Dutta 2020), the auto-correlation function of the \ion{H}{i} intensity map (Dib \& Burkert 2005) or the $\Delta$-variance spectrum (Elmegreen et al. 2001). Elmegreen et al. (2001) computed the power spectrum and the $\Delta$-variance spectrum of the \ion{H}{i} intensity for the LMC. They found two distinct power-law regimes with a transition that occurs at $\approx 250-300$ pc. They were unable to explore the dependence of the shape of the $\Delta$-variance spectra on the galactic star formation activity as their study was restricted to the case of a single galaxy. Dutta \& Bharadwaj (2013) and Dutta et al. (2013) measured the power spectrum of the \ion{H}{i} intensity for a number of the THINGS galaxies. While their approach and data sample overlaps with ours, they favored fitting the entire spectrum of each galaxy with a single power-law function. Their approach could be entirely valid over specific spatial ranges in each galaxy, however, there are instances where a single power-law fit cannot be justified (e.g., see the case of NGC 3184 in Fig. 2 of Dutta \& Bharadwaj 2013). Furthermore, Dutta et al. (2013) did not find any correlation between the exponent of the power spectra and some of the galactic properties they have considered, such as the inclination, the \ion{H}{i} and dynamical masses of the galaxy, and the surface density of the SFR. At first glance, our results might seem to contradict those of Combes et al. (2012) for M33 and Szotkowski et al. (2019) in the Large and Small Magellanic Clouds. Those authors found a steep spectrum on "small" scales and a shallower spectrum at "larger" scales. However, because of the relative proximity of these galaxies, the small scales in these works refer to scales that are not resolved in the THINGS sample. The shallow slopes they find at "larger" scales are similar to those we measure in our study over the same range of spatial scales.

Another approach for studying the structure of the ISM relies on the identification of discrete structures in galactic disks and on quantifying their statistical properties. This approach has been employed to detect \ion{H}{i} holes in the Milky Way (Ehlerov\'{a} \& Palou\v{s} 2005,2013). Ehlerov\'{a} \& Palou\v{s} (2013) measured the size distribution of shells in the Leiden, Argentina, Bonn \ion{H}{i} survey and found that it can be fit with a power-law function $\left(dN/dR_{shell}\right) \propto R_{shell}^{-\xi\approx -2.6}$. Oey \& Clarke (1997) noted that if the \ion{H}{i} shells are the results of feedback from massive stars, then a relation exists between $\xi$ and the exponent of the power-law function that describes the luminosity distribution of OB associations, $\phi\left(L\right) \propto L^{-\eta}$, such that $\xi=2\eta-1$. Ehlerov\'{a} \& Palou\v{s} (2013) already noted that the value of $\xi \approx 2.6$ they have derived implies that $\eta \approx 1.8,$ which is close to the value that is derived from observations ($\approx 2$, McKee \& Williams 1997). Dib et al. (2009) found that the orientations of the main axis of molecular clouds in the outer Galaxy are correlated on spatial scales that are approximately the expected sizes of supernova remnants that are found in these regions of the Galaxy. The results of Dib et al. (2009) and Ehlerov\'{a} \& Palou\v{s} (2013) clearly suggest that feedback processes from massive stars play an important role in shaping the structure of the ISM in the Milky Way. In nearby galaxies, Bagetakos et al. (2011) searched for \ion{H}{i} shells in a subsample of the THINGS galaxies using both a simple morphological selection and a selection based on localized expansion velocity. They found that the size distribution of the \ion{H}{i} shells in each of the THINGS galaxies peaks at a few to several hundred parsecs (see Fig. 3 in their paper), which is broadly similar to the position and width of the bump observed in the $\Delta$-variance spectrum displayed in Fig.~\ref{fig2} and Fig.~\ref{fig3}. Using numerical simulations, Yadav et al. (2017) showed that multiple supernovae remnants can merge to form large bubbles with sizes that range between $\approx 100$ pc and $700$ pc, which is very similar to the range of $L_{sf}$ values we find in this work.  

Figure \ref{fig15} (top subpanel) displays the median size of the \ion{H}{i} shells ($R_{sh,med}$) from Bagetakos et al. (2011) plotted as a function of the galactic SFR. The similarity between the SFR-$R_{sh,med}$ and the SFR-$L_{sf}$ relations is striking. Like in the case of the $L_{sf}$, $R_{sh,med}$ shows no significant dependence on the SFR for value of the SFR $\lesssim 0.5$ M$_{\odot}$ yr$^{-1}$ and a positive correlation at higher values. The lower subpanel of Fig.~\ref{fig15} displays $R_{sh,med}$ plotted as a function of $L_{sf}$.  A clear one-to-one correlation exists between these two quantities, and this confirms the supernovae feedback origin of the characteristic scale $L_{sf}$ that is measured from the $\Delta$-variance spectra. With the exception of one outlier galaxy, the values of $R_{sh,med}$ seem to be systematically higher than those of $L_{sf}$. A plausible explanation is that because the \ion{H}{i} holes in Bagetakos et al. (2011) were entirely identified by eye, the selection favored the identification of the largest holes, and some of the smaller \ion{H}{i} holes went undetected, especially those that could be elongated or deformed. 

\section{Conclusions}\label{conclusions}

We analyzed the structure of the \ion{H}{i} gas using the order zero-maps of 33 galaxies taken from the THINGS survey (Walter et al. 2008). In order to characterize the \ion{H}{i} gas structure, we calculated the $\Delta$-variance spectrum (St\"{u}tzki et al. 1998; Ossenkopf et al. 2008). Most spectra possess common features that include a bump at scales of a few to several hundred parsec, a first self-similar regime at intermediate spatial scales, and a transition to a second, steeper, self-similar regime at larger spatial scales. When extrapolating the first power law to smaller scales and subtracting it from the observed spectra, we were able to measure the position of the maximum deviation between the spectra and the underlying power law, $L_{sf}$. We find that $L_{sf}$, whose values range from one to several hundred parsecs, correlates with the galactic SFR for SFR values $\gtrsim 0.5$ M$_{\odot}$ yr$^{-1}$. Below this value, $L_{sf}$ is independent of the SFR. We also find a strong correlation between the value of $L_{sf}$ for each galaxy and the median size of \ion{H}{i} shells measured by Bagetakos et al. (2011). Both findings clearly suggest that $L_{sf}$ is a measure of the characteristic size of the \ion{H}{i} shells in the THINGS galaxy. The first similar regime is observed to extend from beyond the bump up to a spatial scale of $\approx 0.5 R_{25}$ , and it can be described by a power law whose exponent ranges from $\approx 0.1$ to 1 with a mean value of $\approx 0.55$. This exponent is compatible with the occurrence of compressible supersonic turbulence, which governs the dynamics of the cold component of the \ion{H}{i} gas. On larger spatial scales (i.e., $\gtrsim 0.5 R_{25}$), the structure of the \ion{H}{i} gas can be characterized by a second power law whose exponent is found to vary between $\approx 0.5$ and $2.5$ with a mean value of $1.47$. We find that the values of $\alpha_{2}$ do not correlate with the galactic SFR. This regime corresponds to the dynamics of the gas being governed by subsonic to transonic turbulence. It therefore is a signature of emission from the warm component of the \ion{H}{i} gas\footnote{The velocity dispersions of the \ion{H}{i} gas are observed to vary from 12-15 km s$^{-1}$ in the central regions of galaxies and plateau at 5-7 km$^{-1}$ in the outer regions (Dib et al. 2006; Ianjamasimanana et al. 2015). For warm gas with temperatures in the range 5000 to 10000 K, this implies sound speeds in the range $\approx 5-10$ km s$^{-1}$, and hence turbulence is most likely to lie in the subsonic to transonic regime.}. The transition point between the two self-similar regimes, $L_{tr}$ , is found to correspond to a spatial scale of $\approx 0.4-0.5 R_{25}$. Interestingly, in most THINGS galaxies, this scale is about the size of the molecular disk, and this is probably an indication of where most of the cold \ion{H}{i} gas resides. 

Earlier work on the scale of molecular clouds (scales of 5 to 50 pc) using the $\Delta$-variance technique has allowed us to uncover characteristic scales of  $\approx 1$ pc in massive star-forming regions such as Cygnus X (Dib et al. 2020). These scales are thought to be associated with the sizes of hubs where stellar clusters form. With the current data from the THINGS survey, it is not possible to probe the connection between what we observe in the \ion{H}{i}, and particularly, its cold component with the structure of molecular clouds. Future observations with the Square Kilometer Array will allow us to start probing scales that are about 4-20 pc for galaxies located at distances of $\approx [1-7]$ Mpc (Tolstoy et al. 2010). Combined with high-resolution observations of molecular clouds both in the Galaxy and in nearby galaxies using the Atacama Large Millimeter Array (ALMA), we will be able to probe the link between the structure observed in the cold \ion{H}{i} gas and that seen in the submillimeter and in molecular line transitions. This will allow us to explore the effects of feedback in the pre-supernova phase on the structure of the ISM in greater detail.     

\begin{acknowledgements}

This work is dedicated to the city of Beirut and to its inhabitants for their resilience in the face of adversity. We thank the referee for a careful reading of the paper and for constructive comments. S. D. would like to thank Volker Ossenkopf-Okada for useful discussions on the calculation of the $\Delta$-variance spectrum and Andreas Schruba and Robert Kennicutt for useful discussions on the star formation rates in nearby galaxies. We also thank Bruce Elmegreen for commenting on an earlier version of this paper and Elias Brinks and Fabian Walter for clarifications on the THINGS data. We thank Oscar Agertz and Florent Renaud for making data from their simulation available. JAB was supported by Comision Nacional de Ciencias Y Tecnologia through Fondecyt Grant No. 11170083.
 
\end{acknowledgements}

%

\begin{thebibliography}{}

\bibitem[Adler (1996)] {adler96} Adler, D. S., Westpfahl, D. J. 1996, AJ, 111, 735   
\bibitem[Agertz (2013)] {agertz13} Agertz, O., Kravtsov, A. V., Leitner, S. N., Gnedin, N. Y. 2013, ApJ, 770, 25
\bibitem[Agertz (2015)] {agertz15} Agertz, O., Kravtsov, A. V. 2015, ApJ, 804, 18
\bibitem[Agertz (2020)] {agertz21} Agertz, O., Renaud, F., Feltzing, S. et al. 2021, MNRAS, 503, 5826
\bibitem[Aouad (2020)] {aouad20} Aouad, C. J., James, P. A., Chilingarian, I. V. 2020, MNRAS, 496, 5211
\bibitem[Bacchini (2019)] {bacchini19} Bacchini, C., Fraternalli, F., Iorio, G., Pezzulli 2019, A\&A, 622, 64 
\bibitem[Bacchini (2020)] {bacchini20} Bacchini, C., Fraternalli, F., Iorio, G. et al. 2020, A\&A, 641, 70 
\bibitem[Bagetakos (2011)] {bagetakos11} Bagetakos, I., Brinks, E., Walter, F. et al. 2011, ApJ, 141, 23
\bibitem[Bensch (2001)] {bensch01} Bensch, F., Stutzki, J., Ossenkopf, V. 2001, A\&A, 366, 636
\bibitem[Begum (2010)] {begum10} Begum, A., Chengalur, J. N., Bhardwaj, S. 2006, MNRAS, 372, 33
\bibitem[Bertram (2015)] {bertram15} Bertram, E., Klessen, R. S., Glover, S. C. O. 2015, MNRAS, 451, 196
\bibitem[Bolatto (2008)] {bolatto08} Bolatto, A. D., Leroy, A. K., Rosolowsky, E. et al. 2008, ApJ, 685, 948
\bibitem[Bolatto (2013)] {bolatto13} Bolatto, A. D., Wolfire, M., Leroy, A. K. 2013, ARA\&A, 51, 207
\bibitem[Boomsma (2008)] {boomsma08} Boomsma, R., Oosterloo, T. A., Fraternalli, F. et al. 2008, A\&A, 490, 555
\bibitem [Braine (2001)] {brain01} Braine, J., Duc, P.-A., Lisenfeld, U. et al. 2001, A\&A, 378, 51
\bibitem[Braine (2010)] {braine10} Braine, J., Gratier, P., Kramer, C. et al. 2010, A\&A, 520, 107    
\bibitem[Braun (1997)] {braun97} Braun, R. 1997, ApJ, 484, 637
\bibitem[Burkhart (2013)] {burkhart13} Burkhart., B., Lazarian, A., Ossenkopf, V., Stutzki, J. 2013, ApJ, 771, 123
\bibitem[Cannon (2012)] {cannon12} Cannon, J. M., O'Leary, E. M., Weisz, D. R. et al. 2012, ApJ, 704, 1538
\bibitem [Casasola 2017] {casasola17} Casasola, V., Cassar\`{a}, L. P., Bianchi, S. et al. 2017, A\&A, 605, 18
\bibitem[Chamandy (2020)] {chamandy20} Chamandy, L., Shukurov, A. Galaxies, 8, 56
\bibitem[Chepurnov (2010)] {chepurnov10} Chepurnov, A., Lazarian, A., Stanimirovi\'{c}, S. et al. 2010, ApJ, 714, 1398
\bibitem[Clemens (2000)] {clemens00} Clemens, M. S., Alexander, P., Green, D. A. 2000, MNRAS, 312, 236
\bibitem[Combes (1988)] {combes88} Combes, F., Dupraz, C., Casoli, F., Pagani, L. 1988, A\&A, 203, 9
\bibitem[Combes (2012)] {combes12} Combes, F., Boquien, M., Kramer, C. et al. 2012, A\&A, 539, 67
\bibitem[de Avillez (2005)] {deavillez05} de Avillez, M. A., Breitschwerdt, D. 2005, A\&A, 436, 585
\bibitem[de Blok (2008)] {deblok08} de Blok, W. J. G., Walter, F., Brinks, E. et al. 2008, AJ, 136, 2648
\bibitem[Deshpande (2000)] {deshpande00} Deshpande, A., Dwarakanath, K., Goss, W. 2000, ApJ, 543,227
\bibitem[Dib (2004)] {dib04} Dib, S., Burkert, A. 2004, Ap\&SS, 292, 135
\bibitem[Dib (2005)] {dib05a} Dib, S., Burkert, A. 2005, ApJ, 630, 238 
\bibitem[Dib (2005)] {dib05b} Dib, S. 2005, PhD thesis, University of Heidelberg, Germany
\bibitem[Dib (2006)] {dib06} Dib, S., Bell, E., Burkert, A. 2006, ApJ, 638, 797
\bibitem[Dib (2008)] {dib08} Dib, S., Brandenburg, A., Kim, J., Gopinathan, M., Andr\'{e}, P. 2008, ApJ, 678, L105
\bibitem[Dib (2009)] {dib09} Dib, S., Walcher, C. J., Heyer, M., Audit, E., Loinard, L. 2009, MNRAS, 398, 1201
\bibitem[Dib (2011a)] {dib11a} Dib, S., Piau, L., Mohanty, S., Braine, J. 2011, MNRAS, 415, 3439 
\bibitem[Dib (2011b)] {dib11b} Dib, S. 2011, ApJ, 737, L20
\bibitem[Dib (2012)] {dib12} Dib, S., Helou, G., Moore, T. J. T., Urquhart, J. S., Dariush, A. 2012, ApJ, 758, 125
\bibitem[Dib (2013)] {dib13} Dib, S., Gutkin, J., Brandner, W., Basu, S. 2013, MNRAS, 436, 3727 
\bibitem[Dib (2017)] {dib17} Dib, S., Hony, S., Blanc, G. 2017, MNRAS, 469, 1521
\bibitem[Dib (2019)] {dib19} Dib, S., Henning, S. 2019, A\&A, 629, 135
\bibitem[Dib (2020)] {dib20} Dib, S., Bontemps, S., Schneider, N. et al. 2020, A\&A, 642, 177 
\bibitem[Dickey (1993)] {dickey93} Dickey, J. M., Brinks, E. 1993, ApJ, 405, 153
\bibitem[Dickey (2001)] {dickey01} Dickey, J. M., McClure-Griffiths, N. M., Stanimirovi\'{c}, S. et al. 2001, ApJ, 561, 264
\bibitem[Dutta (2013a)] {dutta13a} Dutta, P., Bharadwaj, S. 2013, MNRAS, 436, L49
\bibitem[Dutta (2013b)] {dutta13b} Dutta, P., Begum, A., Bharadwaj, S., Chengalur, J. N. 2013, New Astron. 19, 89
\bibitem[Eden (2020)] {eden20} Eden, D. J., Moore, T. J. T., Plume, R. et al. 2020, MNRAS, 2995 
\bibitem[Ehlerova (1996)] {ehlerova96} Ehlerov\'{a}, S., Palou\v{s}, J. 1996, A\&A, 313, 478 
\bibitem[Ehlerova (2013)] {ehlerova05} Ehlerov\'{a}, S., Palou\v{s}, J. 2005, A\&A, 437, 101
\bibitem[Ehlerova (2013)] {ehlerova13} Ehlerov\'{a}, S., Palou\v{s}, J. 2013, A\&A, 550, 23
\bibitem[Elia (2018)] {elia18} Elia, D., Strafella, F., Dib, S. et al. 2018, MNRAS, 481, 509     
\bibitem[Elmegreen (2001)] {elmegreen01} Elmegreen, B. G., Kim, S., Staveley-Smith, L. 2001, ApJ, 548, 749
\bibitem[Elmegreen (2004)] {elmegreen04} Elmegreen, B. G., Scalo, J. 2004, ARA\&A, 42, 211
\bibitem[Elmegreen (2006)] {elmegreen06} Elmegreen, B. G.; Elmegreen, D. M., Chandar, R. er al. 2006, ApJ, 644, 879
\bibitem[Elmegreen (2011)] {elmegreen11} Elmegreen, B. G. 2011, ApJ, 737, 10
\bibitem[Fattahi (2018)] {fattahi18} Fattahi, A., Navarro, J. F., Frenk, C. S. et al. 2018, MNRAS, 476, 3816
\bibitem[Federrath (2008)] {federrath08} Federrath, C., Klessen, R. S., Schmidt, W. 2008, ApJ, 688, 79
\bibitem[Fraternalli (2002)] {fraternalli02} Fraternalli, F., van Moorsel, G., Sancisi, R., Oosterloo, T. 2002, AJ, 123, 3124  
\bibitem[Franco (2002)] {franco02} Franco, J., Kim, J., Alfaro, E, J., Hong, S. S. 2002, ApJ, 570, 647
\bibitem[Freeland (2010)] {freeland10} Freeland, E., Sengupta, C., Croston, J. H. 2010, MNRAS, 409, 1518
\bibitem[Gardan (2007)] {gardan07} Gardan, E., Braine, J., Schuster, K. F. et al. 2007, A\&A, 473, 91
\bibitem[Gent (2013)] {gent13} Gent, F. A., Shukurov, A., Fletcher, A. et al. 2013, MNRAS, 432, 1396
\bibitem[Ghosh (2018)] {ghosh18} Ghosh, S., Jog, C. J. 2018, New Astron. 63, 38
\bibitem[Gonzalez (2012)] {gonzalezlopez12} Gonz\'{a}lez-L\'{o}pezlira, R. A., Pflamm-Altenburg, J., Kroupa, P. 2012, ApJ, 761, 124
\bibitem[Gonzalez (2013)] {gonzalezlopez13} Gonz\'{a}lez-L\'{o}pezlira, R. A., Pflamm-Altenburg, J., Kroupa, P. 2013, ApJ, 770, 85
\bibitem[Grasha (2019)] {grasha19} Grasha, K., Calzetti, D., Adamo, A. et al. 2019, MNRAS, 483, 4707
\bibitem[Gratier (2010)] {gratier10} Gratier, P., Braine, J., Rodriguez-Fernandez, N. J. et al. 2010, A\&A, 512, 68
\bibitem[Gratier(2017) ] {gratier17} Gratier, P., Braine, J., Schuster, K. et al. 2017, A\&A, 600, 27
\bibitem[Green (1993)] {green93} Green, D. A. 1993, MNRAS, 262, 327
\bibitem[Guibert (1974)] {guibert74} Guibert, J. 1974, A\&A, 30, 353
\bibitem[Hanasz (2003)] {hanasz03} Hanasz, M., Lesch, H. 2003, A\&A, 412, 331
\bibitem[Heiles (1979)] {heiles79} Heiles, C. 1979, ApJ, 229, 533
\bibitem[Helfer (2003)] {helfer03} Helfer, T. T., Thornley, M. D., Regan, M. W. et al. 2003, ApJS, 145, 259
\bibitem[Heyer (2004)] {heyer04} Heyer, M. H., Brunt, C. M. 2004, ApJ, 615, 45
\bibitem[Heintz (2020)] {heintz20} Heintz, E., Bustard, C., Zweibel, E. G. 2020, ApJ, 891, 157
\bibitem[Heitsch (2009)] {heitsch09} Heitsch, F., Putman, M. E. 2009, ApJ, 698, 1485
\bibitem[Hodge (2006) ]{hodge06} Hodge, J. A., Desphande, A. A. 2006, ApJ, 646, 232 47
\bibitem[Hollenbach (1971)] {hollenbach71} Hollenbach, D. J., Werner, M. W., Salpeter, E. E. 1971, ApJ, 163, 165
\bibitem[Holwerda (2013)] {holwerda13} Holwerda, B. W., Pirzkal, N., de Block, W. J. G., Blyth, S.-L. 2013, MNRAS, 435, 1020
\bibitem[Hony (2015)] {hony15} Hony, S., Gouliermis, D. A., Galliano, F. et al. 2015, MNRAS, 448, 1847
\bibitem[Ianjamasimanana (2015)] {ianjamasimanana15} Ianjamasimanana, R., de Blok, W. J. G., Walter, F. et al. 2015, AJ, 150, 47
\bibitem[Jog (1984)] {jog84} Jog, C. J., Solomon, P. M. 1984, ApJ, 276, 114
\bibitem[Khoperskov (2015)] {khoperskov15} Khoperskov, S. A., Bertin, G. 2015, MNRAS, 451, 2889
\bibitem[Kim (2007)] {kim07} Kim, S., Park, C. 2007, ApJ, 663, 244
\bibitem[Koch (2020)] {koch20} Koch, E. W., Chiang, I.-Da, Utomo, D. et al. 2020, MNRAS, 492, 2663 
\bibitem[Kowal (2007)] {kowal07} Kowal, G., Lazarian, A., Beresnyak, A. 2007, ApJ, 658, 423
\bibitem[Krumholz (2009)] {krumholz09} Krumholz, M. R., McKee, C. F., Tumlinson, J. 2009, ApJ, 693, 216
\bibitem[Laralopez (2013)] {laralopez13} Lara-L\'{o}pez, M., L\'{o}pez-Sanchez, A. R., Hopkins, \'{A}. M. et al. 2013, ApJ, 764, 178 
\bibitem[Lazarian (2000)] {lazarian00} Lazarian, A., Pogosyan, D. 2000, ApJ, 537, 720     
\bibitem[Leroy (2008)] {leroy08} Leroy, A. K., Walter, F., Brinks, E. et al. 2008, ApJ, 136, 2782 
\bibitem[Leroy (2009)] {leroy09} Leroy, A. K., Walter, F., Bigiel, F. et al. 2009, AJ, 137, 4670
\bibitem [Li (2021)] {li21} Li, Z., Krumholz, M. R., Wisnioski, E. et al. 2021, MNRAS, 504, 5496
\bibitem[Lin (1966)] {lin66} Lin, C. C., Shy, F. 1966, PNAS, 55, 229
\bibitem[Lipnicky (2018)] {lipnicky18} Lipnicky, A., Chakrabarti, S., Chang, P. 2018, MNRAS, 481, 2590 
\bibitem[Marchuk (2018a)] {marchuk18a} Marchuk, A. A., Sotnikova, N. Y. 2018, MNRAS, 475, 4891
\bibitem[Marchuk (2018b)] {marchuk18b} Marchuk, A. A. 2018, MNRAS, 476, 3591 
\bibitem[Marcolini (2003)] {marcolini03} Marcolini, A., Brighenti, F., D'Ercole, A. 2003, MNRAS, 345, 1329
\bibitem[Martin (2015)] {martin15} Martin, P. G., Blagrave, K. P. M., Lockman, F. J. et al. 2015, ApJ, 809, 153
\bibitem[Marziani (2003)] {marziani03} Marziani, P., Dultzin-Hacyan, D., D'Onofrio, M., Sultentic, J. N. 2003, AJ, 125, 1897
\bibitem[Mayer (2006)] {mayer06} Mayer, L., Mastropietro, C., Wadsley, J. et al. 2006, MNRAS, 369, 1021
\bibitem[Mckee (1997)] {mckee97} McKee, C. F., Williams, J. P. 1997, ApJ, 476, 144
\bibitem[Miville-Deschenes] {miville10} Miville-Desch\^{e}nes, M.-A., Martin, P., Abergel, A. 2010, A\&A, 518, 104
\bibitem[Miville-Deschenes] {miville03} Miville-Desch\^{e}nes, M.-A., Joncas, G., Falgarone, E., Boulanger, F. 2003, A\&A, 411, 109
\bibitem[Mouschovias (2009)] {mouschovias09} Mouschovias, T. Ch., Kunz, M. W., Christie, D. A. 2009, MNRAS, 397, 14
\bibitem[Murgia (2002)] {murgia02} Murgia, M., Craspi, A., Moscadelli, L., Gregorini, L. 2002, A\&A, 385, 412
\bibitem[Nandakumar (2020)] {nandakumar20} Nandakumar, M., Dutta, P. 2020, MNRAS, 496, 1803
\bibitem[Oey (1997)] {oey97} Oey, M. S., Clarke, C. J. 1997, MNRAS, 289, 570
\bibitem[Ohlin (2019)] {ohlin19} Ohlin, L., Renaud, F., Agertz, O. 2019, MNRAS, 485, 388
\bibitem[Ossenkopf (2001)] {ossenkopf01} Ossenkopf, V., Klessen, R. S., Heitsch, F. 2001, A\&A, 379, 1005
\bibitem[Ossenkopf (2008)] {ossenkopf08} Ossenkopf, V., Krips, M., Stutzki, J. 2008 A\&A, 485, 917 
\bibitem[Park (2016)] {park16} Park, G., Koo, B.-C., Kang, J.-h. et al. 2016, ApJ, 827, 27
\bibitem[Parker (1967)] {parker67} Parker, E. N. 1967, ApJ, 149, 535 
\bibitem[Patra (2020a)] {patra20a} Patra, N. N. 2020a, MNRAS, 499, 2063
\bibitem[Patra (2020b)] {patra20b} Patra, N. N. 2020b, MNRAS, 495, 2867
\bibitem[Paturel (2003)] {paturel03} Paturel, G., Petit, C., Prugniel, P. et al. 2003, A\&A, 412, 45
\bibitem[Pokhrel (2020)] {pokhrel20} Pokhrel, N. R., Simpson, C. E., Bagetakos, I. 2020, AJ, 160, 66
\bibitem[Rathjen (2021)] {rathjen21} Rathjen, T.-E., Naab, T., Girichidis, P. et al. 2021, MNRAS, 886
\bibitem[Renaud (2009)] {renaud09} Renaud, F., Boily, C., Naab, T., Theis Ch. 2009, ApJ, 706, 67
\bibitem[Renaud (2021a)] {renaud20a} Renaud, F., Agertz, O., Read, J. I. et al. 2021a, MNRAS, 503, 5846
\bibitem[Renaud (2021b)] {renaud20b} Renaud, F., Agertz, O., Andersson, E. P. et al. 2021b, MNRAS, 503, 5868 
\bibitem[Rodrigues (2016)] {rodrigues16} Rodrigues, L. F. S., Sarson, G. R., Shukurov, A. et al. 2016, ApJ, 816, 2
\bibitem[Rosen (1995)] {rosen95} Rosen, A., Bregman, J. N. 1995, ApJ, 440, 634
\bibitem[Rowles (2011)] {rowles11} Rowles, J., Froebrich, D. 2011, MNRAS, 416, 294
\bibitem[Santillan (1999)] {santillan99} Santillan, A., Franco, J., Martos, M., Kim, J. 1999, ApJ, 515, 657
\bibitem[Shadmehri (2012)] {shadmehri12} Shadmehri, M., Khajenabi, F. 2012, MNRAS, 421, 841
\bibitem[Shetty (2008)] {shetty08} Shetty, R., Ostriker, E. C. 2008, ApJ, 684, 978  
\bibitem[Schulz (2015)] {shulz15} Shulz, C., Pflamm-Altenburg, J., Kroupa, P. 2015, A\&A, 582, 93
\bibitem[Silich (2006)] {silich06} Silich, S., Lozinskaya, T., Moiseev A. et al. 2006, A\&A, 448, 123
\bibitem[Stanimirovic (1999)] {stanimirovic99} Stanimirovi\'{c}, S., Staveley-Smith, L., Dickey, J. M. et al. 1999, MNRAS, 302, 417 
\bibitem[Steyrleithner (2020)] {steyrleithner20} Steyrleithner, P., Hensler, G., Boselli, A. 2020, MNRAS, 494, 1114
\bibitem[Stilp (2013)] {stilp13} Stilp, A. M., Dalcanton, J. J., Skillman, E. et al. 2013, ApJ, 773, 88 
\bibitem[Stutzki (1998)] {stutszki98} St\"{u}tzki, J., Bensch, F., Heithausen, A. et al. 1998, A\&A, 336, 697
\bibitem[Suad (2019)] {suad19} Suad, L. A., Caiafa, C. F., Cichowolski, S., Arnal, E. M. 2019, A\&A, 624, 43  
\bibitem[Sun (2006)] {sun06} Sun, K., Kramer, C., Ossenkopf, V. et al. 2006, A\&A, 451, 539
\bibitem[Sutherland (1993)] {sutherland93} Sutherland, R. S., Dopita, M. A. 1993, ApJS, 88, 253    
\bibitem[Szotkowski (2019)] {szotkowski19} Szotkowski, S., Yoder, D., Stanimirovi\'{c}, S. et al. 2019, ApJ, 887, 111
\bibitem[Tolstoy (2010)] {tolstoy10} Tolstoy, E., Battaglia, G., Beck, R., et al. 2010, Proceedings of the workshop "Astronomy with Megastructures. Joint Science with the E-ELT and SKA". Eds I. Hook, D. Rigopoulou, S. Rawlings, and A. Karastergiou, Crete University Press, 161
\bibitem[Tosaki (2007)] {tosaki07} Tosaki, T., Shioya, Y., Kuno, N. et al. 2007, PASJ, 59, 33
\bibitem[Vollmer (2004)] {vollmer04} Vollmer, B., Balkowski, C., Cayatte, V. et al. 2004, A\&A, 41, 35
\bibitem[Walter (2001)] {walter01} Walter, F., Taylor, C. L., H\"{u}nttenmeister, S. et al. 2001, AJ, 121,  727
\bibitem[Walter (2008)] {walter08} Walter, F., Brinks, E., de Blok, W. J. G. et al. 2008, AJ, 136, 2563 
\bibitem[Wang (2015)] {wang15} Wang, H.-H., Lee, W.-K., Taam, R. E. et al. 2015, ApJ, 800, 106
\bibitem[Weidner (2004)] {weidner04} Weidner, C., Kroupa, P., Larsen, S. 2004, MNRAS, 350, 1503
\bibitem[Weisz (2009)] {weisz09} Weisz, D., Skillman, E. D., Cannon, J. M. et al. 2009, ApJ, 704, 1538
\bibitem[Wolfire (2008)] {wolfire08} Wolfire, M. G., Tielens, A. G. G. M., Hollenbach, M., Kauffman, M. G. 2008, ApJ, 690, 384
\bibitem[Yadav (2017)] {yadav17} Yadav, N., Mukherjee, D., Sharma, P., Nath, B. B. 2017, MNRAS, 465, 1720     
\bibitem[Yahia (2020)] {yahia20} Yahia, H., Schneider, N., Bontemps, S., Bonne, L., Attuel, G., Dib, S. et al. 2021, A\&A, 649, 33
\bibitem[Zhang (2012)] {zhang12} Zhang, H.-X., Hunter, D., Elmegreen, B. G. 2012, ApJ, 754, 29  
\bibitem[Zielinsky (1999)] {zielinsky99} Zielinsky, M., St\"{u}tzki, J. 1999, A\&A, 347, 630

\end{thebibliography}
%

\begin{appendix}

\section{Importance of deprojecting inclined galaxies}\label{appa}

\begin{figure*}
\centering
\includegraphics[width=0.26\textwidth]{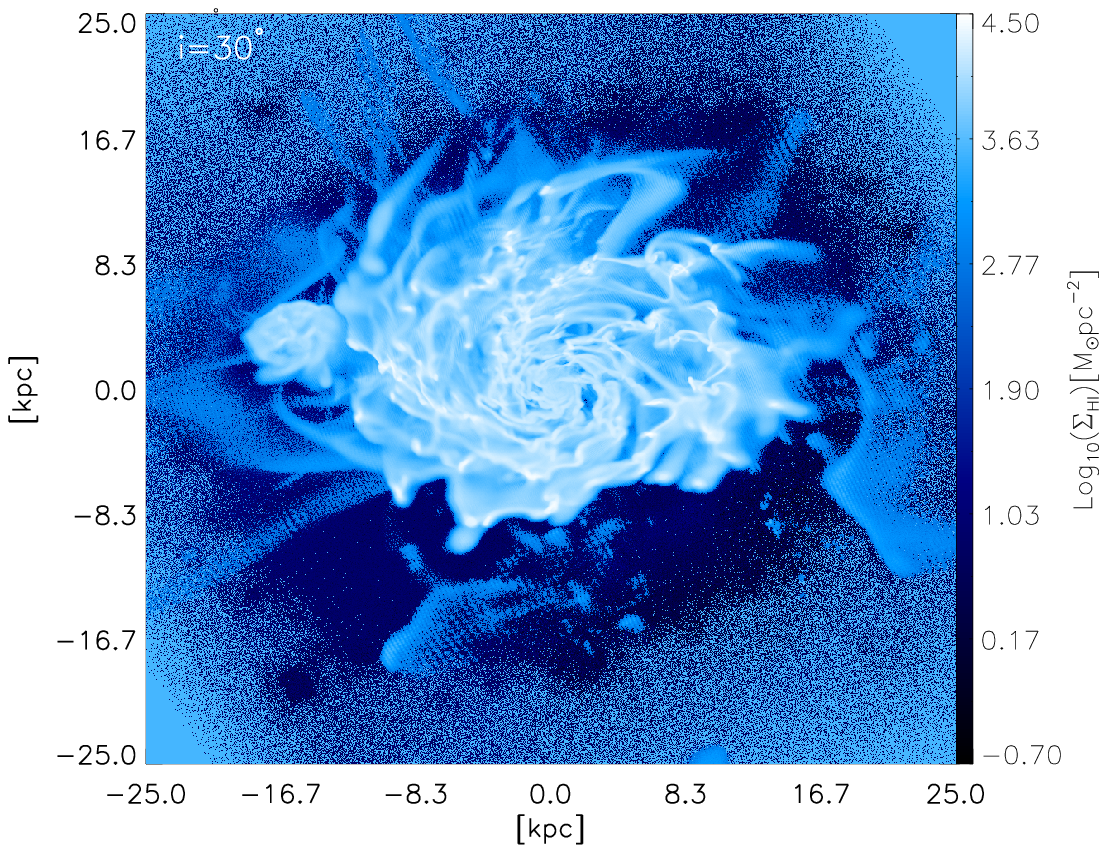}
\hspace{1.5cm}
\includegraphics[width=0.26\textwidth]{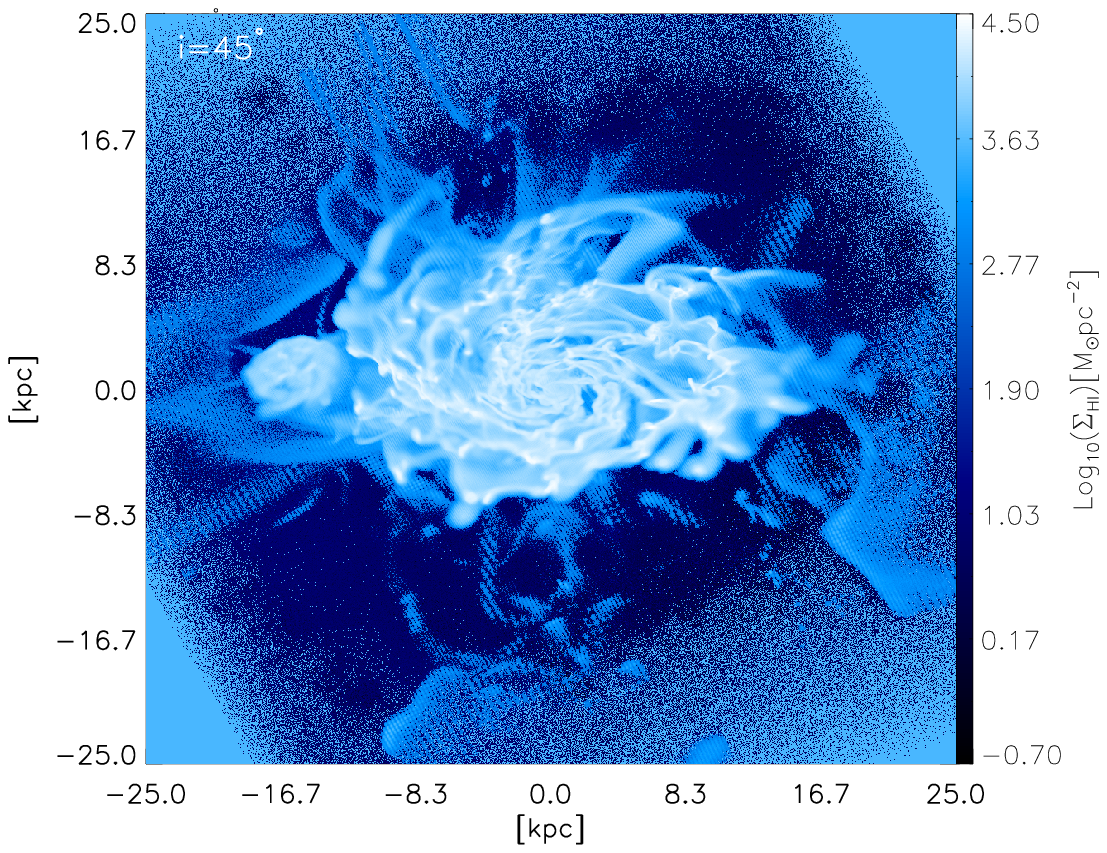}
\hspace{1.5cm}
\includegraphics[width=0.26\textwidth]{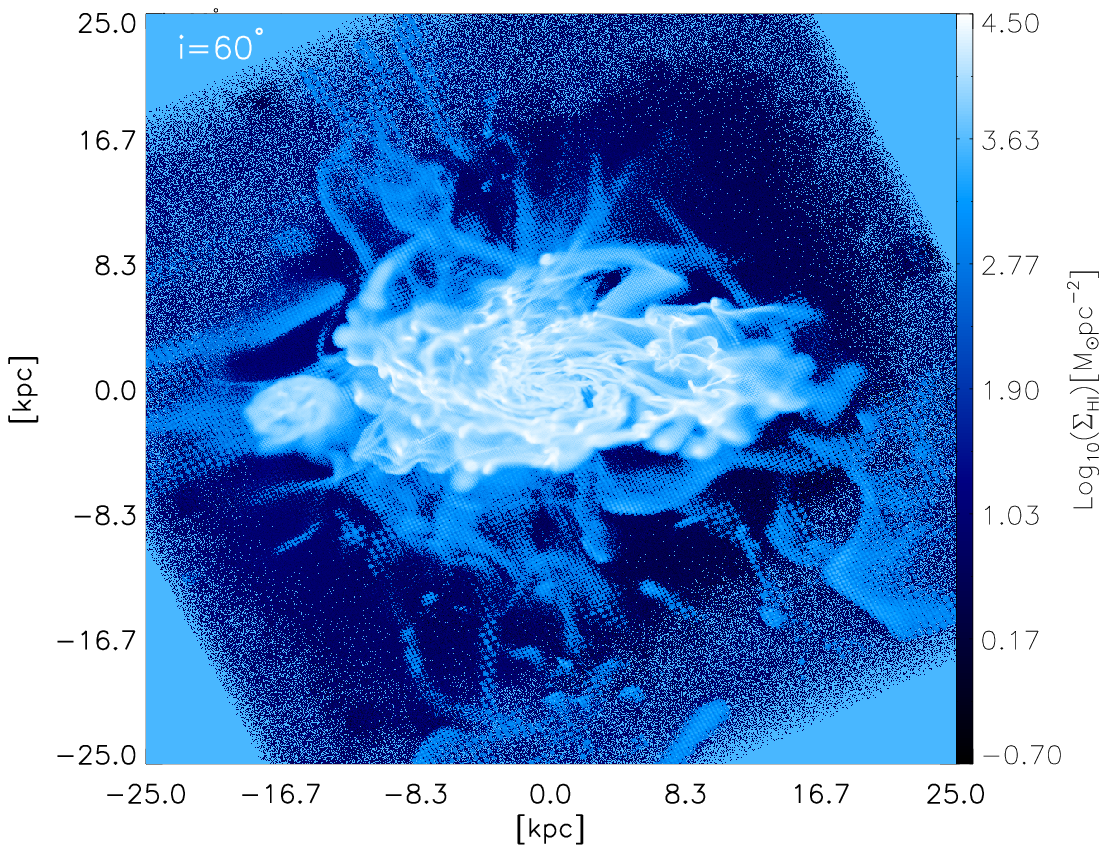}\\
\vspace{0.5cm}
\includegraphics[width=0.26\textwidth]{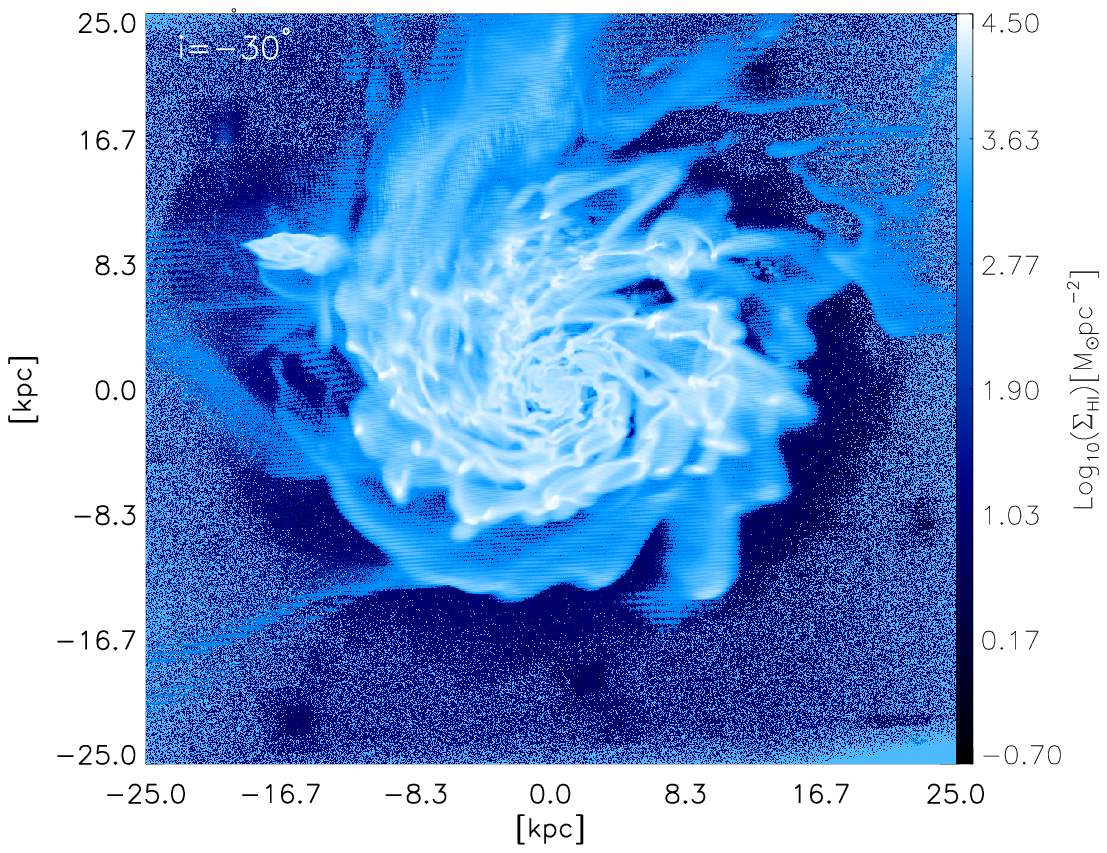}
\hspace{1.5cm}
\includegraphics[width=0.26\textwidth]{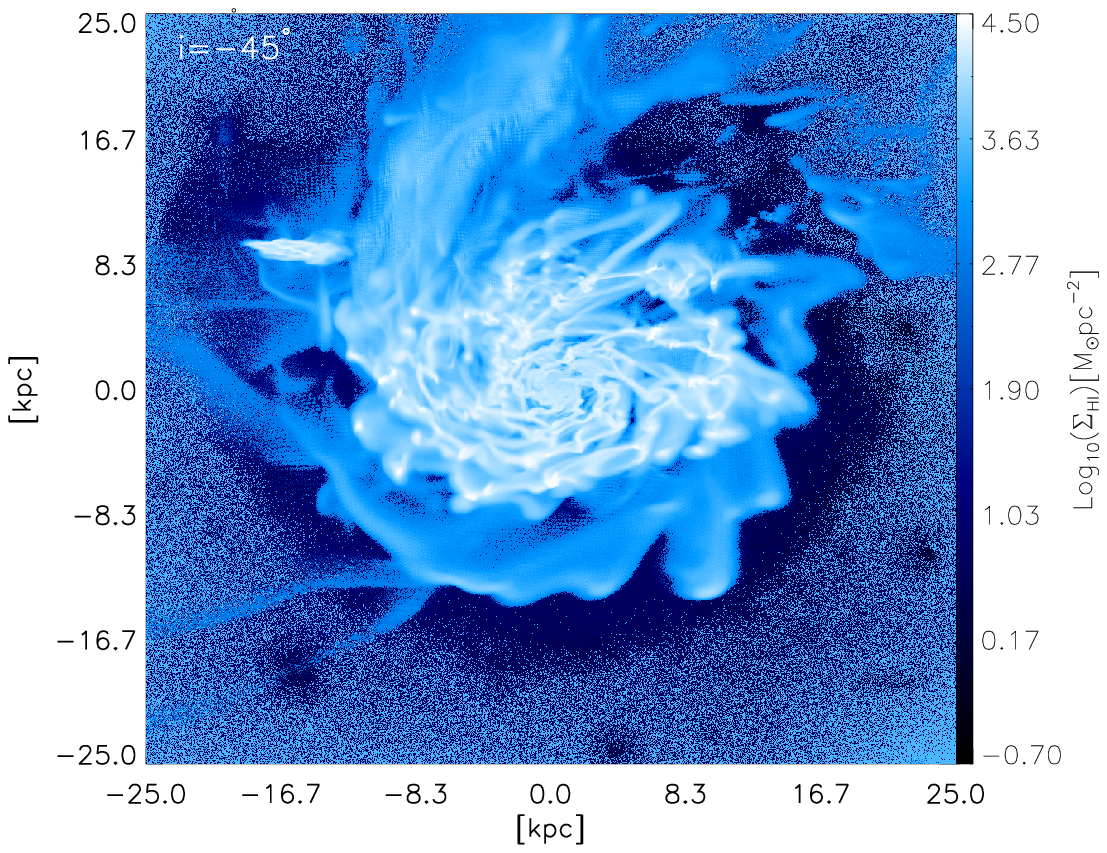}
\hspace{1.5cm}
\includegraphics[width=0.26\textwidth]{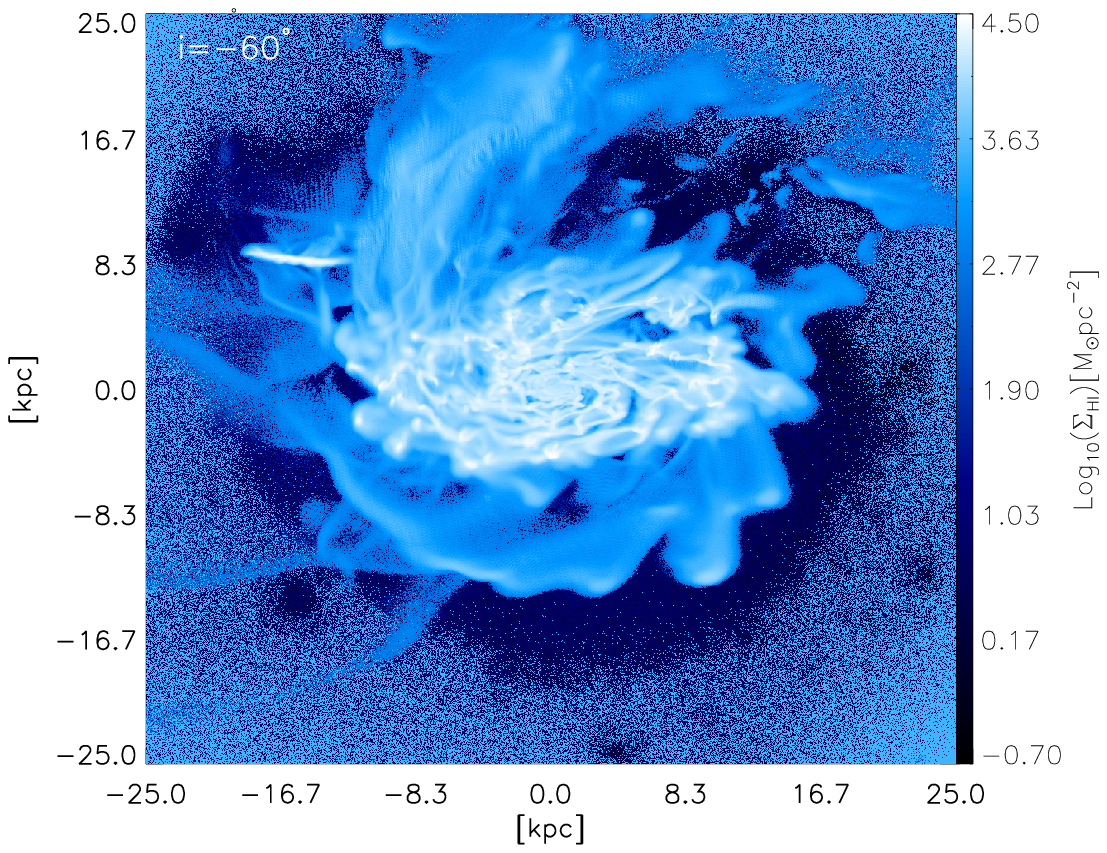}\\
\vspace{0.5cm}
\caption{View of the \ion{H}{I} surface density of the simulated galaxy for various inclination angles. All maps correspond to the snapshot at $t=t_{0}$  at the original resolution of 50 pc.}
\label{figapp1}
\end{figure*}

\begin{figure}
\centering
\includegraphics[width=\columnwidth]{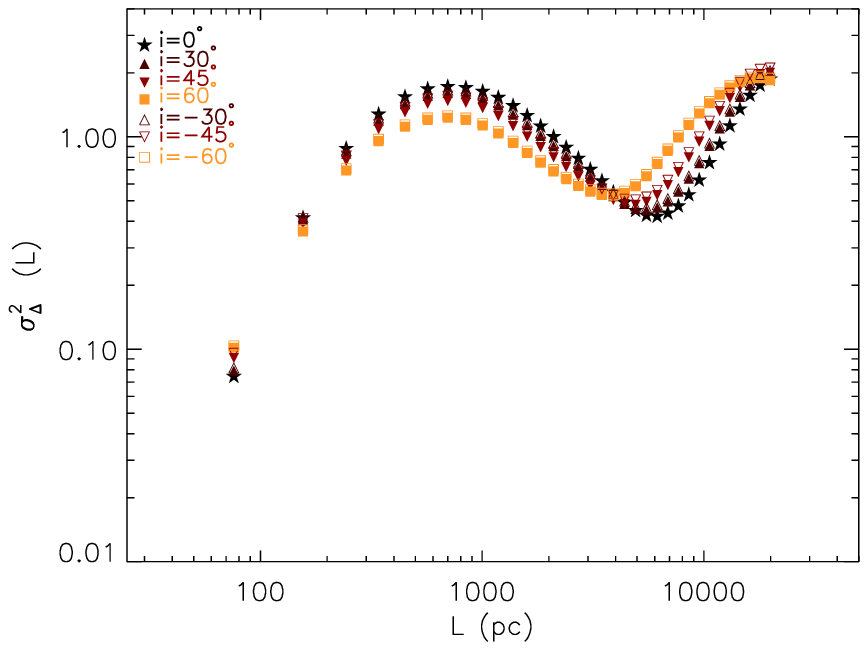}
\caption{$\Delta$-variance spectra for the simulated galaxy at the same snapshot ($t=t_{0})$ and resolution ($50$ pc), but for different inclination angles.}     
\label{figapp2}
\end{figure}

Here, we show how the inclination of a galaxy can impact the measured $\Delta$-variance spectrum. We use the same fiducial snapshot as discussed in Sect.~\ref{interpret} (at $t=t_{0}$, resolution=50 pc), but we consider several inclinations of the \ion{H}{i} disk, $i_\ion{H}{i}$. Fig.~\ref{figapp1} displays the \ion{H}{i} column density maps when the galaxy is inclined by $30^{\circ},45^{\circ},\text{ and } 60^{\circ}$ (top row) and $-30^{\circ}, -45^{\circ},\text{and }-60^{\circ}$ (bottom row). All maps were constructed using the original spatial resolution of 50 pc. The corresponding $\Delta$-variance spectra are shown in Fig.~\ref{figapp2}, along with the case where the galaxy is seen face-on. Fig.~\ref{figapp2} shows that the position of the bump and dip are relatively unaffected. When the galaxy is more inclined, \ion{H}{i} clouds and holes become more elongated, but their sizes, and consequently, their size distributions, are not excessively affected. Nonetheless, the depth of the dip is diminished at higher inclination. This simply reflects the effect of having to integrate the column density of the \ion{H}{i} gas along a larger distance, which then lowers the contrast with the ambient gas. A direct consequence of the dip becoming less pronounced is that the shape of the $\Delta$-variance at larger spatial scales becomes shallower.  

\section{Dependence of the shape of the $\Delta$-variance spectra on galactic stellar mass}\label{appb}

\begin{figure*}
\centering
\includegraphics[width=0.49\textwidth]{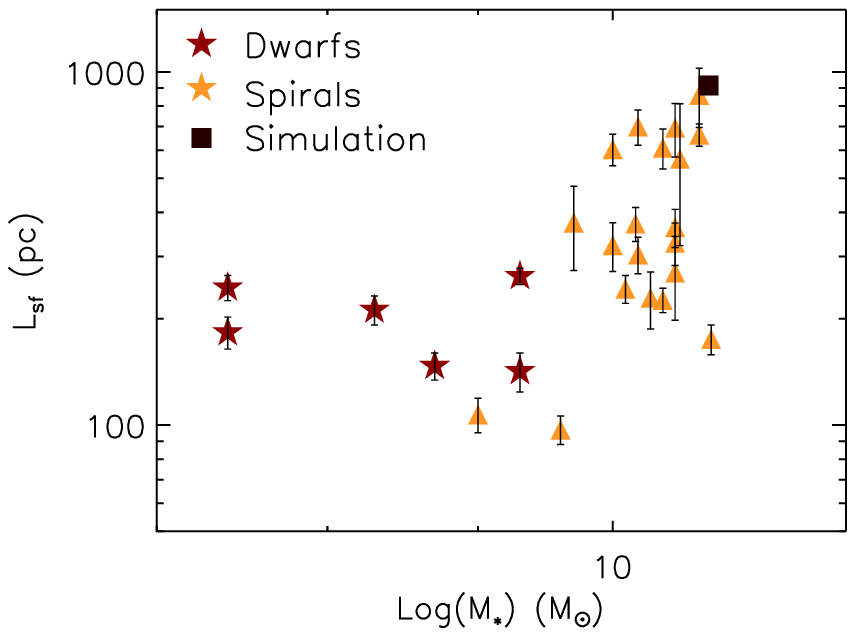}
\includegraphics[width=0.49\textwidth]{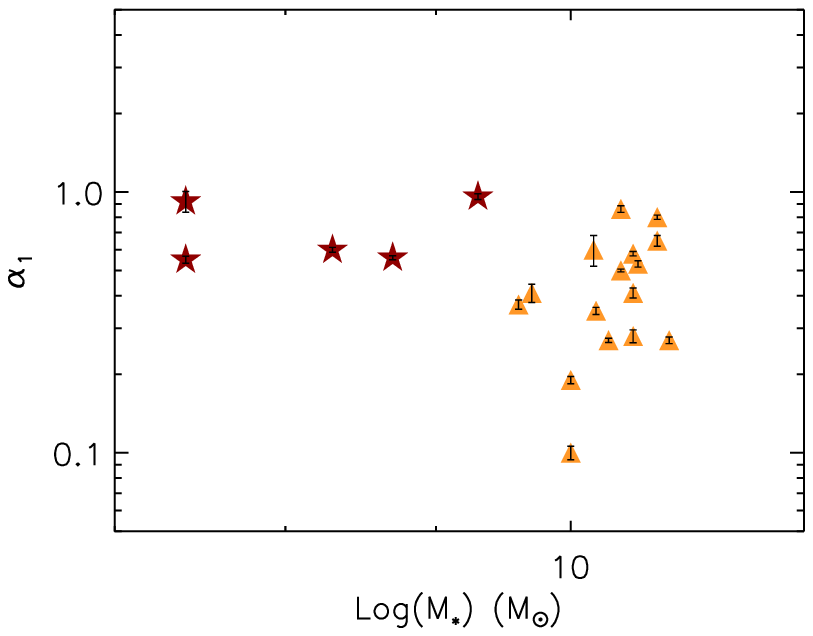}\\
\includegraphics[width=0.49\textwidth]{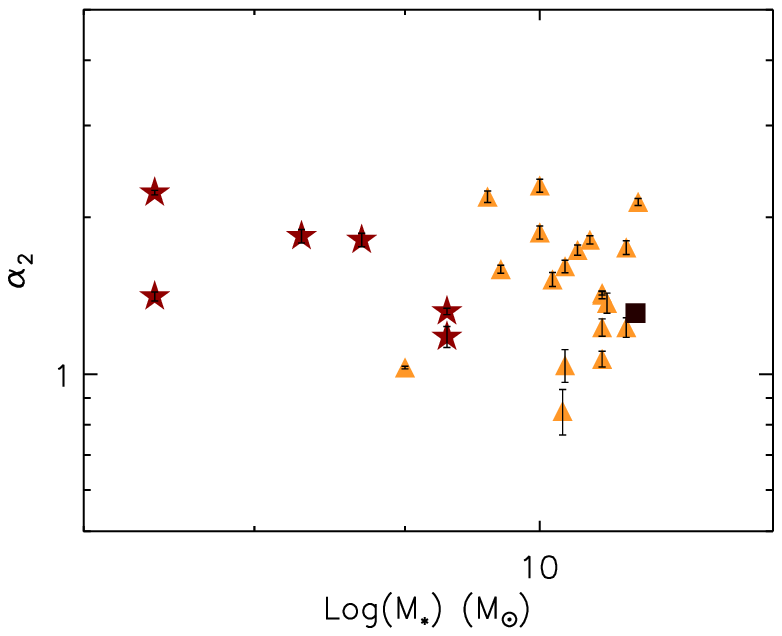}
\includegraphics[width=0.49\textwidth]{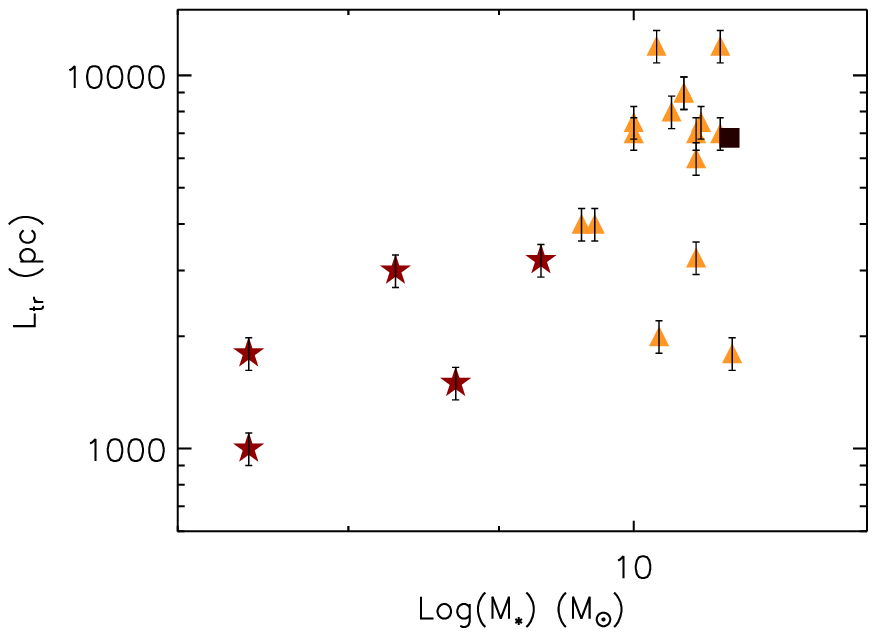}
\caption{Dependence of $L_{sf}$, $\alpha_{1}$, $\alpha_{2}$, and $L_{tr}$ on the galactic stellar mass, $M_{*}$.}     
\label{figapp3}
\end{figure*}

In \S.~\ref{correlation} we have discussed how the main characteristics of the $\Delta$-variance spectra (i.e., $L_{sf}$, $\alpha_{1}$, $\alpha_{2}$, and $L_{tr}$) depend on the galactic SFR. The question now is whether these quantities may also depend on the galactic stellar mass $M_{*}$ and if, when a relation exists between any of the spectral characteristics and M$_{*}$,  the correlation is tighter than with the SFR. Figure \ref{figapp3} displays the dependence of $L_{sf}$, $\alpha_{1}$, $\alpha_{2}$, and $L_{tr}$ on $M_{*}$. Clearly, all four quantities display the same dependence (or lack of it) on $M_{*}$ as on the SFR with a similar amount of scatter as with the SFR. For low values of $M_{*}$, the value of $L_{sf}$ is weakly dependent on $M_{*}$, whereas for high values of $M_{*}$, a positive correlation exists between the two quantities, and it is similar to the correlation found between $L_{sf}$ and the SFR (Fig.~\ref{fig10}). Neither $\alpha_{1}$ nor $\alpha_{2}$ display a dependence (or possibly a week anticorrelation) on $M_{*}$, which echoes the same trend as was found between them and the SFR (Fig.~\ref{fig11}). The transition scale between the two self-similar regimes, $L_{tr}$, displays a positive correlation with $M_{*}$, which is very similar to one observed between $L_{tr}$ and the SFR (Fig.~\ref{fig12}, top subpanel). The same degree of dependence of the $\Delta$-variance characteristics on $M_{*}$ and the SFR is a clear manifestation that galaxies in the THINGS sample follow the fundamental relation between $M_{*}$ and the SFR (e.g., Lara-L\'{o}pez et al. 2013).

\section{ $\Delta$-variance of an exponential disk }\label{appc}

\begin{figure*}
\centering
\includegraphics[width=0.3\textwidth]{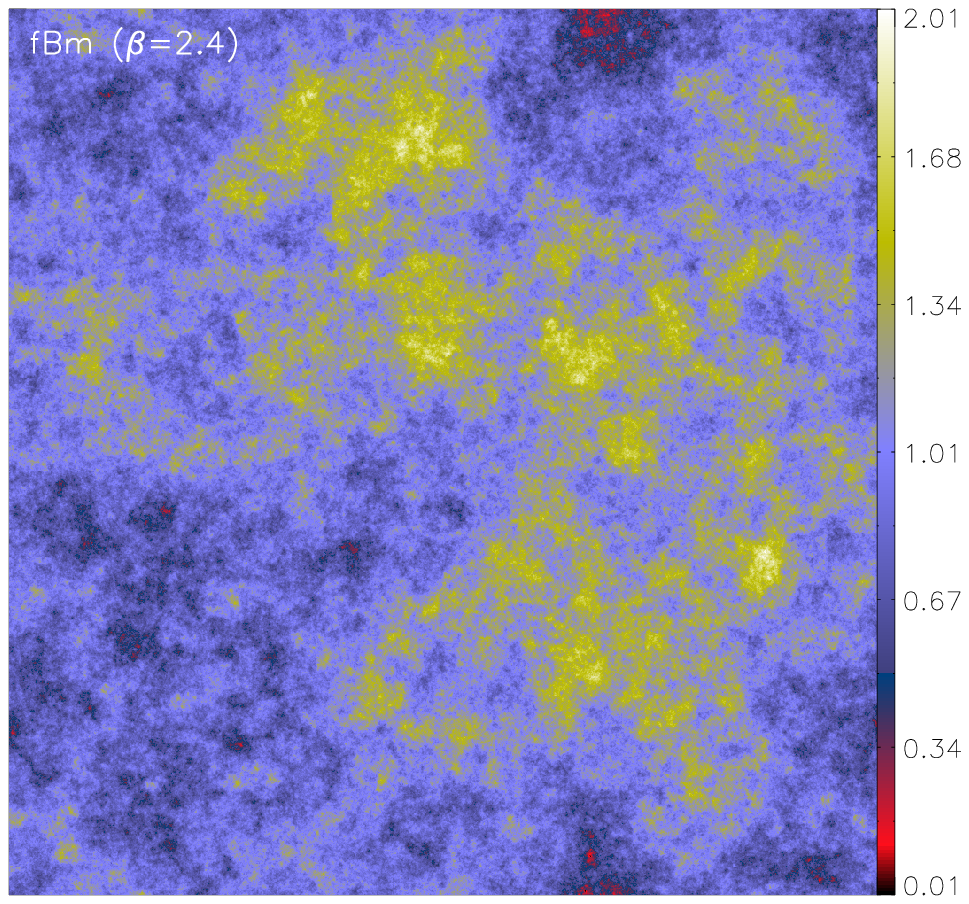}
\hspace{0.3cm}
\includegraphics[width=0.3\textwidth]{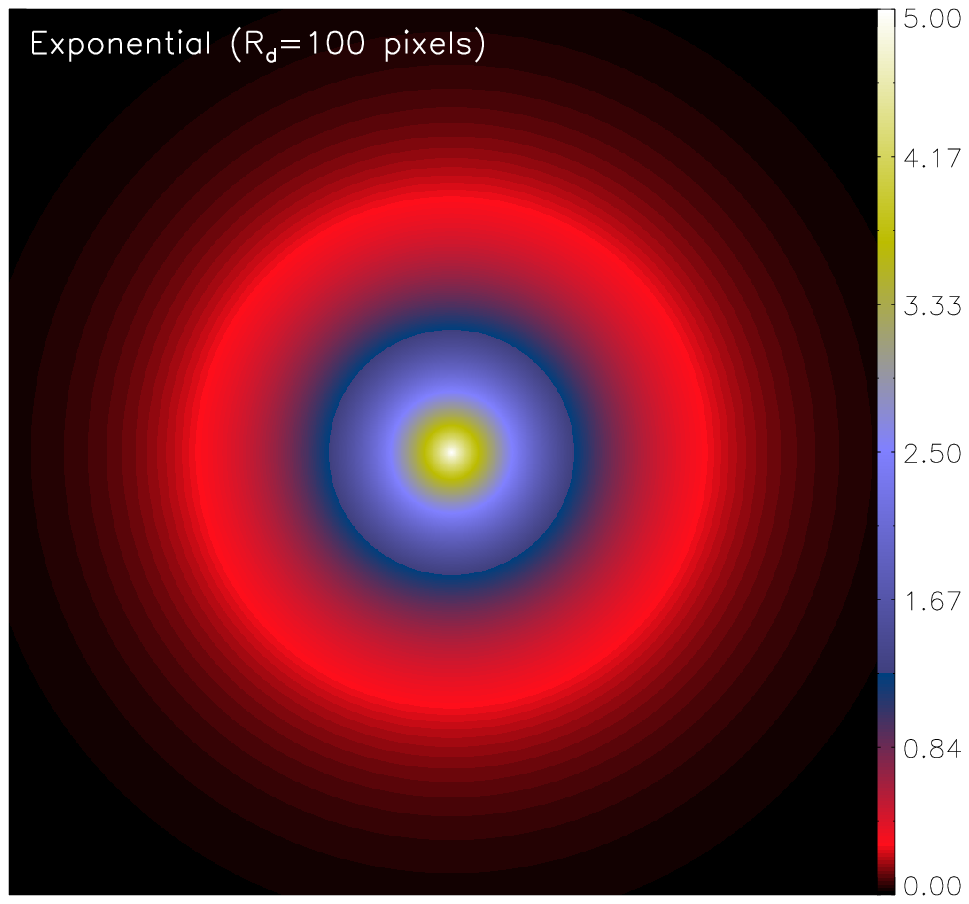}
\hspace{0.3cm}
\includegraphics[width=0.3\textwidth]{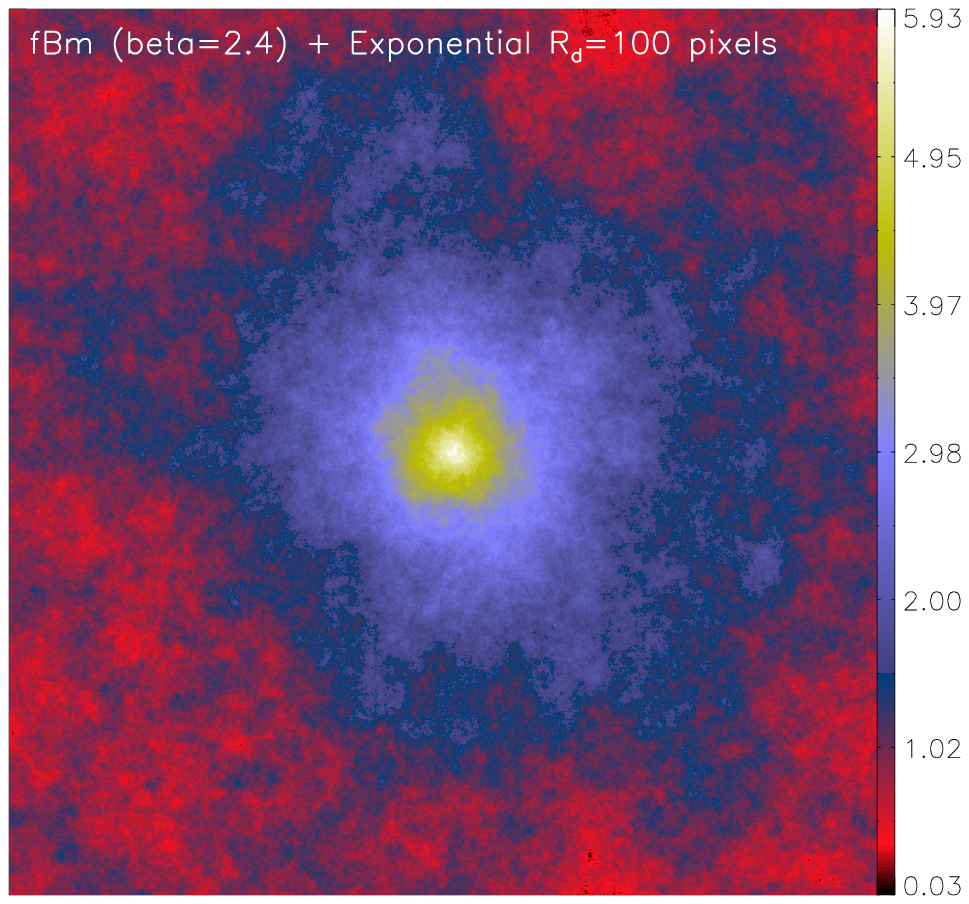}\\
\includegraphics[width=0.32\textwidth]{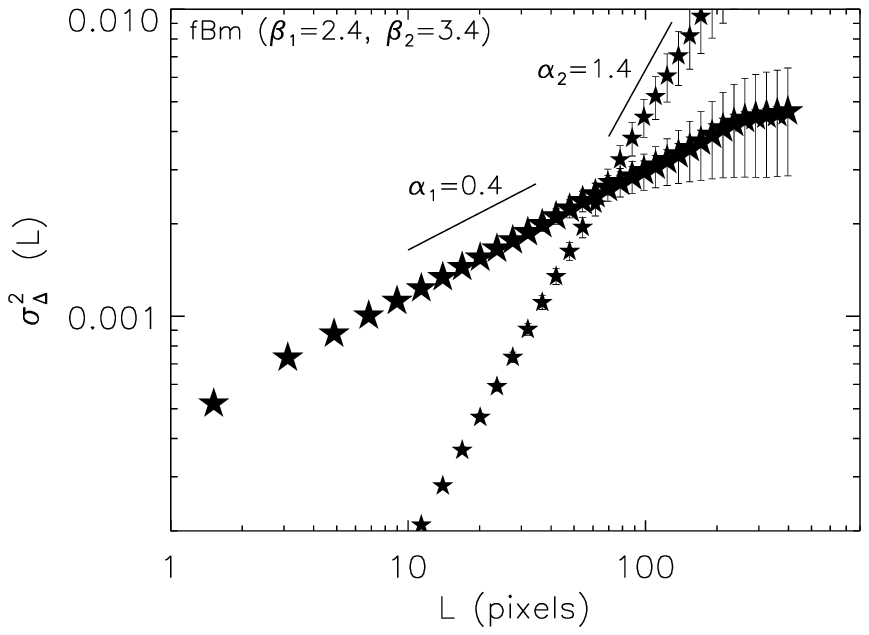}
\includegraphics[width=0.32\textwidth]{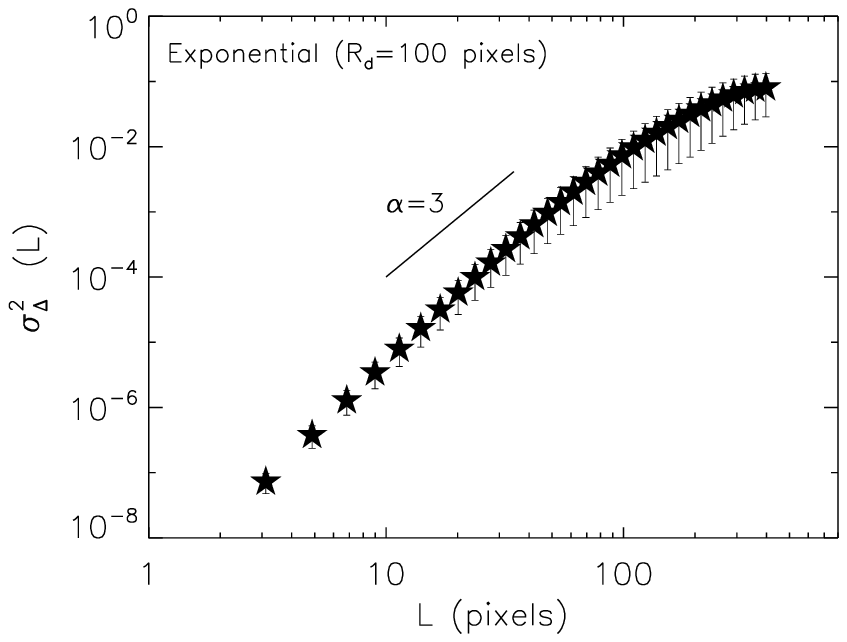}
\includegraphics[width=0.32\textwidth]{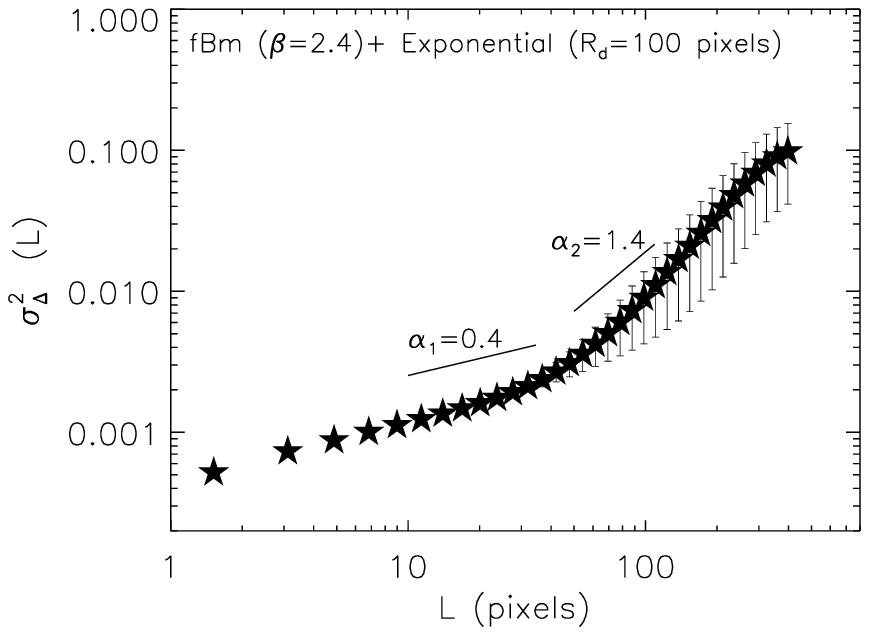}
\caption{Maps of a fBm with $\beta=2.4$ (top left), of a decaying exponential with a decay length scale of $100$ pixels (top mid), and of a mixed model with the same decaying exponential overlaid onto the fBm with $\beta=2.4$. The maps have a resolution of $1000\times1000$ pixels. All fBms used here are shifted to positive values by adding an arbitrary constant and normalized by their mean values. The bottom panels display the corresponding $\Delta$-variance spectra. An additional spectrum corresponding to an fBm with $\beta=3.4$ is shown in the bottom left subpanel (corresponding map not shown). The values of $\alpha=0.4, 1.4,$ and $3$ are not fits to the spectra, but are shown as a reference to guide the eye.}     
\label{figapp4}
\end{figure*}

Even though we discard the idea that the broken power-law $\Delta$-variance spectrum can be due to the existence of an exponential disk on the basis that an exponential profile is not observed for the THINGS galaxies in the 21 cm \ion{H}{i} emission line, we show here how the existence of an exponential disk can generate a broken power-law $\Delta$-variance spectrum. All images displayed in the top row of Fig.~\ref{figapp4} have a resolution of $1000\times1000$ pixels. In the top left subpanel, we show the image of an fBm with $\beta=2.4$. The corresponding $\Delta$-variance spectrum is displayed in the bottom left subpanel. The $\Delta$-variance spectrum in this case is well approximated with a power-law function whose exponent $\alpha \approx \beta-2 \approx 0.4$. The top and bottom middle subpanels display the case of an exponential disk with a profile proportional to $\exp(-R/R_{d}),$ where $R_{d}$ is the radial length scale, taken here to be 100 pixels. The corresponding $\Delta$-variance spectrum is very steep and can be described with a power law whose exponent is $\approx 3$. The scale corresponding to $R_{d}=100$ pixels does not appear in a very prominent way in the spectrum. The combination of an exponential disk (with $R_{d}=100$ pixels) overlaid on an fBm image (with $\beta=2.4$) is displayed in the top right subpanel in Fig.~\ref{figapp4}. The $\Delta$-variance spectrum of this composite image displays a broken power law (bottom right subpanel). On small scales, the spectrum is dominated by the contribution from the fBm, and on large scales, by the structure of the exponential disk. We caution that even if the total gas (\ion{H}{i}+H$_{2}$) profile is a perfect decaying exponential, the application of the $\Delta$-variance spectrum to the total gas column density could be misleading with respect to the true underlying structural and dynamical properties of each component of the gas if any of the neutral and molecular components does not possess an exponential profile.

\end{appendix}

\end{document}